\documentclass[a4paper,fleqn,usenatbib,useAMS]{mnras}

\usepackage{graphicx}
\usepackage{subfig}
\usepackage{cleveref}
\usepackage{pdflscape}
\usepackage{verbatim} 
\usepackage{amssymb}
\usepackage[intlimits]{amsmath}
\usepackage{multirow}
\usepackage[dvipsnames]{xcolor}

\usepackage{txfonts} 
\usepackage{tabulary} 
\usepackage[normalem]{ulem} 
\makeindex \citeindextrue
\newcommand{\rmu}{\,rad\,m$^{-2}$\,} 

\title[Faraday RM and polarization of 20 AGN jets]{Parsec-scale Faraday rotation and polarization of 20 active galactic nuclei jets}
\author[E. V. Kravchenko, Y. Y. Kovalev and K. V. Sokolovsky]{E. V. Kravchenko$^{1,2}$\thanks{Contact e-mail: \href{mailto:evgenia.v.kravchenko@gmail.com}{evgenia.v.kravchenko@gmail.com}}, Y. Y. Kovalev$^{1,3}$ and K. V. Sokolovsky$^{1,4,5}$\\
$^{1}$Lebedev Physical Institute, Astro Space Center, Profsoyuznaya 84/32, Moscow 117997, Russia\\
$^{2}$Pushchino Radio Astronomy Observatory ASC Lebedev, Pushchino 142290, Moscow region, Russia\\
$^{3}$Max-Planck-Institut f\"ur Radioastronomie, Auf dem H\"ugel 69, Bonn D-53121, Germany\\
$^{4}$IAASARS, National Observatory of Athens, 15236 Penteli, Greece\\
$^{5}$Sternberg Astronomical Institute, Moscow State University, Universitetsky pr. 13, Moscow 119991, Russia\\}

\date{Accepted 2017 January 4. Received 2017 January 1; in original form 2016 October 28}
\pubyear{2017}

\begin{document}
\label{firstpage}
\pagerange{\pageref{firstpage}--\pageref{lastpage}}
\maketitle

\begin{abstract}
We perform polarimetry analysis of 20 active galactic nuclei (AGN) jets using the Very Long Baseline Array (VLBA) at 1.4, 1.6, 2.2, 2.4, 4.6, 5.0, 8.1, 8.4, and 15.4~GHz. The study allowed us to investigate linearly polarized properties of the jets at parsec-scales: distribution of the Faraday rotation measure (RM) and fractional polarization along the jets, Faraday effects and structure of Faraday-corrected polarization images. Wavelength--dependence of the fractional polarization and polarization angle is consistent with external Faraday rotation, while some sources show internal rotation. The RM changes along the jets, systematically increasing its value towards synchrotron self-absorbed cores at shorter wavelengths.
The highest core RM reaches 16,900~\rmu\ in the source rest frame for the quasar 0952+179, suggesting the presence of highly magnetized, dense media in these regions. The typical RM of transparent jet regions has values of an order of a hundred \rmu.
Significant transverse rotation measure gradients are observed in seven sources.
The magnetic field in the Faraday screen has no preferred orientation, and is observed to be random or regular from source to source. Half of the sources show evidence for the helical magnetic fields in their rotating magnetoionic media.
At the same time jets themselves contain large--scale, ordered magnetic fields and tend to align its direction with the jet flow. The observed variety of polarized signatures can be explained by a model of spine--sheath jet structure.

\end{abstract}


\section{Introduction}

Magnetic fields play an important role in launching, acceleration and collimation of relativistic jets of active galactic nuclei (\citealt{blandford_znajek_77}). 
Theoretical models and numerical simulations suggest that a rotating accretion disk and ergosphere around a supermassive black hole will produce magnetized plasma outflows. Twisted magnetic fields will thread these flows \citep[e.g.][]{meier_etal01,vlahakis_konigl_04}.
While these models predict strong, ordered magnetic fields close to the central engine \citep[e.g.][]{blandford_znajek_77,2005ApJ...625...60N}, downstream in the jet these fields may be randomized, dissipated or tangled by different types of instabilities \citep[e.g.][]{1994MNRAS.267..629I,2004ApSS.293..117H}.
Despite numerous theoretical and observational studies, the exact geometry and main properties of the jet magnetic fields are still not known.

Faraday rotation and depolarization at radio wavelength \citep[e.g.][]{1997ApJ...483L...9U,zavala_taylor_03,osullivan_gabuzda_09,hovatta_etal12,osullivan_etal12,farnes_etal14,2016A...586A.117P}, together with relativistic and opacity effects complicate direct analysis of the jet properties. 
Multiwavelength full-Stokes very long baseline interferometric (VLBI) observations are needed to consider and account for these effects.

Complex linear polarization $\Pi$ is defined as \begin{equation}\Pi = Q + \mathrm{i}U = Pe^{2\mathrm{i}\chi} = mIe^{2\mathrm{i}\chi},\end{equation} where $I$, $Q$ and $U$ are the measured Stokes parameters, $P$ is the modulus of $\Pi$, $\chi$ is the observed electric vector position angle (EVPA), and $m$ is observed degree of polarization, $m = \sqrt{Q^2 + U^2}/I^2$.
The depolarization and Faraday rotation change the intrinsic polarization properties of a source:
the observed polarization angle is rotated with respect to the intrinsic one ($\chi_0$) by an amount ${\mathrm{RM}} = \mathrm{d}\chi/\mathrm{d}(\lambda^2)$, while the degree of polarization undergoes depolarization or even repolarization and become wavelength--dependent.
Here RM denotes Faraday rotation measure \citep{burn66} in radians per squared metre. RM is proportional to the magnetic field component parallel to the line of sight (LoS), $\bf{B_{\parallel}}$, and to the particle volume density $n_e$ along the path $\bf{dl}$:
\begin{equation}{\mathrm{RM}} \propto \int\limits_{\mathrm{LoS}} n_e \bf{B_{\parallel}} \cdot d\bf{l}.\end{equation}
In the simplest case, when the Faraday rotation occurs external to the jet magneto-ionic medium, the degree of polarization does not experience depolarization and RM can be defined as: \begin{equation}\chi = \chi_0 + {\mathrm{RM}} \lambda ^{2}.\end{equation}
Meanwhile, observed significant wavelength--dependent behaviour of the fractional polarization in the AGN jets \citep[e.g.][]{conway_etal_74} suggests presence of a depolarization and various Faraday effects, e.g. differential Faraday rotation and Faraday dispersion \citep{sokoloff_etal98}.
Thus, multiwavelength observations allow the determination of Faraday effects and RM, and hence enable one to study intrinsic orientation of the jet magnetic field. 

\begin{table*}
 \centering
 \begin{minipage}{175mm}
  \caption{Target sources and RM frequency ranges.}
\begin{tabular}{@{}cccclccccccccc@{}}
  \hline
    IAU name & Other & Redshift & Optical & Observational & \multicolumn{9}{c}{Observational frequencies (GHz)} \\
   (B1950.0) & name & & class & epoch &1.4&1.6&2.2&2.4&4.6&5.0&8.1&8.4&15.4 \\
 \hline
 0148$+$274  & & 1.260 &   QSO  & 2007--03--01&L&L&L&L&H&H&H&H&H\\
 0342$+$147 && 1.556 &   QSO  & 2007--06--01 &L&L&L&L&M&M&MH&MH&H\\
 0425$+$048 &OF~42& 0.517$^\mathrm{a}$ &   AGN  & 2007--04--30 &L&L&L&L&M&M&MH&MH&H\\
 0507$+$179 && 0.416 &   AGN  & 2007--05--03    &L&L&L&L&M&M&MH&MH&H\\
 0610$+$260 & 3C~154 & 0.580 &   QSO  & 2007--03--01  &-&-&-&-&-&-&-&-&-\\
 0839$+$187 && 1.272 &   QSO  & 2007--06--01   &L&L&L&L&H&H&H&H&-\\
 0952$+$179 && 1.478 &   QSO  & 2007--04--30 &L&L&L&L&M&M&MH&MH&H\\
 1004$+$141 && 2.707 &   QSO  & 2007--05--03    &L&L&L&L&H&H&H&H&-\\
 1011$+$250 && 1.636 &   QSO  & 2007--03--01 &-&-&-&-&M&M&M&M&-\\
 1049$+$215 && 1.300 &   QSO  & 2007--06--01   &L&L&L&L&M&M&MH&MH&H\\
 1219$+$285 &W~Comae& 0.161$^\mathrm{b}$ & BLL & 2007--04--30 &L&L&L&L&H&H&H&H&H\\
 1406$-$076 && 1.493 &   QSO  & 2007--05--03   &L&L&LM&LM&M&MH&H&H&H\\
 1458$+$718 &3C~309.1& 0.904 &   QSO  & 2007--03--01  &L&L&L&L&M&M&MH&MH&H\\
 1642$+$690 && 0.751 &   QSO  & 2007--04--30 &L&L&L&L&M&M&MH&MH&H\\
 1655$+$077 && 0.621 &   QSO  & 2007--06--01  &L&L&L&L&H&H&H&H&H\\
 1803$+$784 && 0.680 &   QSO  & 2007--05--03   &L&L&L&L&M&MH&MH&MH&H\\
 1830$+$285 && 0.594 &   QSO  & 2007--03--01  &-&-&M&M&M&M&-&-&-\\
 1845$+$797 &3C~390.3& 0.056 &   AGN  & 2007--06--01  &-&-&-&-&-&-&-&-&-\\
 2201$+$315 && 0.298 &   QSO  & 2007--04--30 &L&L&L&L&-&-&H&H&H\\
 2320$+$506 && 1.279 &   QSO  & 2007--05--03   &L&L&L&L&-&H&H&H&H\\
\hline
\end{tabular}
\medskip

The value of RM was estimated at different frequency intervals separately due to the considered linear dependence of EVPA vs. $\lambda^2$, namely Low (1.4~GHz to 2.4~GHz), Middle (2.2~GHz to 5.0~GHz) and High (4.6~GHz to 15.4~GHz).
Redshifts and optical classes are from \citet{veron_veron_10}. VLBI position can be found in \citet{sokolovsky_etal11}. 
Observational frequencies are given in GHz.
$^\mathrm{a}$ -- Spectroscopic redshift obtained by \citet{afanasiev_etal03}. $^\mathrm{b}$ -- Photometric redshift, see \citet{finke_etal08}.
\label{t_1}
\end{minipage}
\end{table*}

The most likely source of Faraday rotation is the ionized medium, located in close proximity to the jet. 
It might be dense gas clouds, surrounding AGN jets and producing large observed RMs \citep{1999MNRAS.308..955J,2010AA...518A..33M}.
However, outer jet layers \citep{zavala_taylor_04} or the jet sheath \citep{asada_etal02,2007ApJ...662..835M} are more plausible. 
Many pieces of observational evidence favour this suggestions. 
One of these is the tendency for opaque VLBI cores increase their RM values towards shorter wavelengths \citep[e.g.][]{zavala_taylor_04,jorstad_etal07,osullivan_gabuzda_09,algaba13}. 
Owing to the opacity effects (known as ``core shift''), this behaviour is expected, since the $\tau\approx1$ surface moves towards the central engine with increasing frequency, where the magnetic field is more ordered and stronger \citep{2011MNRAS.418L..79T,zamaninasab_etal14}. 
Assuming the outer jet layer is a source of Faraday rotation, lower RMs at longer wavelengths may arise from a decrease in electron density and magnetic field strength with distance from the central engine.
Another sign of the outer jet media to be responsible for Faraday rotation is observed RM gradients across the jets \citep{asada_etal02,2008ApJ...681L..69G,osullivan_gabuzda_09,zamaninasab_etal13}, which can be a signature of the helically-shaped jet magnetic field \citep{blandford93,broderick_mckinney_10}.
Some sources exhibit RM variability during their active state, accompanied by emergence of a new jet feature \citep{taylor_00,lico_etal14,giroletti_etal15,2016MNRAS.462.2747K}. 
In this aspect, RM variations may reflect changes in the accretion rate, thus are tightly connected with the jet.
Rapid variations in RM downstream in the jet \citep{asada_etal08} also favour this suggestion.

Despite a vast amount of observed studies, it is not clear precisely what is the source of Faraday rotation in AGN jets.
In this work we investigate the Faraday effects and study the magnetic field structure in 20 AGN jets at parsec-scales using full polarization Very Long Baseline Array (VLBA) observations conducted in the frequency range from 1.4 to 15.4~GHz. 
The study complements the works by \citet{farnes_etal14}, \citet{2016A...586A.117P} and \citet{2016ApJ...825...59A} who investigated sources at kpc--scale in a comparable frequency range of 1~GHz to 15~GHz, as well as the numerous individual sources \citep[e.g.][]{asada_etal08,asada_etal10} and sample studies \citep[e.g.][]{zavala_taylor_03,osullivan_gabuzda_09,hovatta_etal12}.
Application of the VLBI technique allows us (i) to minimize the level of depolarization within the telescope beam compared to single-dish and connected-interferometer observations \citep[e.g.][]{osullivan_etal12} and (ii) study separately jet regions with different synchrotron opacity.

This paper is structured as follows: observations and data reduction are described in Section~\ref{s:obs}.
In Section~\ref{s:frm_res} we give our results of Faraday effects study and Faraday rotation measurements, the description of the proposed method for the reconstruction of the spatial magnetic field geometry and results of its application. 
Section~\ref{sum} summarizes our findings.
Through the paper the model of flat $\Lambda$CDM cosmology was used with $H_0$ = 68~km s$^{-1}$, $\Omega_M$ = 0.3, and $\Omega_{\Lambda}$ = 0.7 \citep[][]{planck_15}. 
The position angles are measured from north through east.
The spectral index $\alpha$ is defined as $S \propto \nu^{\alpha}$, where $S$ is the flux density, observed at frequency $\nu$.

\begin{figure}
 \centering
 \includegraphics[width=0.34\textwidth]{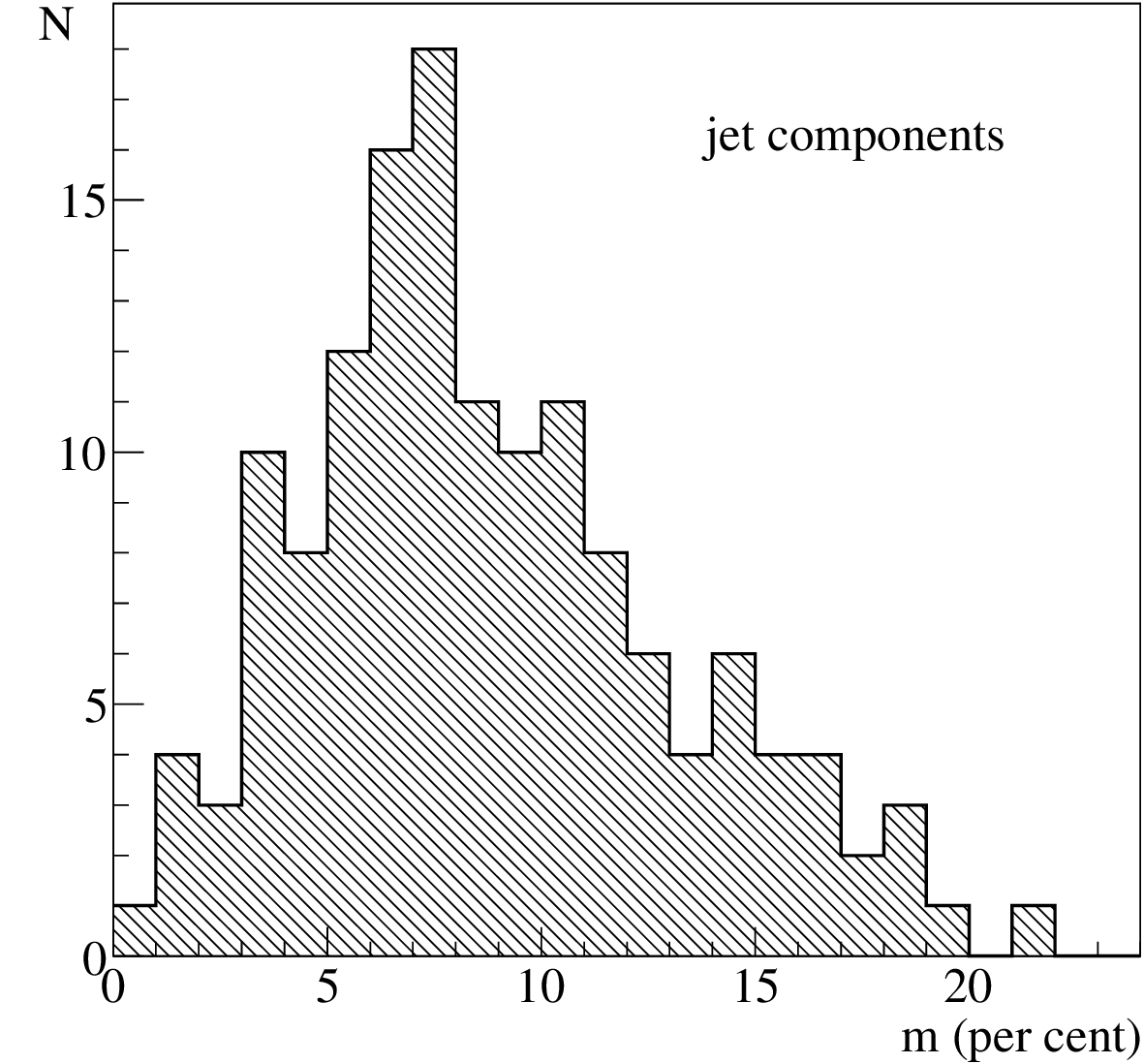}
 \includegraphics[width=0.34\textwidth]{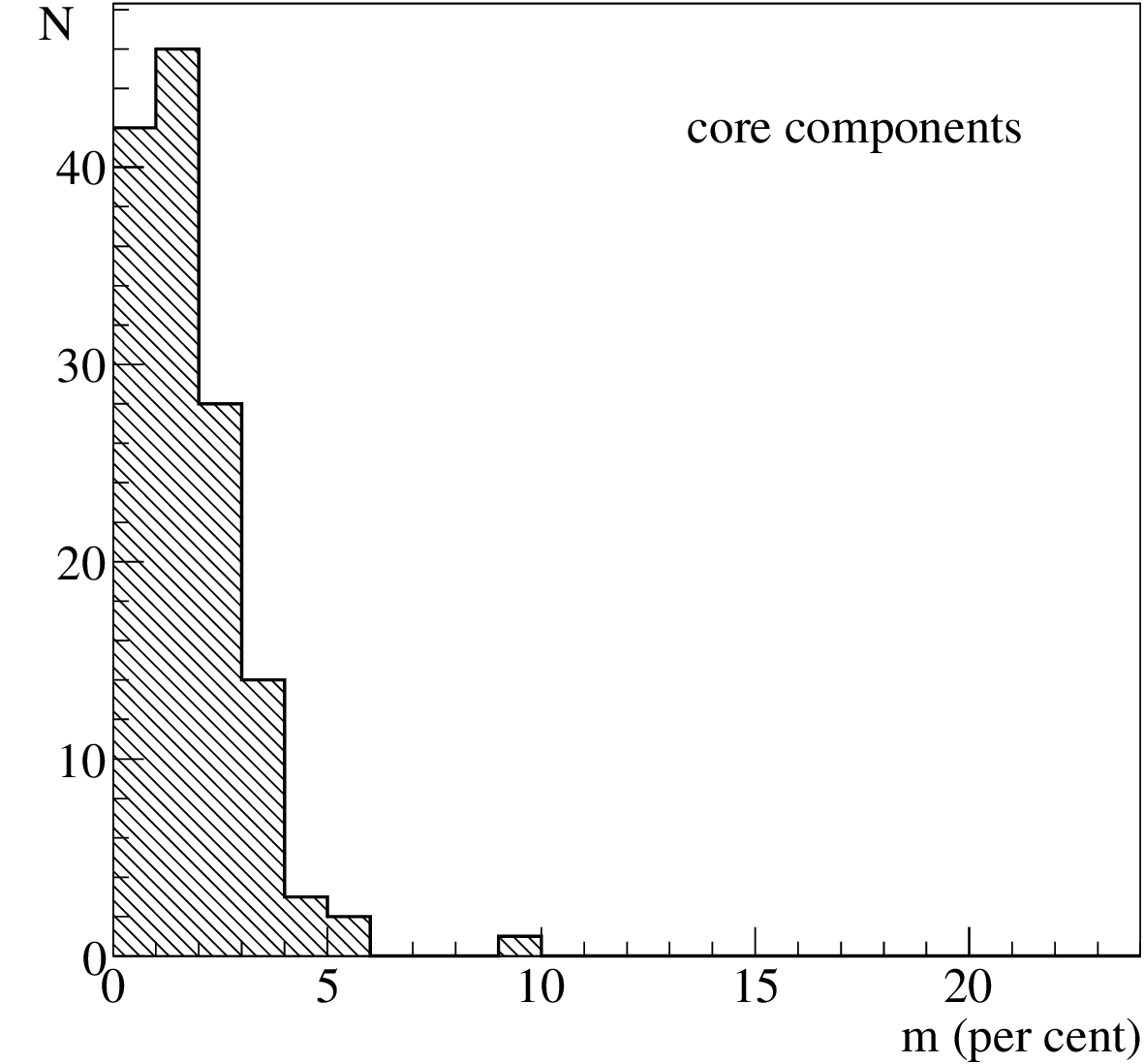}\\
 \caption{Distribution of measured polarization degree for all sources in the core and jet components.  The number of appearance of each source in the Figure corresponds to the number of frequency bands, where polarized flux was detected, and can reach a maximum value of nine bands.
 The sources, with no significant polarized flux detected, are not presented. The definition 
 of core and jet components are given in Section~\ref{s:obs1}.
 \label{pistat}}
\end{figure}

\begin{table}
  \caption{Galactic rotation measures.\label{gal_rm}}
  \begin{center}
  \begin{tabular}{cr@{}c@{}lr@{}c@{}lr@{}c@{}l}
  \hline
   Source&\multicolumn{3}{c}{NVSS}&\multicolumn{3}{c}{Optically}&\multicolumn{3}{c}{Averaged}\\
   &\multicolumn{3}{c}{}&\multicolumn{3}{c}{thin regions}&\multicolumn{3}{c}{over map}\\
  \hline
0148$+$274&$-$82&$\pm$&3&$-$90&$\pm$&4&$-$91&$\pm$&5\\
0342$+$147&6&$\pm$&8&14&$\pm$&4&13&$\pm$&4\\
0425$+$048&35&$\pm$&11&42&$\pm$&4&42&$\pm$&4\\
0507$+$179&$-$29&$\pm$&6&$-$65&$\pm$&4&$-$53&$\pm$&4\\
0610$+$260&44&$\pm$&7&&$-$&&&$-$&\\
0839$+$187&38.2&$\pm$&1.7&35&$\pm$&4&32&$\pm$&5\\
0952$+$179&$-$5.5&$\pm$&1.0&$-$8&$\pm$&4&$-$9&$\pm$&5\\
1004$+$141&12&$\pm$&2&3&$\pm$&4&7&$\pm$&5\\
1011$+$250&$-$36&$\pm$&12&&$-$&&&$-$&\\
1049$+$215&8.8&$\pm$&1.9&$-$2&$\pm$&4&0&$\pm$&5\\
1219$+$285&15&$\pm$&13&$-$1&$\pm$&4&2&$\pm$&4\\
1406$-$076&$-$10&$\pm$&7&$-$2&$\pm$&4&$-$2&$\pm$&5\\
1458$+$718&63.1&$\pm$&0.9&66&$\pm$&4&40&$\pm$&5\\
1642$+$690&$-$11.5&$\pm$&1.3&$-$5&$\pm$&4&$-$5&$\pm$&5\\
1655$+$077&14.0&$\pm$&1.9&41&$\pm$&4&40&$\pm$&5\\
1803$+$784&$-$66.5&$\pm$&0.9&$-$62&$\pm$&4&$-$65&$\pm$&6\\
1830$+$285&30&$\pm$&3&&$-$&&19&$\pm$&8\\
1845$+$797&$-$5&$\pm$&2&&$-$&&&$-$&\\
2201$+$315&$-$96&$\pm$&3&$-$108&$\pm$&5&$-$103&$\pm$&7\\
2320$+$506&$-$78&$\pm$&5&$-$55&$\pm$&5&$-$57&$\pm$&10\\
\hline
\end{tabular}
\end{center}
 RM values are quoted in \rmu. NVSS RM are obtained by \citet{taylor_etal09}.
 RMs given in the third column are estimated under assumption that the electric vector position 
 angle maintains its direction in the optically thin regions over the whole frequency range.
 Values in the last column are averaged over the whole RM maps across 1.4~GHz to 2.4~GHz range.
 \end{table}

\section{Observations and data reduction}\label{s:obs}
\subsection{VLBA Observations}\label{s:obs1}

We selected twenty sources from the geodetic VLBI database \citep{fey_charlot_97,petrov_etal09}, showing large frequency-dependent core shifts at cm-band and having bright jet features suitable as reference points for aligning images obtained at different frequencies. 
\citet{sokolovsky_etal11} used this sample to investigate the core shift effect concluding that cores of all the selected sources are well described by the \citet{blandford_konigl_79} model with the shift being $\propto\nu^{-1}$.
Accordingly, we define the radio core (or simply the core) at a given frequency as the surface with optical depth $\tau\approx1$ being the apparent base of the jet \citep{marscher08}. 
See also duscussion in sect.~\ref{s:3dmet}.
Targets were observed with the Very Long Baseline Array during four sessions in March--June 2007 \citep{sokolovsky_etal11} quasi-simultaneously at 9 frequencies in the dual-polarization mode. 
According to the IEEE nomenclature, L, S, C, X, and $\mathrm{K}_\mathrm{u}$ (noted as U below) 
frequency bands were used. Each band consists of four 8 MHz-wide frequency channels (IFs) per polarization. 
L, S, C and X bands were split into two sub-bands (two IFs in each sub-band) centered at 1.41, 1.66, 2.28, 2.39, 4.60, 5.00, 8.11, and 8.43 GHz respectively. 
In the following analysis these sub-bands were processed independently. 
At U-band all four IFs were combined into a single sub-band centered at 15.4 GHz in order to provide sensitivity similar to that of individual sub-bands at lower frequencies.
This division into sub-bands allows us to decrease the effect of signal depolarization over the bandwidth. 
The resulting width of a channel is 32~MHz at 15.4~GHz and 16 MHz at other frequencies.
The total on-source observing time for each target was about one hour per band.

The dual-polarization observations recorded data at each station with 2-bit sampling and a total rate of 256~Mbps.
The data were correlated at Socorro (NM, USA) with averaging time of 2~s.

\subsection{Data Processing}

Data calibration and imaging, as well as model fitting were done in a standard manner for each  
sub-band independently using the Astronomical Image Processing System (AIPS, \citealt{bridle_greisen_94}) and 
Difmap \citep{shepherd_etal94,shepherd_97}. Calibration steps are described by \citet{sokolovsky_etal11} and
the AIPS Cookbook\footnote{\url{http://www.aips.nrao.edu/cook.html}} in details. 
The amplitude calibration accuracy in the 1.4~GHz to 15~GHz frequency range is estimated to be $\sim 5$~per~cent \citep{sokolovsky_etal11}.
The errors in Stokes $I$ flux density, $\sigma_\mathrm{I}$, and linearly polarized flux density, $\sigma_\mathrm{P}$, are calculated by considering absolute calibration uncertainties and the map rms noise.

Polarization leakage parameters of the antennas (D-terms) were determined within AIPS with the task LPCAL \citep{1995AJ....110.2479L}. 
D-terms were examined for time stability \citep{roberts_etal94,gomez_etal02} using the two polarizations of each IF at all antennas.
Corrections were calculated at L- and S-bands by averaging and minimizing the R-L phase offset over four epochs; 
at C- and X-bands by comparing the R-L offset over one year time interval; and at U-band by connecting our observations with the D-term phase solutions from 10-year MOJAVE program data \citep{lister_etal09,lister_etal13}. 
Thus, accuracy of leakage parameters approaches $2^\circ$ (about 1 per~cent) at L- and S-bands and $1^\circ$ (about 0.5 per~cent) at other frequencies. 

Absolute EVPA calibration at L- and S-bands was done using the EVPAs of 3C~286 and at C-, X-, and U-bands using 
the EVPAs of 3C~273 and 4C~$+$39.25 from the VLA Monitoring 
Program\footnote{\url{http://www.vla.nrao.edu/astro/calib/polar/}} \citep{taylor_myers_00}, 
the University of Michigan Radio Astronomy Observatory monitoring program, and from the MOJAVE program. 
Absolute calibration error is assessed as the difference in corrected EVPA between pairs of sub-bands, 
and at U-band it is assessed as the deviation of the corrected EVPA from the EVPA of calibrators taken from the monitoring programs.
The resulting absolute EVPA calibration error is $4^\circ$ and $2^\circ$ for L-, S- and C-, X-, U-bands respectively.
Final calibration error of absolute vector position angle comprises errors from instrumental and absolute calibrations and is estimated to be $5^\circ$ at low frequencies (L- and S-bands), $2.5^\circ$ at middle frequencies (C- and X-bands) and $2^\circ$ at U-band.

The uncertainties in fractional polarization, $\sigma_\mathrm{m}$, and EVPA, $\sigma_{\chi}$, are calculated from error propagation theory:
\begin{equation}
\sigma_m = \frac{1}{I}\sqrt{\sigma_\mathrm{P}^2+\sigma_\mathrm{I}^2\frac{P^2}{I^2}},
\end{equation}
and 
\begin{equation}
\sigma_{\chi} = \frac{\sqrt{\sigma_\mathrm{U}^2 Q^2+\sigma_\mathrm{Q}^2 U^2}}{{2(Q^2+U^2)}},
\end{equation}
where $\sigma_\mathrm{Q}$ and $\sigma_\mathrm{U}$ are the uncertainties in the $Q$ and $U$ Stokes flux densities.
The contribution of the instrumental polarization leakage to the Stokes $I$ and linear polarization maps~\citep{roberts_etal94,hovatta_etal12} is considered as
\begin{equation}
\sigma_{Dterm} = \frac{\sigma_{\Delta}}{\sqrt{N_\mathrm{ant} N_\mathrm{IF} N_\mathrm{scan}}}\sqrt{I^2+(0.3 I_\mathrm{peak})^2},
\end{equation}
where $\sigma_{\Delta}$ is the scatter of D-terms, $N_\mathrm{ant}$ is the number of antennas, $N_\mathrm{IF}$ is the number of IFs, $N_\mathrm{scan}$ is the number of VLBA scans on polarization calibrator with independent parallactic angles, and $I_\mathrm{peak}$ is the peak Stokes $I$ flux density of the map. 
The scatter of D-terms in our data is estimated to be 0.01, 0.005 and 0.002 for 1.4~GHz to 2.4~GHz, 4.6~GHz to 8.4~GHz and 15.4~GHz frequencies. 
The number of antennas is 10, the number of scans on 3C~286, OJ~287 and 4C~$+$39.25 is 3, and the number of IFs is 4 for 15.4~GHz and 2 for other frequencies. 
This error is added to the all corresponding rms errors in quadrature.

\begin{table*}
\begin{minipage}{175mm}
\caption{Measured Stokes $I$ flux density and polarization degree ($m$) of Gaussian core and jet components, peak Stokes $I$ flux density ($I_\mathrm{peak}$) and rms noise ($\sigma_\mathrm{I}$) in the image. The full table is available online. \label{frac_p1}}
\begin{center}
\begin{tabular}{@{}lcr@{$\pm$}lr@{$\pm$}lr@{$\pm$}lr@{$\pm$}lr@{$\pm$}lc@{}}
\hline
Source&Frequency&\multicolumn{2}{c}{$m_\mathrm{core}$}&\multicolumn{2}{c}{$I_\mathrm{core}$}&\multicolumn{2}{c}{$m_\mathrm
{jet}$}&\multicolumn{2}{c}{$I_\mathrm{jet}$}&\multicolumn{2}{c}{$I_\mathrm{peak}$}&$\sigma_\mathrm{I}$\\
&(GHz)&\multicolumn{2}{c}{(per~cent)}&\multicolumn{2}{c}{(mJy)}&\multicolumn{2}{c}{(per~cent)}&\multicolumn{2}{c}{(mJy)}&\multicolumn{
2}{c}{(mJy beam$^{-1}$)}&(mJy beam$^{-1}$)\\
\hline
0148$+$274
 & 1.4 & 1.62 & 0.03 & 541.01 & 0.24 & 5.34 & 0.03 & 590.93 & 0.22 & 590.93 & 0.25 & 0.14\\
 & 1.7 & 2.78 & 0.05 & 498.41 & 0.18 & 5.65 & 0.05 & 493.27 & 0.17 & 519.92 & 0.18 & 0.13\\
 & 2.3 & 3.52 & 0.07 & 399.89 & 0.31 & 5.44 & 0.06 & 379.67 & 0.25 & 465.08 & 0.31 & 0.20\\
 & 2.4 & 3.24 & 0.07 & 389.79 & 0.23 & 5.11 & 0.06 & 379.78 & 0.20 & 464.93 & 0.20 & 0.22\\
 & 4.6 & 1.37 & 0.06 & 357.65 & 0.22 & 3.89 & 0.08 & 219.84 & 0.29 & 357.65 & 0.22 & 0.17\\
 & 5.0 & 1.19 & 0.04 & 350.04 & 0.12 & 3.42 & 0.08 & 199.79 & 0.20 & 350.04 & 0.12 & 0.15\\
 & 8.1 & 1.73 & 0.08 & 336.49 & 0.09 & 3.28 & 0.13 & 112.71 & 0.08 & 336.49 & 0.09 & 0.16\\
 & 8.4 & 1.92 & 0.07 & 354.87 & 0.10 & 3.15 & 0.13 & 110.44 & 0.06 & 354.87 & 0.10 & 0.14\\
 & 15.4 & 1.31 & 0.07 & 273.49 & 0.13 & \multicolumn{2}{c}{$<$2.23} & 32.00 & 0.09 & 294.34 & 0.12 & 0.15\\
0342$+$147
 & 1.4 & \multicolumn{2}{c}{$<$0.46} & 182.47 & 0.13 & 5.36 & 0.26 & 96.27 & 0.11 & 182.47 & 0.13 & 0.16\\
 & 1.7 & 0.93 & 0.13 & 197.34 & 0.15 & 7.06 & 0.30 & 74.04 & 0.14 & 197.34 & 0.15 & 0.15\\
 & 2.3 & 0.46 & 0.12 & 246.47 & 0.15 & 7.86 & 0.44 & 58.91 & 0.12 & 246.47 & 0.15 & 0.20\\
 & 2.4 & 0.56 & 0.15 & 256.57 & 0.12 & 8.11 & 0.54 & 57.03 & 0.17 & 256.57 & 0.11 & 0.21\\
 & 4.6 & 1.77 & 0.08 & 274.57 & 0.19 & 6.80 & 0.54 & 32.35 & 0.12 & 295.37 & 0.19 & 0.17\\
 & 5.0 & 1.80 & 0.06 & 275.70 & 0.25 & 8.91 & 0.57 & 30.08 & 0.14 & 299.11 & 0.25 & 0.14\\
 & 8.1 & 2.49 & 0.08 & 293.40 & 0.26 & 9.87 & 0.74 & 19.74 & 0.12 & 306.44 & 0.27 & 0.17\\
 & 8.4 & 2.68 & 0.06 & 294.70 & 0.16 & 9.54 & 0.61 & 19.82 & 0.09 & 318.33 & 0.16 & 0.15\\
 & 15.4 & 4.19 & 0.07 & 290.99 & 0.27 & 17.55 & 2.18 & 8.80 & 0.08 & 292.04 & 0.26 & 0.16\\
\hline
\end{tabular}
\end{center}

\medskip
 The core and jet regions correspond to the positions of $\tau\sim1$ and $\tau\ll1$ jet components respectively, which are presented in the sources through all observed frequencies. Positions of these components are indicated by colour circles in Fig.~\ref{mfcore}.
 Upper limit on fractional polarization is three times the averaged pixel value, located at the position of corresponding component on a linear polarization map.
\end{minipage}
\end{table*}

Depolarization due to non-zero bandwidth \citep{gardner_whiteoak_66} is characterized by the EVPA variation across the receiving band, $\Delta\nu$:
\begin{equation}
\Delta \chi=-{\mathrm{RM}}\lambda^2{\frac{\Delta\nu}{\nu}}. 
\end{equation}
It is noticeable at lower frequencies only, where it amounts to $\approx1\fdg5$ for the typical RM values in AGN jets~\citep[e.g.][]{zavala_taylor_03}.
At higher frequencies the resulting EVPA smearing is well within calibration errors.

To model the structure of the source, a number of circular or elliptical two-dimensional Gaussian components were fitted to the fully calibrated visibility data in the u--v plane, using the task \textit{modelfit} in DIFMAP. 
The VLBI core was identified as the bright component at the apparent jet base at each frequency and further we refer to it as ``core component". 
To analyse properties of the optically thin jet, we choose single, bright jet component, that could be identified through all frequencies.
Hereinafter we refer to it as ``jet component".
More details of model fitting are given in \citet{sokolovsky_etal11}.

\subsection{RM Map Reconstruction Procedure}
\label{s:rm_reconstr}

The dependence of EVPA on $\lambda^{2}$ was considered at `Low' (1.4~GHz to 2.4~GHz), `Middle' (2.2~GHz to 5.0~GHz) and `High' (4.6~GHz to 15.4~GHz) frequency intervals separately for all sources because of sparse observational frequency coverage, changing opacity in core region, and presence of Faraday effects. 
The RM is estimated as a linear slope of EVPA with $\lambda^2$ at these frequency ranges, presented in Table~\ref{t_1}.

To match the resolution of images obtained at different frequencies, the images were restored with the beam corresponding to uniform weighting of the data at the lowest frequency of the Low, Middle, or High range, respectively (see Table~\ref{t_1}).
The self-calibration technique used to obtain high quality images unfortunately loses information about absolute position of a source. 
Therefore, the positions of optically thin jet components were used to align images at different frequencies. See details of this technique, tests for possible biases and comparison with other methods in, e.g. \citet{kovalev_etal08,pushkarev_etal12,kutkin_etal14,2014MNRAS.441.1899F}. 

We used only pixels with polarized flux density stronger than three times the polarization error (see \citealt{taylor_zavala_10} for a thorough discussion and references). 
The resolution of the $n\pi$-wrap problem was done by minimizing the reduced $\chi^2$. Finally, we used the 95~per~cent confidence level for the respective number of degrees of freedom.

\subsection{Galactic and Extragalactic RM}
\label{s:gal_rm}
Observed Faraday rotation occurs at three main locations: (i) the jet itself and its immediate vicinity, (ii) the intergalactic and (iii) Galactic medium. 
All values discussed in this work refer to the first environment, while other two act as foreground Faraday screens and for further analysis their contribution must be removed.

The presence of magnetic fields in the intergalactic medium will cause Faraday rotation \citep[see ][]{akahori_ryu_10,akahori_ryu_11,bernet_etal12}. 
The origin of these fields outside galaxy clusters, as well as their distribution in the cosmic web or redshift dependence is not well understood yet. 
Thus, the contribution of the extragalactic RM component for every source is not reliably estimated and we did not account it in this analysis.

Estimations of RM resulting from plasma in our Galaxy have been performed by, e.g.
\citet{taylor_etal09,mao_etal10,vaneck_etal11,jansson_farrar_12,opperman_etal12,han13}, focusing on reconstruction of Galactic magnetic fields based on analysis of polarized properties of extragalactic radio sources, Galactic pulsars, recombination lines and ionized gas. 
Estimated values of Galactic RM agree well between these works.
Therefore we substracted RMs obtained by \citet{taylor_etal09}\footnote{\url{http://www.ucalgary.ca/ras/rmcatalogue}}  from the observed polarization angles, since they provide RM estimates for all sources in our sample (see Table~\ref{gal_rm}).
We note that this correction is not needed for the RM gradient analysis we perform.

Average uncertainties in these Galactic RMs amount to a few to tens of \rmu\ (see Table~\ref{gal_rm}).
In Table~\ref{rm_1} we show values of EVPA rotation for different radio frequencies for ${\mathrm{RM}}=1$~\rmu. 
An error of 5~\rmu\ in Galactic RM affects EVPA at 1.4~GHz to 2.4~GHz by $13.0^\circ$ to $4.5^\circ$ respectively.
We note that these errors might significantly corrupt the $\chi_0$ values at low frequencies.

Section~\ref{s:gfrm} contains a discussion of the observed EVPA in optically thin components in our sources, which enables us to estimate foreground RMs and compare them with values reported by \citet{taylor_etal09}.

\begin{table*}
 \centering
 \begin{minipage}{120mm}
  \caption{EVPA rotation, $\Delta_{evpa}$, due to rotation measure of 1~\rmu.}
  \begin{tabular}{@{}cccccccccc@{}}
  \hline
   Frequency (GHz) & 1.4 & 1.6 & 2.2 & 2.4 & 4.6 & 5.0 & 8.1 & 8.4 & 15.1 \\
 \hline
 $\lambda$ (cm) & 21.40 & 18.07 & 13.17 & 12.55 & 6.51 & 6.00 & 3.70 & 3.56 & 1.95 \\
 $\Delta_\mathrm{evpa}(^\circ)$ & 2.60 & 1.87 & 0.99 & 0.90 & 0.24 & 0.21 & 0.08 & 0.07 & 0.02 \\
\hline
\end{tabular}
\label{rm_1}
\end{minipage}
\end{table*}

\begin{figure*}
\centering
    \includegraphics[scale=0.26,angle=0]{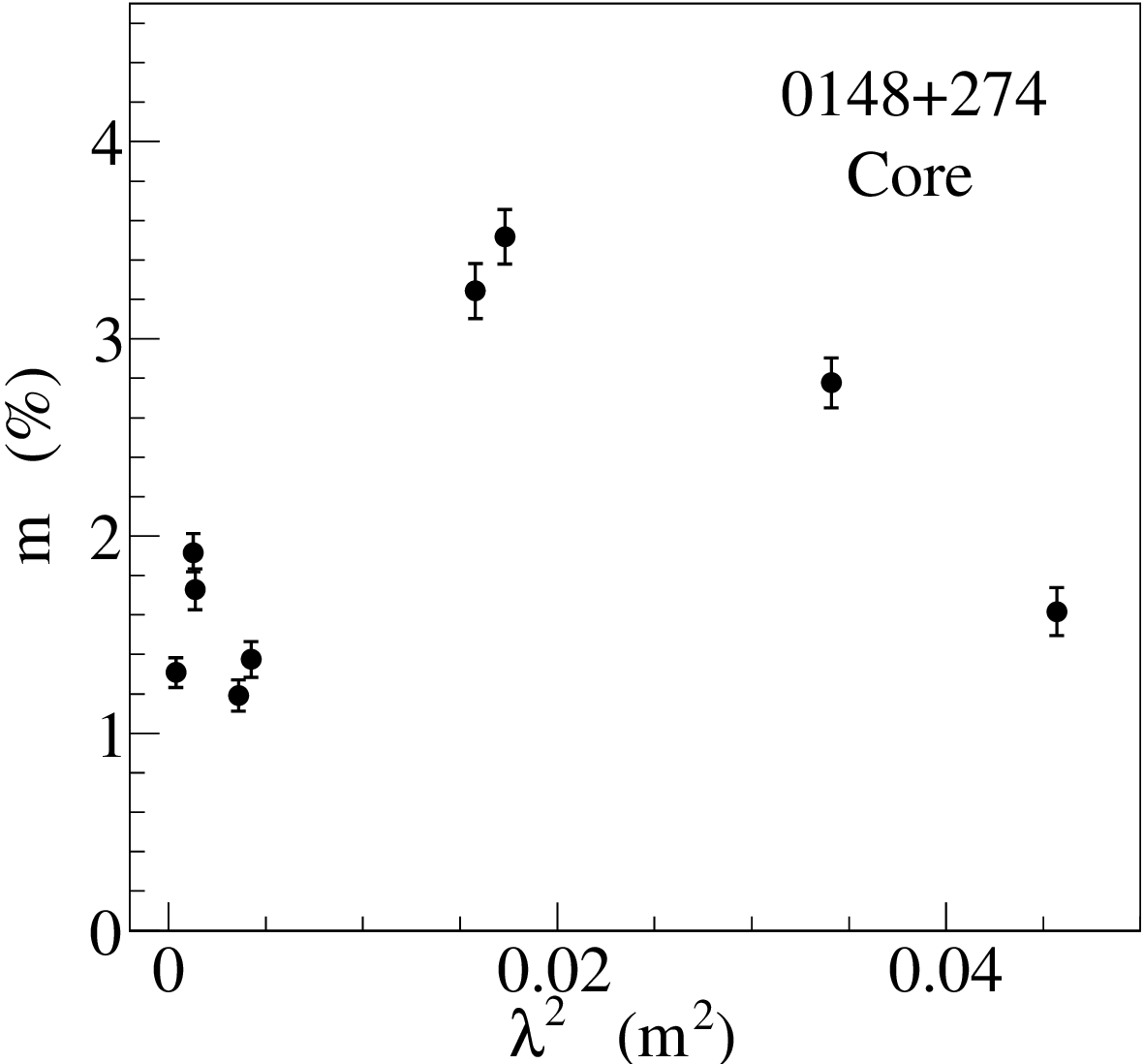}\quad
    \includegraphics[scale=0.31,angle=0]{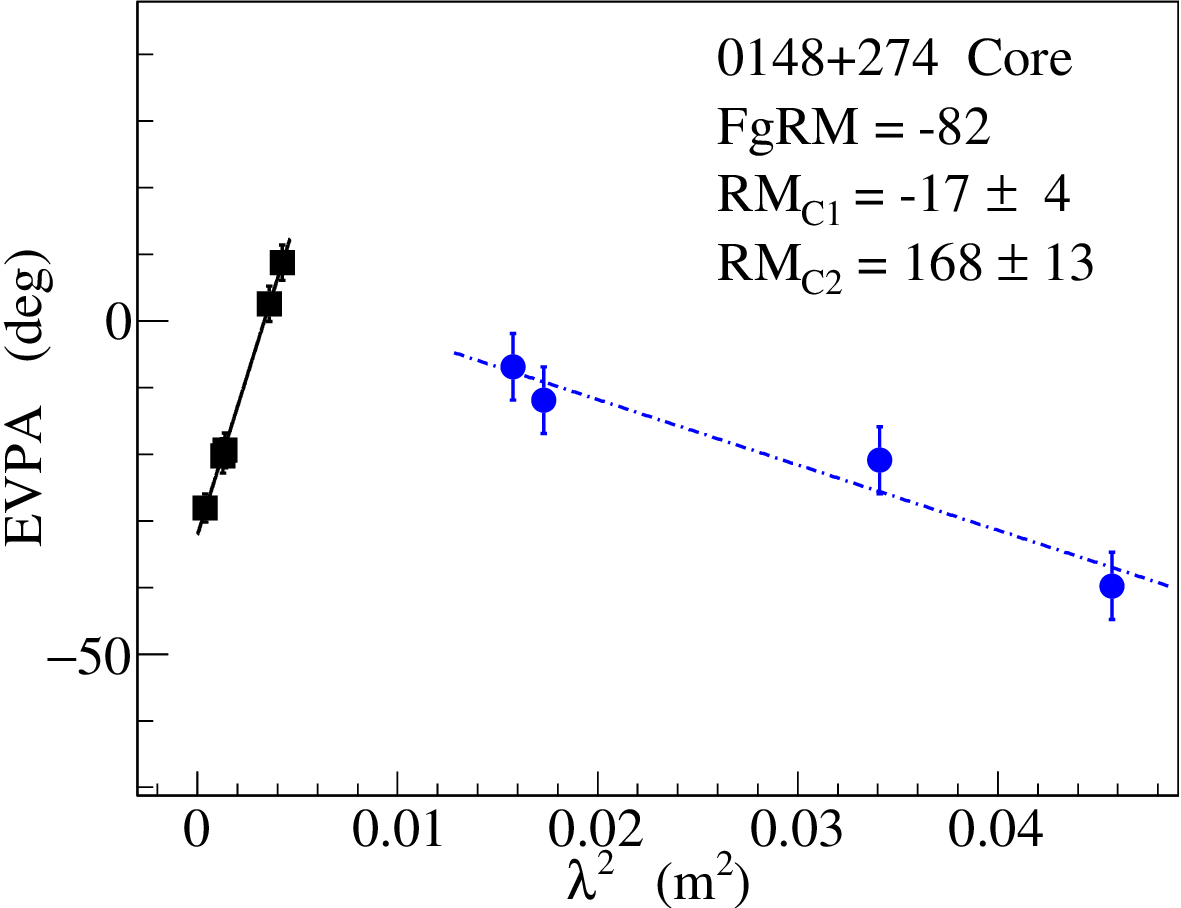} \quad
    \includegraphics[scale=0.26,angle=0]{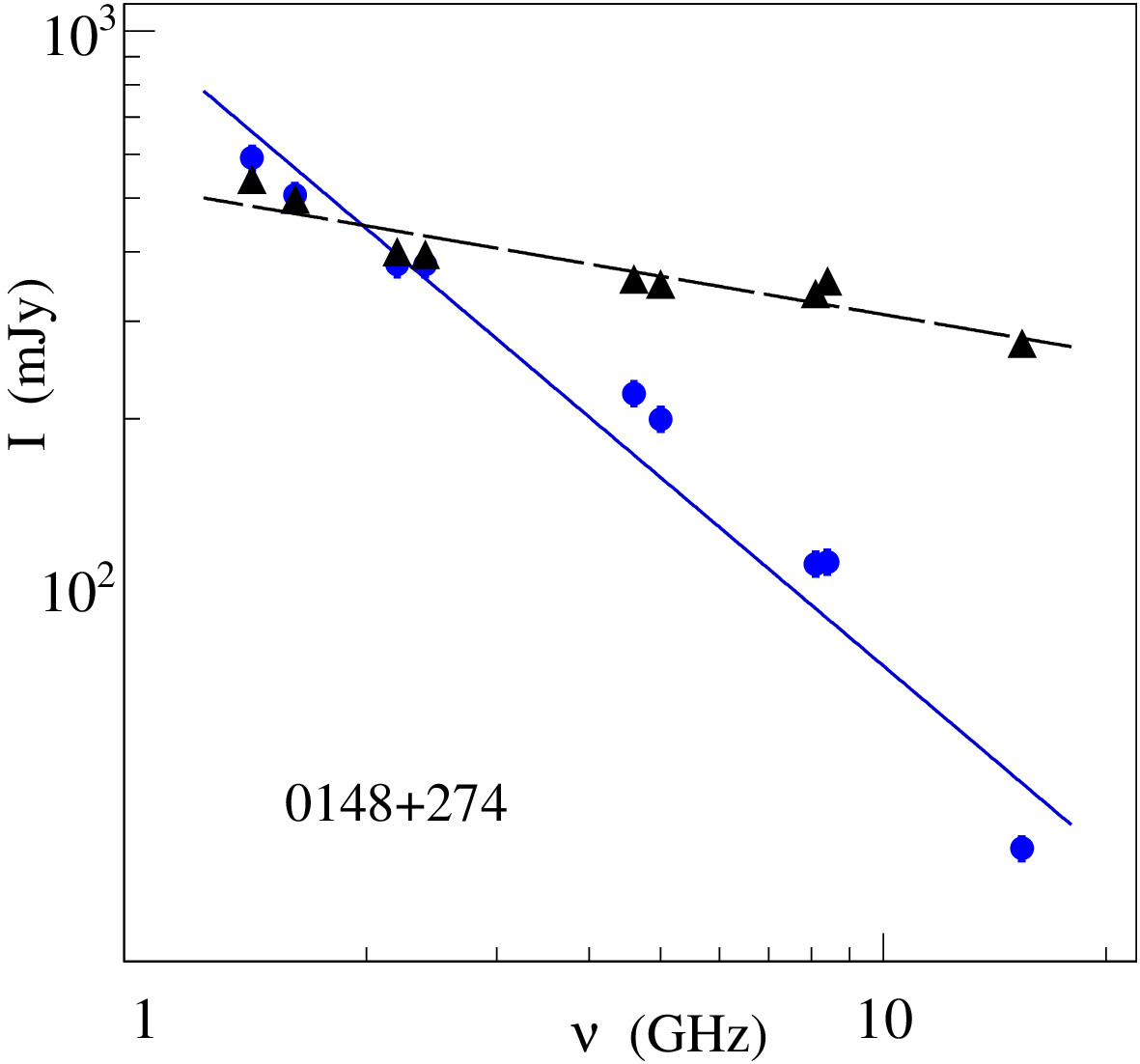} \\
    \includegraphics[scale=0.4,angle=270]{frmgr.0148+274.l1.2007_03_01.eps}\quad
    \includegraphics[scale=0.25,angle=270]{0148+274_trs.eps}\quad
    \includegraphics[scale=0.4,angle=270]{frmgr.0148+274.c1.2007_03_01.eps} \\
    \includegraphics[scale=0.26,angle=0]{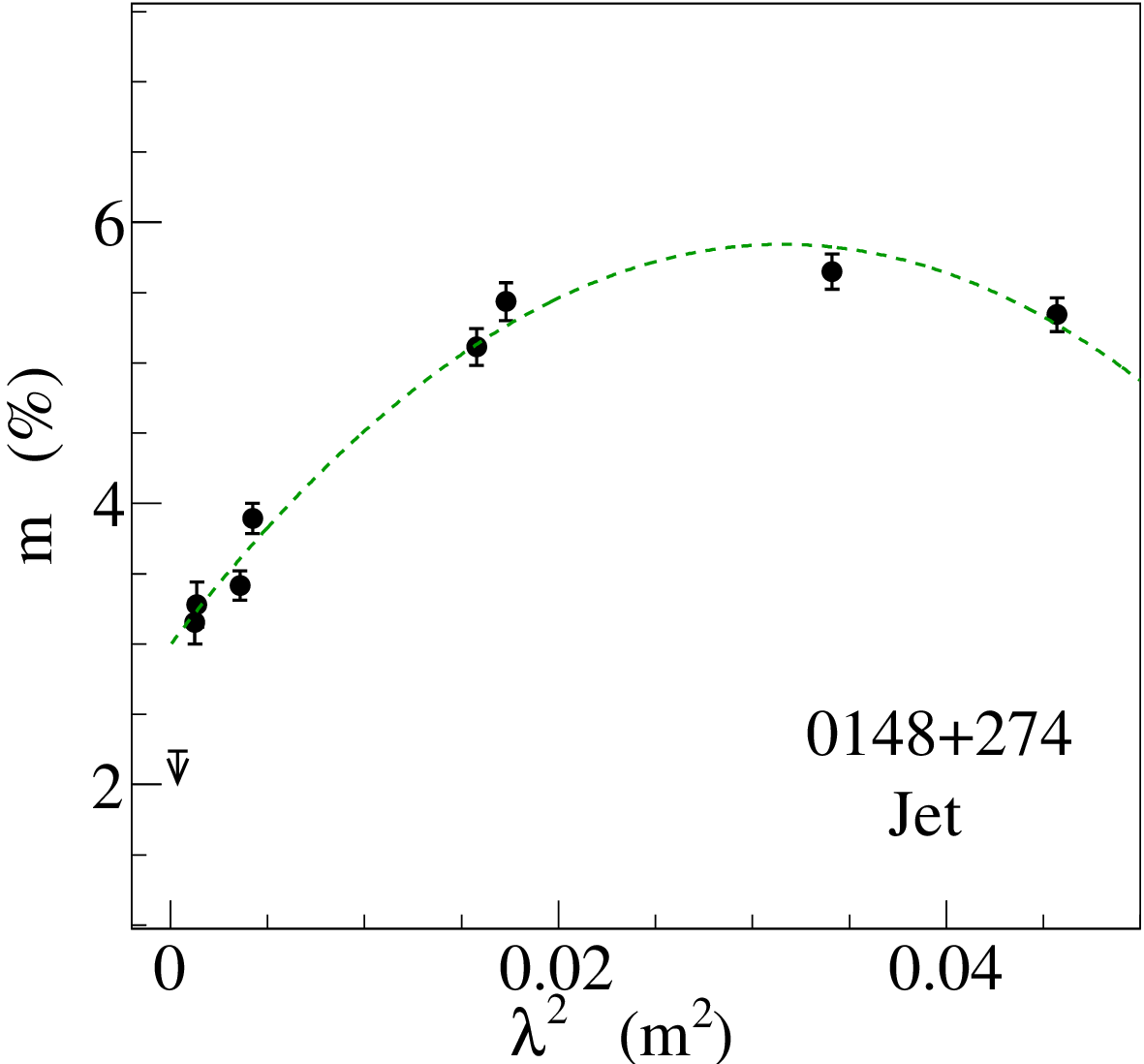} \quad
    \includegraphics[scale=0.31,angle=0]{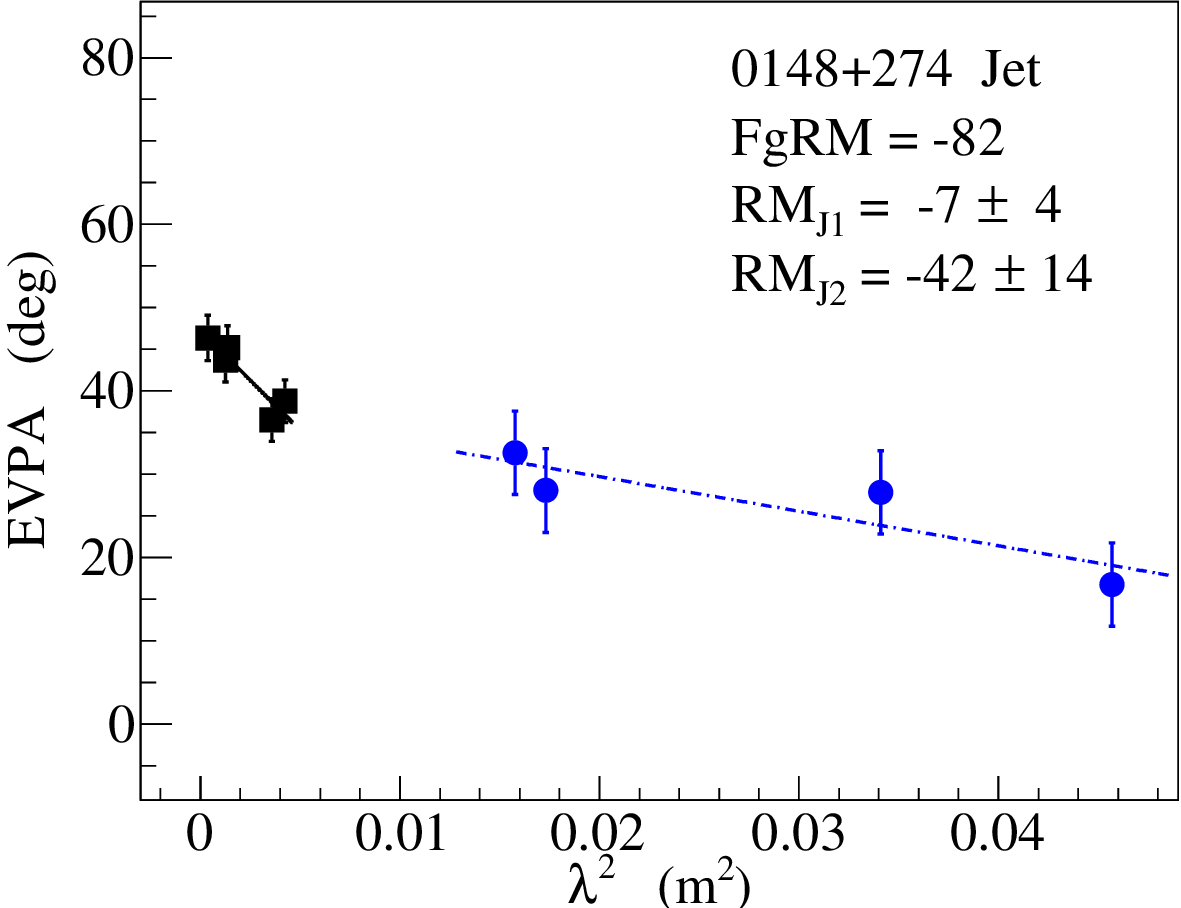}\\
\caption{Source 0148$+$274. (a) 1.4~GHz to 2.4~GHz Faraday RM map in the observer's frame. 
(b) 4.6~GHz to 15.4~GHz RM map. 
Stokes $I$ levels are drawn at (-1,1,2,4...)$\times$ the lowest contour (that is 3$\sigma_\mathrm{S}$ and given in Table~\ref{frac_p1}).
Here and hereafter, $I$ contours and their values are given for the lowest frequency of a given frequency range.
The colour bar is given in \rmu.The dot and letter represent positions of the model-fitted core ("C") and jet ("J") components, EVPA vs. $\lambda^2$ fits shown in the inset; the corresponding degree of polarization and spectrum are given. 
The EVPA value is taken from a polarization map convolved with the beam size of the lowest frequency of the RM range that is given in Table~\ref{t_1}.
The fitted values of RM and foreground RM (``FgRM") are shown and given in \rmu.
The spectrum is fitted by a power-law ($S\propto\nu^{\alpha}$) and is given for the core (black triangles and dashed line) and for the jet (blue circles and solid line) components. 
Upper limits on the degree of polarization correspond to 3$\sigma_\mathrm{m}$ at the position of component on the map.
Fits to the $m$ vs. $\lambda^2$ by Gaussian (solid blue line), Gaussian with constant term (dotted green line), polynomial (dashed red line) and two Gaussians (dashed-dotted violet line) are shown, see Sect.~\ref{s:m_fit} for details.
The solid black line on RM map indicates the slice (``S") across the RM map, taken in the clockwise direction, with the corresponding RM distribution along the slice given in the inset. 
Slices at different frequencies are centred at the position of jet component.
The beam size along the slice is shown by line in the left corner of each plot.
Grey lines surrounding black lines represent a $\pm1\sigma$ interval on observed RM and do not include absolute EVPA calibration error.
The length of slice S1 is 2.01 beams or 15.03 mas, of slice S2 is 1.71 beams or 3.75 mas. RM gradient of the S2 slice is significant.
Results for other targets are presented in the electronic version of the article.
\label{fig_0148}
}
\end{figure*}

\subsection{Model Fitting $m$ vs.\ $\lambda^2$}
\label{s:m_fit}
Faraday rotation can produce frequency-dependent depolarization or even repolarization \citep[e.g.][]{burn66,sokoloff_etal98}.
To study the polarized properties of the sources, we consider a number of effects and used them to model observed $m$ vs. $\lambda^2$ dependencies.
Meawnhile linear fit was applied to EVPA vs. $\lambda^2$ in ranges given in Table~\ref{t_1}.

Detailed discussion of various Faraday effects is given by, e.g. \citet{sokoloff_etal98} and \citet{farnes_etal14}.
To fit the data we follow suggestion of \citet{farnes_etal14} and model $m$ vs. $\lambda^2$ by four functions, which reproduce the wavelength-dependence of Faraday effects quite well. 
Namely these are: Gaussian, Gaussian with a constant term, two Gaussians, and  polynomial, given by following equations respectively
\begin{equation}
\label{eq_par1}
m = a_0 \mathrm{exp}\Big[ \frac{-(\lambda-a_1)^2}{2a_2} \Big], 
\end{equation}
\begin{equation}
m = a_0 \mathrm{exp}\Big[ \frac{-(\lambda-a_1)^2}{2a_2} \Big] + a_3, 
\end{equation}
\begin{equation}
m = a_0 \mathrm{exp}\Big[ \frac{-(\lambda-a_1)^2}{2a_2} \Big] + a_3 \mathrm{exp}\Big[ \frac{-(\lambda-a_4)^2}{2a_5} \Big], 
\end{equation}
\begin{equation}
\label{eq_par4}
m = a_0 \lambda^{a_1}.
\end{equation}
Coefficients $a_i$ in the equations are to be solved for during the model fitting.

For the analysis, we consider both the core and jet components. Meanwhile, optically thick synchrotron emission could produce more complex EVPA vs. $\lambda^2$ signatures, thus can not be well modelled by these parametric functions \citep[see][]{pacholczyk_swihart_67,fukui_73,jones_odell_77b}.

The Bayesian Information Criterion (BIC,~\citealt{schwarz_78}) is used to compare the goodness of fit of models with different number of parameters. 
The BIC is given by:
\begin{equation}
\mathrm{BIC} \approx -2\mathrm{ln}L+k\mathrm{ln}(N),
\end{equation}
where ln$L$ is the log-likelihood of the data given the model, $k$ is the number of parameters of the model, and $N$ is the sample size.
Assuming Gaussian noise, the \textit{Likelihood} of the model $\theta$ is defined as the joint probability to fit the data as
\[L = \mathrm{max}(p(y|\theta)),\]
\begin{equation}
p(y|\theta) = \prod_{i=1}^{N} \frac{1}{\sqrt{2\pi}\sigma_i}\mathrm{exp}\Big(-\frac{(y_i-y_i^M(\theta))^2}{2\sigma_i^2}\Big) 
\end{equation}
where $y=\{y_1,...,y_N\}$ is a set of measurement points with uncertainties $\{\sigma_i,...,\sigma_N\}$, and $\theta=\{\theta_1,...,\theta_k\}$ is a set of model parameters, that predict the measurements as $\{y_1^M(\theta),...,y_N^M(\theta)\}$.
We favour the model that has the lowest value of BIC. 
The results of the fit and discussion is given in Section \ref{s:res_m}.

\begin{figure}
 \centering
 \includegraphics[width=0.33\textwidth]{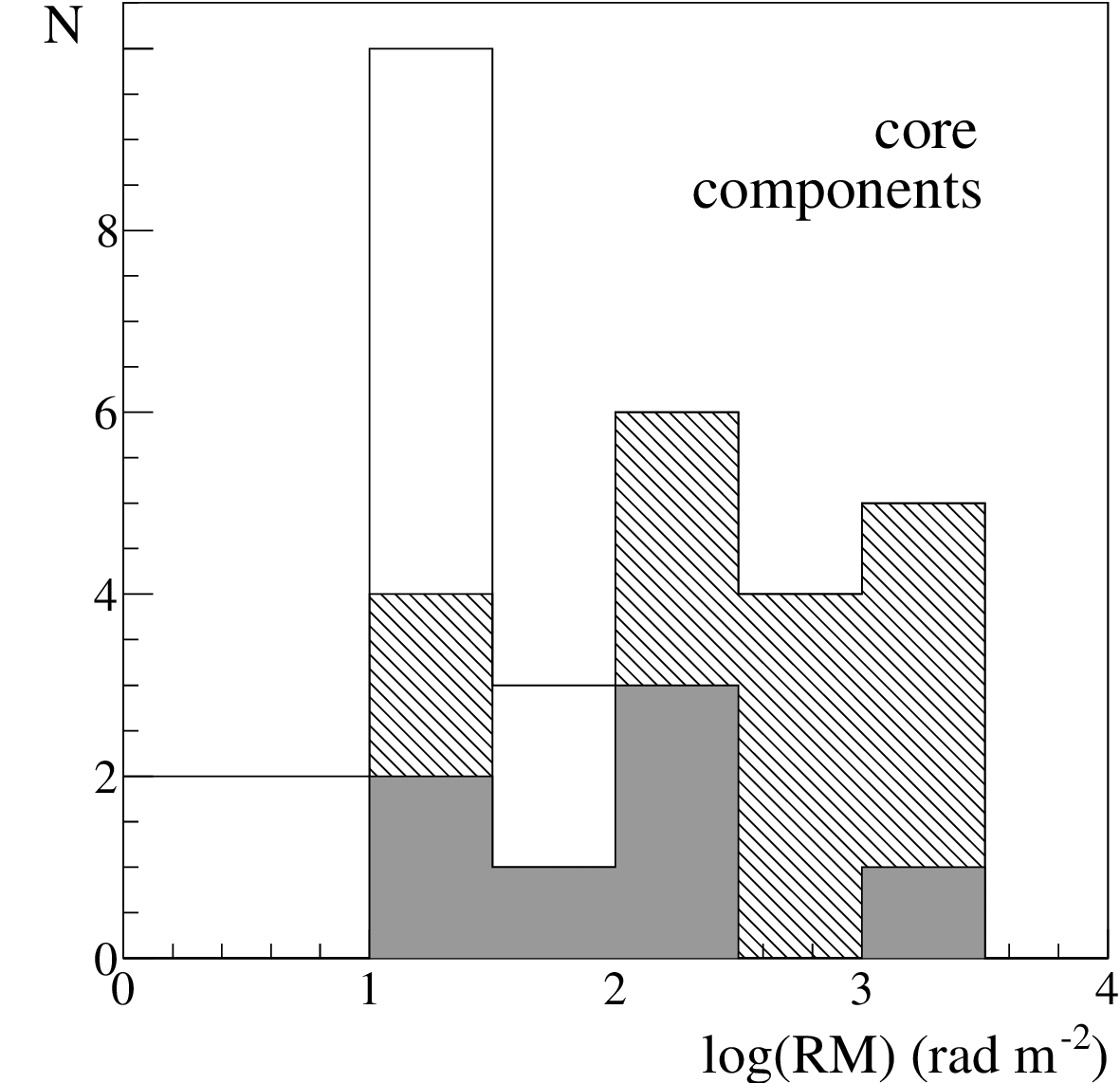}
 \includegraphics[width=0.33\textwidth]{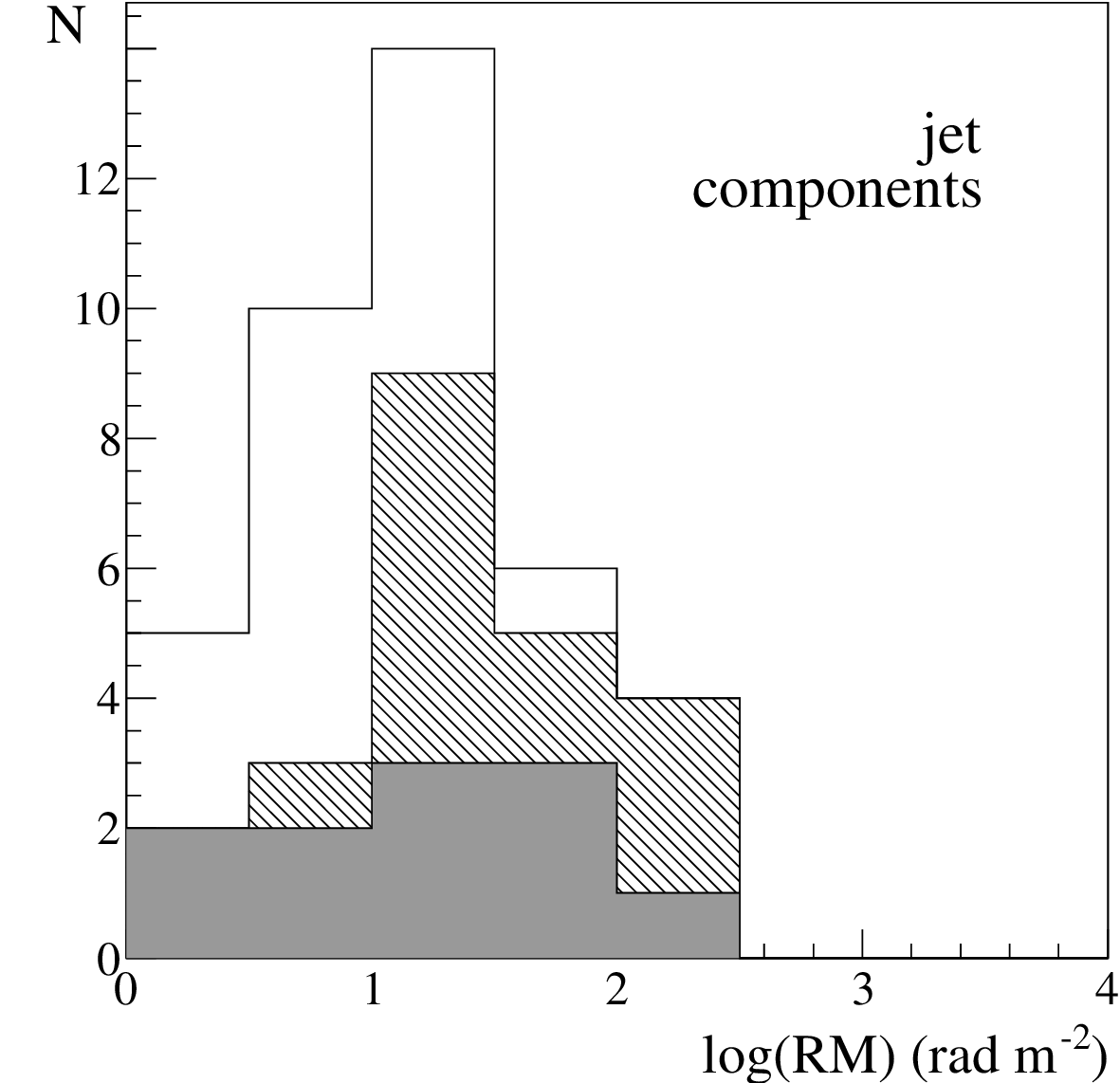}
 \caption{Distribution of the observed RM values for all sources in the modelled core (top) and jet (bottom) components. White filling corresponds to 1.4~GHz to 2.4~GHz RM frequency range, grey filling~-- 2.4~GHz to 8.4~GHz, and oblique filling to the 4.6~GHz to 15.4~GHz. Details are given in Table~\ref{t_1} and Table~\ref{frm}.
 \label{rmdist}
 }
\end{figure}

\section{Results and Discussion}
\label{s:frm_res}
We first outline general results, then consider each source individually.

\begin{table*}
 \centering
 \begin{minipage}{175mm}
  \caption{Results for the fitted depolarization models and Faraday--corrected electric field orientation.}
\begin{tabular}{@{}cccccccc@{}}
  \hline
    Source & Core & Core & Jet & Jet & RM & Helical & Jet \\
     & model & physical law & model & physical law & gradient & signatures & EVPA\\
 \hline
 0148$+$274 & peaked & multiple components$^1$ & peaked & anomalous & Y & aberration & misaligned \\
 0342$+$147 & power$^2$ & Tribble & power & Tribble & N & $\dots$ & misaligned \\
 0425$+$048 & power & Burn & Gaussian$^3$ & low depolarization & N & $\dots$ & unclear \\
 0507$+$179 & two Gaussians$^4$ & multiple components & Gaussian + constant$^5$ & Ros. -- Mantov. & N & $\dots$ & misaligned \\
 0610$+$260 & $\dots$ & $\dots$ & Gaussian + constant & Tribble & $\dots$ & $\dots$ & $\dots$ \\
 0839$+$187 & Gaussian + constant & Ros. -- Mantov. & Gaussian & low depolarization & N & $\dots$ & misaligned \\
 0952$+$179 & peaked & multiple components & Gaussian + constant & Tribbe & Y & 90\degr\ jumps & unclear \\
 1004$+$141 & Gaussian + constant & Ros. -- Mantov. & Gaussian + constant & Tribble & Y & $\dots$ & aligned \\
 1011$+$250 & $\dots$ & $\dots$ & power & Burn & N & $\dots$ & misaligned \\
 1049$+$215 & peaked & multiple components & Gaussian & low depolarization & N & $\dots$ & misaligned \\  
 1219$+$285 & peaked & multiple components & Gaussian & low depolarization & Y & $\dots$ & unclear \\
 & & /anomalous & & &  & & \\
 1406$-$076 & Gaussian& Tribble & Gaussian &  low depolarization & N & $\dots$ & misaligned \\
 1458$+$718 & peaked & multiple components & peaked & anomalous & Y & spine--sheath & misaligned \\
  & & /anomalous & & &  & & \\
 1642$+$690 & peaked & multiple components & Gaussian + constant & anomalous & Y & spine--sheath & aligned \\
 1655$+$077 & two Gaussians & multiple components& Gaussian + constant & Tribble & N & $\dots$ & aligned \\
 1803$+$784 & power & anomalous & Gaussian + constant & low depolarization & N & $\dots$ & aligned \\
 1830$+$285 & $\dots$ & $\dots$ & Gaussian & Tribble & N & $\dots$ & misaligned \\
 1845$+$797 & $\dots$ & $\dots$ & $\dots$ &$\dots$ & $\dots$ & $\dots$ & $\dots$ \\
 2201$+$215 & $\dots$ & unclear & Gaussian & Tribble & Y & $\dots$ & misaligned \\
 2320$+$506 & Gaussian + constant & Burn/Tribble & power & Tribble & N & aberration & misaligned \\
 
\hline
\end{tabular}
\medskip

$^1$ multiple Faraday rotation measure or jet components. $^{2,3,4,5}$ are the fitted models given by equations from~(\ref{eq_par1}) to ~(\ref{eq_par4}).
Fitted parametric models are described in Section~\ref{s:m_fit}. Physical depolarization laws, corresponding to the fitted models, are listed in Section~\ref{s:m_fit}. The observed significant RM gradients are given in Table~\ref{t:grads}. Under ``helical structure'' we assume polarization structures, arising in a presence of a helical magnetic fields, e.g. spine--sheath structure. The orientation of the jet EVPA is given relative to the local jet direction and is shown in Table~\ref{t:evpa_jet}.
Discussion of each source individually is given in Section~\ref{s:inds}.
\label{t_results}
\end{minipage}
\end{table*}

\subsection{Spectral Index}
\label{s:spi}
Spectra of the core and jet components are displayed in Fig.~\ref{fig_0148} with the  power-law fit to the Stokes $I$: $I \propto \nu^{\alpha}$.
Jet components show an optically thin synchrotron spectrum.
Meanwhile, core components show flat or slightly inverted spectra over the whole frequency range, which is attributed to synchrotron self-absorption \citep[e.g.][]{blandford_konigl_79,konigl81,sokolovsky_etal11}.

\begin{figure*}
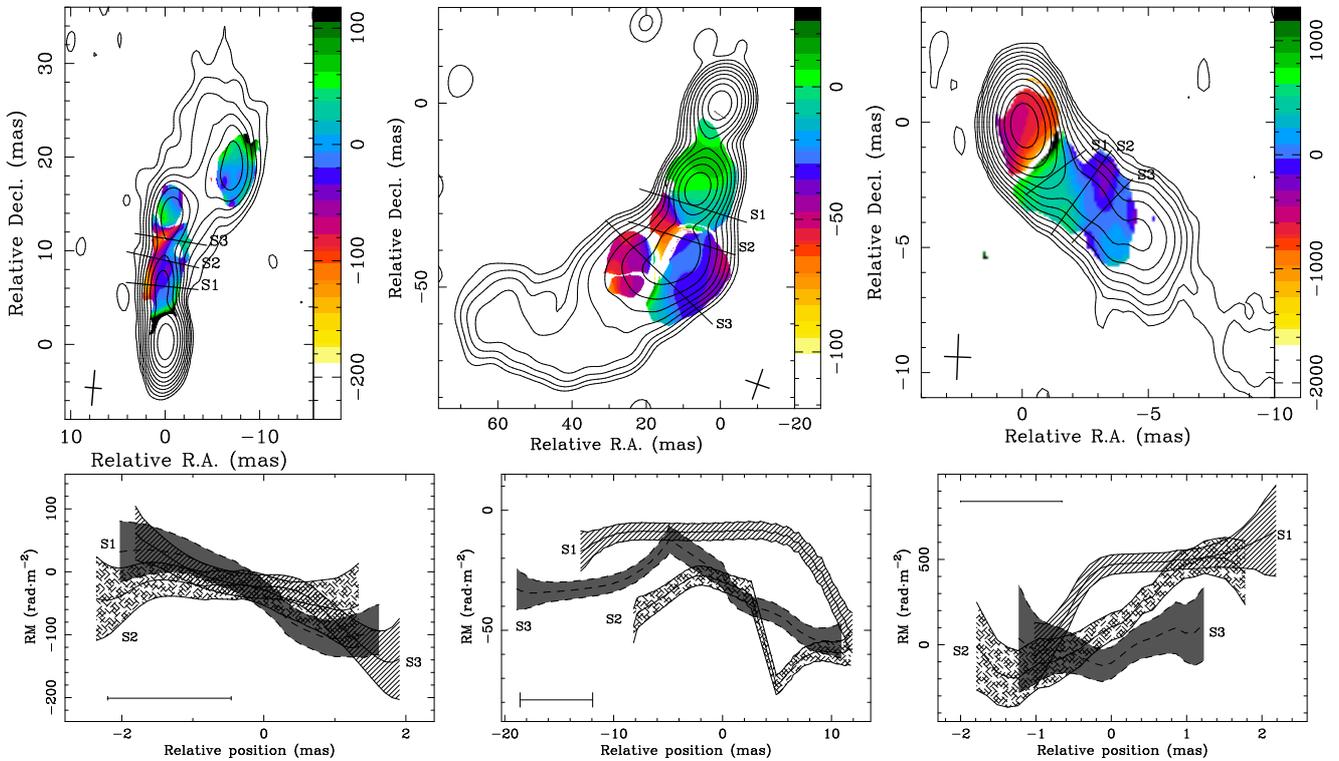

\centering
    \includegraphics[scale=0.38,angle=270]{0952+179_c1_frm.eps}\quad
    \includegraphics[scale=0.36,angle=270]{1458+718_l1_frm.eps}\quad
    \includegraphics[scale=0.36,angle=270]{2201+315_x1_frm.eps}\\
    \includegraphics[scale=0.23,angle=270]{0952+179_c1_trs.eps}\quad
    \includegraphics[scale=0.23,angle=270]{1458+718_l1_trs.eps}\quad
    \includegraphics[scale=0.23,angle=270]{2201+315_x1_trs.eps}\\
\caption{Faraday RM maps and transverse RM slices for 0952$+$179 (left, 4.6~GHz to 8.4~GHz), 1458$+$718 (middle, 1.4~GHz to 2.4~GHz) and 2201$+$315 (right, 8.1~GHz to 15.4~GHz). Slices are centered at the position of ridge line. The other details are the same as in Fig.~\ref{fig_0148}.
Parameters for significant slices are given in Table~\ref{t:grads}. \label{fig_trs}}
\end{figure*}

\subsection{Degree of Linear Polarization}
No significant polarized flux density was detected from two sources, 0610+260 and 1845+797, so they were excluded from the RM analysis.
The distribution of polarization degree in the opaque core and the optically thin jet components for all sources is presented in Fig.~\ref{pistat} and given in Table~\ref{frac_p1}.
The median degree of polarization in the core region is about 1~per~cent, while maximum value does not exceed 8~per-cent, which is expected for an opaque medium \citep[e.g.][]{pacholczyk_swihart_67}.
The median polarization degree in the jet region is about 8~per~cent~and reaches 25~per~cent, which is lower than the value expected in a presence of uniform magnetic fields \citep[e.g.][]{gardner_whiteoak_66}.
These estimates of fractional polarization are similar to measures obtained by analogous studies \citep[e.g.][]{lister_homan_05,jorstad_etal07}.
Clearly, such low observed degree of polarization is caused by depolarization (either internal or external Faraday rotation, beam depolarization) and randomly oriented magnetic fields in these jet regions \citep[see, e.g.][]{2011ApJ...737...42P}.
Observations with higher spectral and spatial resolutions can resolve this issue including space VLBI polarization measurements \citep[e.g.][]{lobanov_etal15,2016ApJ...817...96G}.

\subsection{$m$--$\lambda^2$ Dependence and Faraday Effects}
\label{s:res_m}

All sources posses complex polarized structure. 
Figure~\ref{fig_0148} presents the fractional polarization in the core and jet components versus wavelength squared. 
We model the $m$ vs. $\lambda^2$ dependences by a number of functions, described in Section~\ref{s:m_fit}, and summarize results in Table~\ref{t_results} and present them in Fig.~\ref{fig_0148}.
The majority (12 out of 18) of the sources show a decrease in polarization degree with increasing wavelength. 
An inhomogeneous external Faraday screen is responsible for such behaviour, described by \citet{burn66}, \citet{tribble91}, \citet{2008AA...487..865R}, and \citet{2009AA...502...61M} laws.
This Faraday screen may contain turbulent or systematically varying regular field. Fractional polarization in both these cases possess an analogous behaviour. Unfortunately, our study is not able to distinguish between them.
Six sources exhibit an increase of fractional polarization with wavelength, known as anomalous \citep{sokoloff_etal98} or inverse \citep{homan_12} depolarization.
Four sources possess oscillatory polarization behaviour, expected when a few jet or rotation measure components are blended within the region \citep{conway_etal_74,gl_84}.
Sparse frequency coverage does not allow us to study polarized structure in more detail or even to distinguish among different Faraday effects. 

The listed effects in the transparent jet components are accompanied by a linear EVPA vs. $\lambda^2$ dependence across the full frequency range which indicates a one-component Faraday rotating screen.
At the same time, the majority of the opaque cores exhibit complex EVPA vs. $\lambda^2$ behaviour, resulting from Faraday rotation either in external or internal medium relative to the synchrotron emission region. 
Detailed discussion of individual sources is given in Section~\ref{s:inds}.

\subsection{Faraday Rotation Measures}
\label{s:frm}
We constructed 43 maps of Faraday RM for 18 sources over two or three frequency ranges (as shown in Table~\ref{t_1}) following the steps described in Section~\ref{s:rm_reconstr}. In one third of the sources no Faraday RM estimates could be obtained in the core regions since they break the linear $\lambda^2$ law.
The resulting RM maps with substracted Galactic contribution are presented in Fig.~\ref{fig_0148} together with the EVPA vs. $\lambda^2$ fits.
RM pixel values at the location of transparent jet and at opaque core components are given in Table~\ref{frm}, while the distribution of these values over all frequencies is given in Fig.~\ref{rmdist}.
One can see from Fig.~\ref{rmdist} that high RM values are observed at short wavelengths in the cores only. Thus, we confirm the expected result \citep[e.g.][]{trippe_etal12}, indicating the presence of stronger magnetic fields and denser medium upstream in the jet. 

Using the relation between the observed rotation measure (RM) and the rest frame value ($\mathrm{RM}_0$):
$\mathrm{RM}_0=\mathrm{RM}(1+z)^2$, we estimate the median value of $\mathrm{RM}_0$ averaged over all sources, listed in the bottom row of Table~\ref{frm}.
The highest measured Faraday rotation in the source rest frame is ($-1.69\pm0.03$)$\times10^4$~\rmu\ in the 0952$+$179 core.

We observed a systematic increase of the absolute RM values in optically thin regions with increasing frequency.
Most probably, this is due to beam depolarization at long wavelengths \citep[see simulations by][]{2011ApJ...737...42P}.

\begin{table*}
 \centering
 \begin{minipage}{175mm}
  \caption{Rotation measure results.}
  \begin{tabular}{cr@{}c@{}lr@{}c@{}lr@{}c@{}lr@{}c@{}lr@{}c@{}lr@{}c@{}lr@{}c@{}l}
  \hline
  &\multicolumn{6}{c}{Low frequency range}&\multicolumn{6}{c}{Middle frequency range}&\multicolumn{6}{c}{High frequency range}&\multicolumn{3}{c}{Power}\\
 Source&\multicolumn{3}{c}{$\mathrm{RM}_\mathrm{core}$}&\multicolumn{3}{c}{$\mathrm{RM}_\mathrm{jet}$}&\multicolumn{3}{c}{$\mathrm{RM}_\mathrm{core}$}
 &\multicolumn{3}{c}{$\mathrm{RM}_\mathrm{jet}$}&\multicolumn{3}{c}{$\mathrm{RM}_\mathrm{core}$}&\multicolumn{3}{c}{$\mathrm{RM}_\mathrm{jet}$}&
 \multicolumn{3}{c}{$a$}\\
  \hline
0148$+$274&$-$17&$\pm$&4&$-$7&$\pm$&4&	&$-$&&	&$-$&&	168&$\pm$&13&$-$42&$\pm$&14&1.9&$\pm$&0.2\\
0342$+$147&&$-$&&	9&$\pm$&4&$-$14&$\pm$&17&	0&$\pm$&22&$-$431&$\pm$&52&$-$167&$\pm$&88&6.1&$\pm$&2.9\\
0425$+$048&&$-$&&	7&$\pm$&4&	&$-$&&$-$22&$\pm$&19&	&$-$&&	&$-$&&	&$-$&\\
0507$+$179&$-$30&$\pm$&4&$-$35&$\pm$&4&1035&$\pm$&19&$-$55&$\pm$&17&1055&$\pm$&53&$-$141&$\pm$&57&2.98&$\pm$&0.13$^1$\\
0610$+$260&&$-$&&	&$-$&&	&$-$&&	&$-$&&	&$-$&&	&$-$&&	&$-$&\\ 
0839$+$187&&$-$&&$-$3&$\pm$&4&	&$-$&&	&$-$&&	&$-$&&$-$25&$\pm$&17&	&$-$&\\
0952$+$179&$-$17&$\pm$&4&4&$\pm$&4&	&$-$&&	13&$\pm$&19&$-$2767&$\pm$&53&40&$\pm$&89&2.90&$\pm$&0.14\\
1004$+$141&$-$1&$\pm$&4&$-$9&$\pm$&4&$-$127&$\pm$&17&$-$32&$\pm$&17&&$-$&&&$-$&&4.1&$\pm$&4.6\\
1011$+$250&&$-$&&	&$-$&&	&$-$&&$-$157&$\pm$&29&	&$-$&&	&$-$&&	&$-$&\\
1049$+$215&$-$4&$\pm$&4&$-$11&$\pm$&4&$-$107&$\pm$&17&$-$23&$\pm$&23&$-$547&$\pm$&52&	&$-$&&2.9&$\pm$&0.3\\
1219$+$285&$-$16&$\pm$&4&$-$16&$\pm$&4&	&$-$&&	&$-$&&$-$17&$\pm$&17&$-$27&$\pm$&15&0.0&$\pm$&0.9\\
1406$-$076&30&$\pm$&4&7&$\pm$&4&$-$47&$\pm$&5&2&$\pm$&6&$-$235&$\pm$&17&$-$16&$\pm$&25&1.77&$\pm$&0.12\\
1458$+$718&&$-$&&3&$\pm$&4&	&$-$&&$-$33&$\pm$&17&1153&$\pm$&56&$-$31&$\pm$&53&&$-$&\\
1642$+$690&$-$1&$\pm$&4&	8&$\pm$&4&$-$239&$\pm$&17&$-$2&$\pm$&18&272&$\pm$&50&106&$\pm$&102&4.6&$\pm$&9.6$^1$\\
1655$+$077&24&$\pm$&4&27&$\pm$&4&&$-$&&&$-$&&$-$491&$\pm$&13&$-$4&$\pm$&13&2.54&$\pm$&0.14\\
1803$+$784&$-$8&$\pm$&4&	3&$\pm$&4&21&$\pm$&17&6&$\pm$&18&28&$\pm$&17&11&$\pm$&24&0.9&$\pm$&0.6\\
1830$+$285&&$-$&&	&$-$&&	&$-$&&$-$5&$\pm$&6&	&$-$&&	&$-$&&	&$-$&\\
1845$+$797&&$-$&&	&$-$&&	&$-$&&	&$-$&&	&$-$&&	&$-$&&	&$-$&\\
2201$+$315&$-$44&$\pm$&4&$-$15&$\pm$&4&	&$-$&&	&$-$&&$-$514&$\pm$&50&15&$\pm$&253&1.40&$\pm$&0.08\\
2320$+$506&44&$\pm$&4&19&$\pm$&4&	&$-$&&	&$-$&&$-$1627&$\pm$&18&	&$-$&&2.84&$\pm$&0.07\\
$|$median$|$&17&&&8&&&107&&&22&&&491&&&29&&&2&.5&\\
$|$median${(1+z)}^{2}|$&61&.6&&30&.5&&566&.0&&79&.8&&1460&.5&&129&.1&&&$-$&\\
\hline
\end{tabular}

\smallskip
 The shown RM pixel values were determined at the position of the modelled core and jet components.
 RM is given in \rmu and corrected for the Galactic rotation.
 The calculated power $a$ of the rotation measure fall-off with frequency $|\mathrm{RM}_\mathrm{core}|\sim{\nu}^{a}$ is given in the last column, see for details Sect.~\ref{s:rm-core}.
 All the values are calculated in the observer's frame except for the last row which is given in the sources' rest frame.
 The medians are calculated for absolute RM values. 
 
 $^1$ - RM at the high frequency range has not been considered in the fit.
 \label{frm}
\end{minipage}
\end{table*}

\begin{table*}
 \centering
 \begin{minipage}{175mm}
  \caption{Detected significant transverse rotation measure gradients.}
  \begin{tabular}{cccr@{}c@{}lr@{}c@{}lr@{}c@{}lcc}
 \hline
 Source&Name&Range&\multicolumn{3}{c}{$\mathrm{RM}_1$}&\multicolumn{3}{c}{$\mathrm{RM}_2$}&\multicolumn{3}{c}{$|\Delta\mathrm{RM}|$}&Significance&Width\\
 &&(GHz)&\multicolumn{3}{c}{(\rmu)}&\multicolumn{3}{c}{(\rmu)}&\multicolumn{3}{c}{(\rmu)}&&(beams)\\
  \hline
\multicolumn{14}{c}{Slices at position of model-fitted, transparent jet component, given in Fig.~\ref{fig_0148}}\\
0148$+$274&S2&4.6--15.4&1&$\pm$&22&-62&$\pm$&20&63&$\pm$&30&2.1$\sigma$&1.7\\
0952$+$179&S1&1.4--2.4&-8&$\pm$&6&26&$\pm$&6&34&$\pm$&8&4.1$\sigma$&2.3\\
 &S3&8.1--15.4&65&$\pm$&192&610&$\pm$&177&545&$\pm$&261&2.1$\sigma$&1.3\\
1004$+$141&S2&4.6--8.4&63&$\pm$&35&-41&$\pm$&24&104&$\pm$&42&2.5$\sigma$&1.8\\
1219$+$285&S2&4.6--15.4&60&$\pm$&42&-63&$\pm$&19&123&$\pm$&46&2.7$\sigma$&2.3\\
1642$+$690&S2&4.6--8.4&-153&$\pm$&57&-135&$\pm$&59&17&$\pm$&82&2.9$\sigma^\mathrm{1}$&2.7\\
 &S3&8.1--15.4&-272&$\pm$&258&554&$\pm$&216&826&$\pm$&336&2.5$\sigma$&3.6\\
 2201$+$315&S1&1.4--2.4&-20&$\pm$&7&21&$\pm$&7&41&$\pm$&10&4.1$\sigma$&1.7\\
\multicolumn{13}{c}{Slices at other jet locations, given in Fig.~\ref{fig_trs}}\\
0952$+$179&S1&4.6--8.4&62&$\pm$&43&-139&$\pm$&64&200&$\pm$&77&2.6$\sigma$&1.7\\
	  &S3&4.6--8.4&32&$\pm$&49&-92&$\pm$&40&124&$\pm$&63&2.0$\sigma$&1.8\\
1458$+$718&S1&1.4--2.4&-17&$\pm$&8&-51&$\pm$&8&34&$\pm$&12&2.9$\sigma$&6.7\\
	  &S2&1.4--2.4&-27&$\pm$&7&-72&$\pm$&7&45&$\pm$&10&4.5$\sigma^\mathrm{1}$&6.7\\
	  &S3&1.4--2.4&-12&$\pm$&9&-54&$\pm$&7&42&$\pm$&11&3.7$\sigma^\mathrm{1}$&6.9\\
2201$+$315&S1&8.1--15.4&-64&$\pm$&195&668&$\pm$&267&732&$\pm$&330&2.2$\sigma$&1.3\\
\hline
\end{tabular}
\smallskip

The total change in rotation measure, $|\Delta\mathrm{RM}|$,  is characterized by ${\mathrm{RM}}_1$ and ${\mathrm{RM}}_2$, which are either edge or extreme RM values of the monotonic or sharp transverse RM profiles accordingly. The error in $|\Delta\mathrm{RM}|$ is defined as the square root of $\sigma_{\mathrm{RM}_1}$ and $\sigma_{\mathrm{RM}_2}$ added in quadrature. The significance of a gradient is determined as the total change in RM divided by its error. We note, that we used this value as a quantitative measure for the RM gradient. Meanwhile, qualitative analysis of the  gradient significance is based on two--steps approach, described in Section~\ref{s:trgr}.

$^\mathrm{1}$ -- RM gradient is not monotonic.
 \label{t:grads}
 \end{minipage}
\end{table*}

\subsection{Transverse Faraday RM Gradients}
\label{s:trgr} 
We studied the RM distribution along the cuts made perpendicular to the local jet direction defined by the jet's ridge-line. 
The ridge-line (jet emission centre) was determined by fitting a Gaussian to the image brightness distribution profile (convolved with circular beam), taken at a given radial distance from the core \citep[see description of the technique in, e.g.][]{lobanov_zensus_01}.

During the analysis we omitted errors in the absolute EVPA calibration since they affect only the absolute rotation measure values, while significance of a gradient is determined by relative EVPA changes \citep[e.g.][]{mahmud_etal09,hovatta_etal12}. 
While, to analyze RM profiles at different frequency ranges, these errors together with the uncertainty in Galactic RM (discussed in Section~\ref{s:gal_rm}) should be taken into account.
Complexity of the AGN cores and opacity effects may result in unreliable RM gradients \citep[as, e.g. shown in simulations by][]{broderick_mckinney_10}. 
Thus we analyse only regions located more than 1 beam size downstream from the core, where synchrotron emission is optically thin.

The criterion for a gradient to be significant comprises two steps. 
At first, a line was fitted to the RM values along a slice. 
If the linear slope exceeds 3 times its RMS, the second step is applied. 
It consists of fitting a constant line to the data at the level equal to the weighted average of the RM along the slice. 
If the $\chi^2$ of the second fit is larger than the theoretical value for the appropriate degrees of freedom, the RM gradient was considered to be significant.
We assume the degrees of freedom as the number of pixels along the RM slice minus two. 
The $\chi^2$ test, estimated in this way, carries qualitative information and has no statistical meaning, because adjacent pixel values of the RM image are correlated and considered average rms errors in the Stokes $I$, $Q$ and $U$ are not evenly distributed in the images.
Simulations  (like \citealt{hovatta_etal12} and Pashchenko et al. in preparation) should be used for the statistical estimation of the gradient significance.
Following \citet{broderick_mckinney_10} and \citet{taylor_zavala_10}, we also calculated the significance of a gradient as the total change in RM divided by the average RM error at the edges of a slice.
Because it is a simplified approach, it results in the smaller value of significance, than the value derived by our two--steps approach.
Since our observations allow us to study transverse RM profiles across a few frequency bands simultaneously, we took these slices only at the position of model-fitted transparent jet components, which is identified across all frequencies. 
This enabled us to make a direct comparison of rotation measure gradients obtained at different frequency ranges.

We were able to take transverse slices on 41 RM maps.
Three out of analysed slices have width less than 1 half power beam width (HPBW), nine are $\leq1.5$~HPBW wide, and others moderately resolve the jet (wider than 1.5~HPBW). 
Monte Carlo simulations by \citet{mahmud_etal13} have shown that reliable RM gradients can be observed even when the jet slice is a fraction of  the observing beam. 
Results for all transverse RM profiles are presented in Fig.~\ref{fig_0148}.
For a thorough discussion of confusion between real and spurious RM gradients we refer readers to \citet{taylor_zavala_10,hovatta_etal12,algaba13}.

Significant RM gradients were detected in 8 out of 41 taken slices, namely in 0148$+$274, 0952$+$179, 1004$+$141, 1219$+$285, 1642$+$690 and 2201$+$315 (see Table~\ref{t:grads}). 
Other sources, 1458$+$718, 0952$+$179, and 2201$+$315, show significant RM gradients at location, other than the position of the model-fitted jet component. 
These cuts are presented in Fig.~\ref{fig_trs} and their characteristics are given in Table~\ref{t:grads}.
Transverse RM profiles in the jets of 1458$+$718 and 1642$+$690 are not monotonic. We found in the literature the only source 3C~454.3, having similar RM gradient \citep{hovatta_etal12,zamaninasab_etal13}. Most likely, the helical magnetic field in the bending jet of 3C~454.3 gives rise to such RM profile. It is difficult to say, what causes such sharp gradients in 1458$+$718 and 1642$+$690, meanwhile the both sources show evidences of a helical fields in their jets (see Section~\ref{s:inds}).

\subsection{Rotation Measure in the Core}
\label{s:rm-core}

We see (Figure~\ref{rmdist}, Table~\ref{frm}) an overall tendency of the rotation measure to increase its value with increasing frequency and to break the linear $\lambda^2$ law in the cores \citep[see also VLA results by][]{kravchenko_etal15}. 
Since the observed sources exhibit significant core shifts among compact extragalactic sources \citep{sokolovsky_etal11,kovalev_etal08}, the shift of the photosphere towards central engine will result in noticeable difference of regions probed at the observed frequencies.
Hence, higher values of RMs at these frequencies indicate denser media and stronger line-of-sight magnetic fields \citep{lobanov98,kovalev_etal08,sokolovsky_etal11,2011ApJ...737...42P,pushkarev_etal12}; or more Faraday active material between the observer and the source \citep{2011ApJ...737...42P}. 
Considering the location of the external Faraday screen to be very close to the jet, one may assume a thinner Faraday screen for the VLBI core with increasing wavelength. 
Frequency dependence (and thus distance dependence) of the core RM can be studied using the relation 
\begin{equation}|\mathrm{RM}_\mathrm{core}|\sim{\nu}^{a},\end{equation}
derived in \citet{jorstad_etal07}, where $a$ is a power-low fall-off in the electron density, $n_e$, with distance $r$ from the central engine, \begin{equation}n_e\sim r^{-a}.\end{equation}
These calculations are based on a helical structure of the jet magnetic field and the assumption that the constant toroidal component of the field provides the main contribution to the magnetic field along the line of sight for a spherical or conical outflow.

Table~\ref{frm} summarizes the estimated values of $a$.
Eight of our sources show $a\approx1$ to 3 suggesting a model of a jet with Faraday sheath surrounding a conically/spherically expanding jet. Two of our sources show flatter fall-off and three have higher $a$. These results agree with other, similar measurements  \citep{jorstad_etal07,osullivan_gabuzda_09,trippe_etal12}.
The observed deviations of $a$ can naturally be explained by (i) different geometry of the jet, (ii) flaring activity of the source, that results in enhancement of the poloidal magnetic field component or significant changes of the physical conditions at the jet base; (iii) possible filamentary structure of the Faraday screen, which will be smeared out within a beam at long wavelengths.

In thirteen sources we could measure the core RM at two or three frequency ranges; seven of them show reversals of RM sign (see Fig.~\ref{fig_0148}).
Such behaviour may result from a complex structure of a source: the $\tau\sim1$ surface could be blended by highly polarized optically thin jet emission. 
Another explanation is variations of the the jet viewing angle, resulted in prevailents of different LoS component of the magnetic field. 
Such a possibility may arise within a models of helical magnetic fields \citep{osullivan_gabuzda_09} or relativistic helical motion in the jet \citep{broderick_loeb_09}. Changes in the jet viewing angle may originate from small changes in the jet speed (acceleration or deceleration) or jet bends, which have frequently been observed in AGN jets \citep[e.g.][]{2014AA...571L...2R,2016AJ....152...12L}.

\subsection{Electric Vector -- Jet Direction Alignment}
\label{s:evpa_theta}

\begin{figure*}
 \centering
 \includegraphics[width=0.5\textwidth,angle=270]{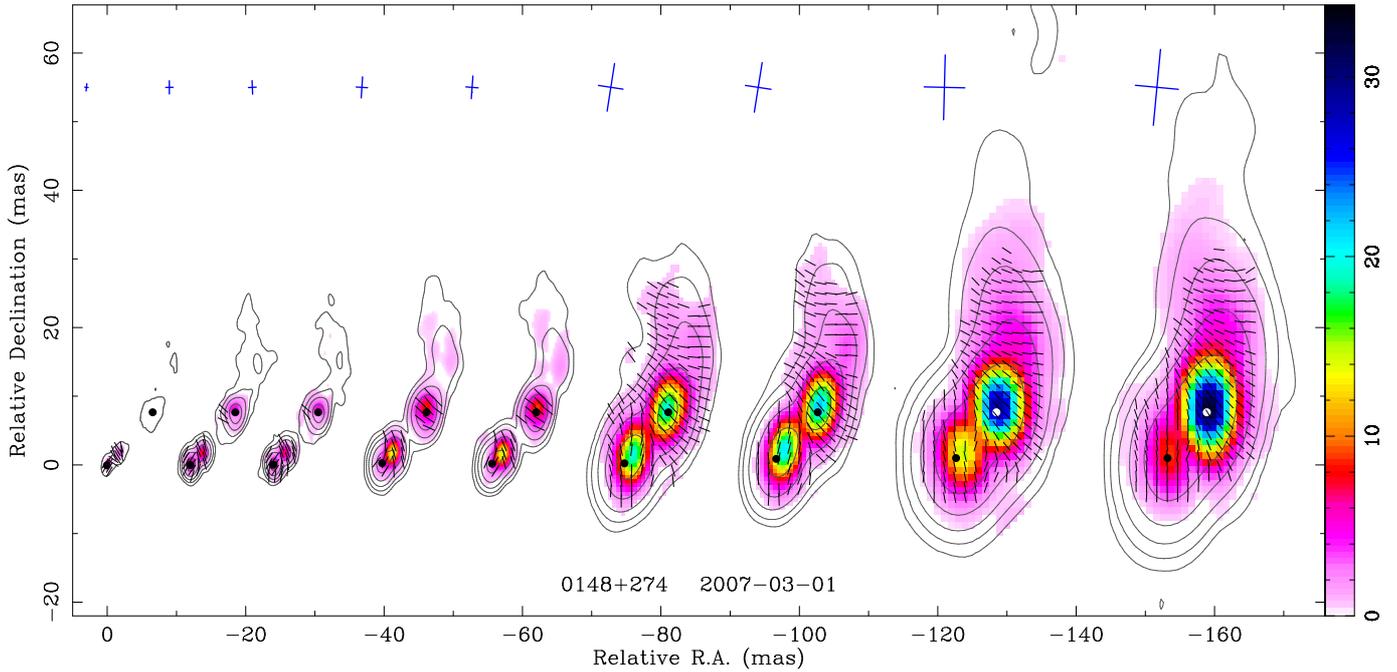}
 \caption{Faraday-corrected maps of electric vector position angle supplemented by linear polarization colour maps for 0148$+$274 in the 1.4~GHz to 15.4~GHz range. 
  Stokes $I$ contour plots together with E-field ticks for 1.4, 1.6, 2.2, 2.4, 4.6, 5.0, 8.1, 8.4, and 15.5~GHz are artificially shifted along R.A.\ or dec.\ axis in the negative direction such that optically thin jet components lay on the same horizontal or vertical line, respectively. 
  Black and white dots mark the positions of the modelled transparent jet and opaque core components.
  $I$ contours begin at the $5\sigma$ level and are plotted in $\times4$ steps.
  Results for other targets are presented in the electronic version of the article.
  \label{mfcore}
 }
\end{figure*}

\begin{figure}
 \centering
 \includegraphics[width=0.38\textwidth]{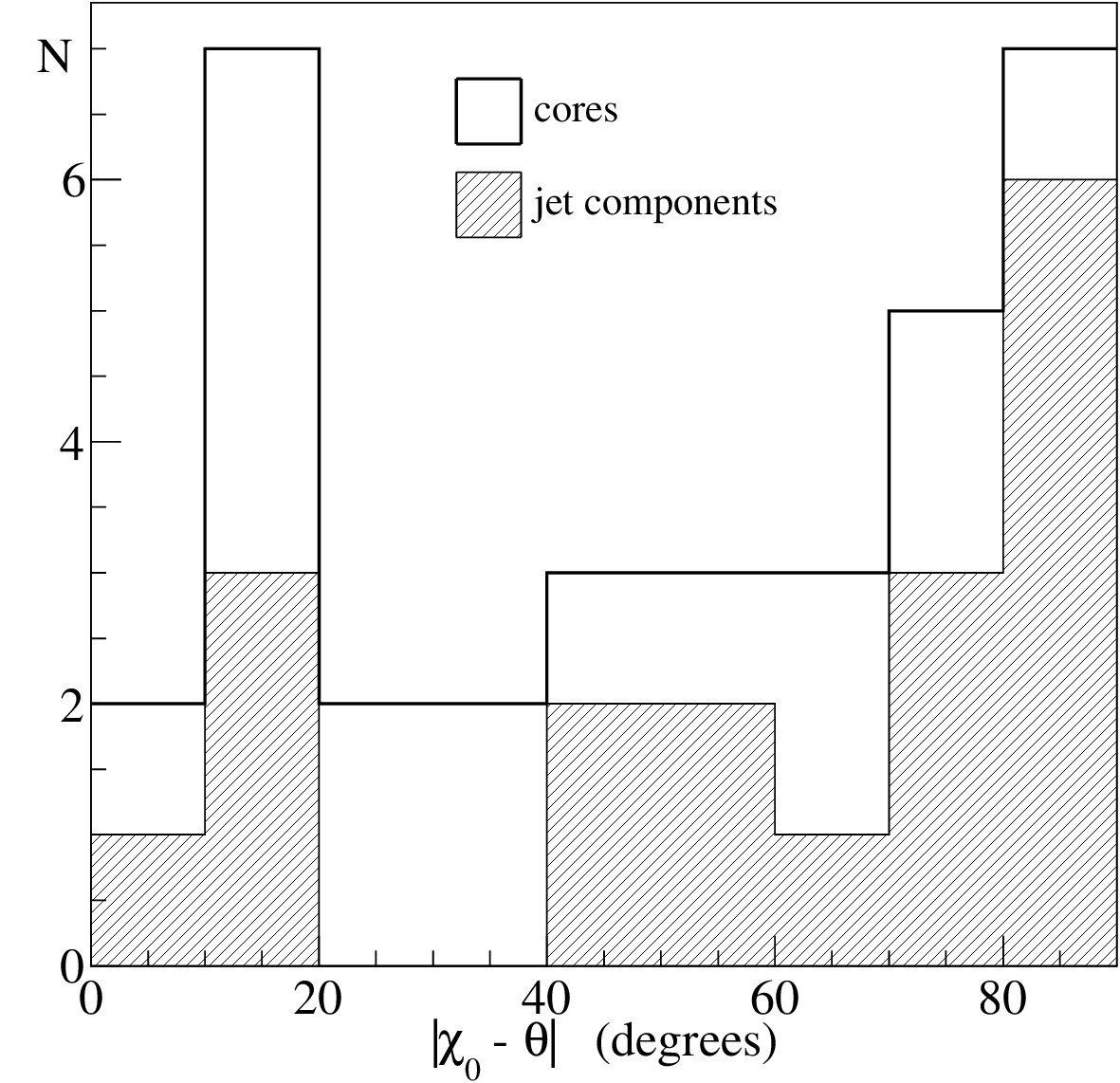}\\
 \caption{Distribution of the difference between the electric vector position angle, $\chi_0$, and the jet direction, $\theta$, at the jet and core components from Table~\ref{t:evpa_jet}.
The thick line represents the total distribution of all components. 
White filling corresponds to the core and hatched filling corresponds to the jet components. 
  \label{evpa_jet}}
\end{figure}

The intrinsic direction of the electric field can be determined after performing the Faraday de-rotation.
Figure~\ref{mfcore} shows Faraday-corrected maps of EVPA across the 1.4~GHz to 15.4~GHz range.
The misalignment between corrected EVPA, $\chi_0$, and the local jet position angle, $\theta$, is shown in Fig.~\ref{evpa_jet} and in Table~\ref{t:evpa_jet}.
These are median values obtained along ridge lines through all frequencies at the positions of model-fitted core and jet components.
There is weak excess of sources with intrinsic polarization angles inclined nearly perpendicular to local jet direction in the transparent jet components. 
Partially, this can be explained by shearing of the tangled magnetic field due to velocity gradient between the jet and the surrounding medium \citep{1980MNRAS.193..439L,laing81}.
The EVPAs in the core region distribute their values at any angle between $0^\circ$ and $90^\circ$ to the jet direction.

Generally, there is no overall trend for EVPA to be either perpendicular or parallel to the local jet direction as expected from theoretical studies \citep[e.g.][]{1980MNRAS.193..439L,lyutikov_etal05,2006MNRAS.367..851C,2009MNRAS.395..524N}.
Similar results also were shown by \citet{pollack_etal03,lister_homan_05,hovatta_etal12} and \citet{agudo_etal14}. 
The absence of preferred alignment of $\chi_0$ with $\theta$ may be caused by a number of effects, like oblique shocks, superposition of multiple components with different spectral or opacity properties within the telescope beam, residual Faraday rotation, asymmetric jets, and others.
Further discussion is given in Section~\ref{s:inds}, where we consider the polarized structure of the each source individually.

\subsection{Jet Polarization Structure}

Generally, the majority of the sources show simple polarized structure in the transparent jet regions, resulting from small depolarization and external nature of the Faraday rotating media. 
This resulted in an overall tendency of the sources to maintain their E-field direction across the full observed bandwidth in these regions.
At the same time, optically thick jet regions show complex behaviour, naturally expected for our sources, since they exhibit large apparent core shifts.

The following complex polarized structures are observed: (i) peaked behaviour of the EVPA with $\lambda^2$, which arises within a model of multiple rotation measure components \citep{gl_84}; (ii) $\sim$ 90\degr\ EVPA flips along the jet -- the possible consequence of a change in the jet viewing angle \citep{lyutikov_etal05}, interaction of the jet with ambient medium \citep{1999ApJ...518L..87A}, relativistic shocks in a jet \citep{1994ApJ...437..108B,1994ApJ...437..122W} or instabilities driven by velocity shear \citep{2004ApSS.293..117H}; (iii) ``spine-sheath" structure \citep{1996ASPC..100..241L,2007ApJ...662..835M}; (iv) EVPA -- jet direction alignments other than 0\degr of 90\degr\ in the core region, which might be caused by flaring activity of the source and consequent emergence of a new jet component \citep[e.g.][]{1994ApJ...437..108B,2016MNRAS.462.2747K}.
In Section~\ref{s:inds} we analyze the polarization structure of each source individually and discuss how the abovementioned effects can modify the intrinsic orientation of the electric field.

From the overall analysis one may conclude, that despite complex observed polarized structure, our targets can well be described by the model of a ``spine-sheath" jet \citep{1996ASPC..100..241L,1999ApJ...518L..87A,2007ApJ...662..835M} which has previously been suggested for individual sources \citep[e.g.][]{2008ApJ...682..798A,asada_etal10}. Within this suggestion, the observed rotation measures originate in the magneto-ionic boundary layer (sheath), located external to the synchrotron emission region of the jet. Results of the Section~\ref{s:inds} show that there is no preferred orientation of the magnetic field in the sheath: it can be both randomized and ordered. In some cases, sheath may carries signatures of the helical field. Meanwhile jet itself (spine) contains well ordered magnetic fields, with the apparent dominance of either poloidal or toroidal component.
Unfortunately, observed B-field orientation is ambiguously connected with the real (intrinsic to the source) orientation. 
A number of factors are responsible for this, like relativistic boosting and abberation \citep{lyutikov_etal05}, or an origin of the observed emission within a part of a jet volume, thus only some fraction of the jet is visible over time.

\subsection{Spatial Magnetic Field Geometry}
\label{s:3dmet}

The bright region visible at the aparent base of the jet in all observed sources is referred to as the VLBI core (see Fig.~\ref{fig_0148}).
The most likely physical interpretation of the core is a surface in a continuous jet flow where the optical depth of synchrotron-emitting jet plasma becomes equal to unity, $\tau \approx 1$, see also discussion by \cite{marscher08}.
Strong support for the $\tau \approx 1$ interpretation is provided by the numerous detection of the significant core shift effect, i.e.\ a systematic change of aparent position of the VLBI core as a function of observing frequency \citep[e.g.][]{kovalev_etal08,2009MNRAS.400...26O,2011Natur.477..185H,sokolovsky_etal11,pushkarev_etal12}. 
This can be naturally understood: while the jet plasma density and magnetic field strength gradually decreas with increasing distance from the central engine, the jet becomes transparent at progressively lower frequencies \citep[see][]{marcaide_shapiro_84,lobanov98,kutkin_etal14}.

Therefore by observing the core at various frequencies we obtain information about the jet properties (including polarization) at various distances from the central engine.
Combining this analysis with the observed polarization angle, corrected for Faraday effects, one may  recover the spatial magnetic field geometry in this region.
This technique gives only rough estimate of the intrinsic structure of the magnetic field, because of a number of factors.
The most significant are (i) the exact geometry of the jet must be known (e.g. jet viewing angle), otherwise only the sky projection of the field geometry is available; (ii) small viewing angle with relativistic and Doppler effects may dramatically distort the intrinsic direction of polarization; (iii) jet curvature which is changing with frequency (thus with distance), which results in frequency-dependency of the jet viewing angle; (iv) the shape of the $\tau\sim1$ surface may change with frequency; (v) opacity changes from frequency to frequency, which may produce 90\degr\ EVPA flips.

\begin{figure}
 \centering
 \includegraphics[width=0.4\textwidth,angle=-90]{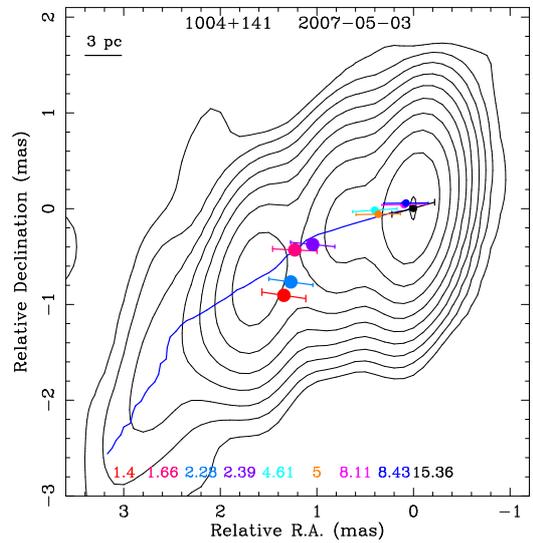}
 \caption{1004$+$141 spatial geometry (sky projection) of the electric vectors (colour sticks), reconstructed by using the technique described in Section~\ref{s:3dmet}.  The frequency range is indicated by numbers and colours on the figures. The solid blue line represents the jet ridge line.
 Relative 15~GHz Stokes $I$ contours are shown.
 Error bars at the end of the sticks represent the $1\sigma$ total EVPA error, while $1\sigma$ uncertainties in the position of model-fitted components are given by filled circles.
 \label{mfcore_3d}
 }
\end{figure}

Because of the lack of information, which could account influence of abovementioned factors, we are limited in studying the spatial magnetic field geometry for our sample. 
Figure~\ref{mfcore_3d} presents an example of application of this approach to 1004$+$141 (sky plane projection) superimposed with the 15.4~GHz Stokes $I$ contours and jet ridge line.
Assuming a typical jet viewing angle of $6^\circ$ \citep{savolainen_etal10,pushkarev_etal15,2014MNRAS.441.1899F}, the detailed view of the 1004$+$141 core region reveals the existence of ordered electric (and thus magnetic) field, keeping its direction for a few hundreds of pc.

Maps of the Faraday-corrected direction of the electric field over the whole source are given in Fig.~\ref{mfcore} in the full frequency range. 
To apply proposed technique we (i) mark the positions of model-fitted core and jet components and (ii) align images at the position of optically thin components in Fig.~\ref{mfcore}.
Overlaying all frequencies will give the spatial geometry of the magnetic field in the cores.
This approach is not applicable in the optically thin regime, where we see the same physical regions through all frequencies. 
Analysis of these regions is made in Sections~\ref{s:evpa_theta} and ~\ref{s:gfrm}.
To study whether these results can be influenced by relativistic or other effects, a combined polarimetric and kinematic analysis should be done, which is the subject of future research.

\subsection{Optically Thin Regions and Galactic Rotation}
\label{s:gfrm}

Results of Sections~\ref{s:frm}, \ref{s:trgr} and \ref{s:rm-core} have shown, that observed Faraday rotation occurs in an external screen, located in the vicinity of AGN. Despite this, we may assume that optically thin jet emission experiences only rotation in the Galactic medium. 
Thus, we can determine the Galactic rotation measure.
This approach also requires that synchrotron emission originates within the same volume in the source to avoid frequency--dependence of the polarization angle.
Under these assumptions, we did not substract Galactic rotation measure and applied linear fit to the observed EVPA vs. $\lambda^2$ over the whole frequency range.
Moreover, we calculated the integrated RM value over the 1.4~GHz to 2.4~GHz RM maps.
The resulting values are presented in Table~\ref{gal_rm} and agree well with the values of \citet{taylor_etal09}.

\citet{2015ApJ...815...49A} recently suggest that the Galactic medium has a complex Faraday structure. 
Despite our sparse frequency coverage in the same region, we assume that this complexity rather arises within the radio sources, than in the Galactic medium. 
This is supported by results of \citet{2016ApJ...829....5L} obtained with the polarization survey of the northen sky at 1.4~GHz and 2.3~GHz.

\begin{table}
  \caption{Misalignment between EVPA in the core, $\chi^\mathrm{core}_0$, and jet components, $\chi^\mathrm{jet}_0$, and the jet direction, $\theta$.\label{t:evpa_jet}}
  \begin{center}
  \begin{tabular}{ccc}
  \hline
  Source&$|\chi^\mathrm{core}_0-\theta|$&$|{\chi}^\mathrm{jet}_0-\theta|$\\
  \hline
0148$+$274 & 15 & 81\\
0342$+$147  &  18 &   75\\
0425$+$048 &   37  & 41\\
0507$+$179  &  72  &  61\\
0610$+$260 & -- & -- \\
0839$+$187 &   82  &  85\\
0952$+$179  &  65   & 15\\
1004$+$141  & 40  & 13\\
1011$+$250  & -- & 86\\
1049$+$215  &  48 &   51\\
1219$+$285  &  18   & 83\\
1406$-$076  &  67  &  87\\
1458$+$718   & 3 &87\\
1642$+$690 &   71&    45\\
1655$+$077  &  27  &  4\\
1803$+$784  &  15 &   15\\
1830$+$285   & -- & 77\\
1845$+$797 & -- & -- \\
2201$+$315 &   55  & 78\\
2302$+$506  & 27   & 55\\
\hline
\end{tabular}
\end{center}
Presented are median values for measurements at all frequencies (in degrees). 
A 1$\sigma$ accuracy of misalignments is 10$^{\circ}$.
A histogram representing this table is given in Fig.~\ref{evpa_jet}.
\end{table}

\subsection{Discussion of Individual Sources}
\label{s:inds}

\subsubsection{0148$+$274}
This quasar shows an rotation measure of $168\pm13$~\rmu\ in the core and $-42\pm14$~\rmu\ in the jet at 4.6~GHz to 15.4~GHz (Fig.~\ref{fig_0148}.1). 
The average RM value over the 1.4~GHz to 2.4~GHz map amounts $-9\pm5$~\rmu. 
The jet EVPA vs. $\lambda^2$ shows a close to linear dependence, however the core region breaks this law. 
The polarization degree does not exceed 6 per~cent anywhere in the jet and shows complex behaviour with $\lambda^2$ both in the core and jet regions. 
Its peaked behaviour in the opaque core may be explained by smearing of two RM components, seen in the EVPA vs. $\lambda^2$ plot, or by superposition of two jet components with different polarization properties. 
A close view of this region in Fig~\ref{mfcore}.1 and Fig.~\ref{fig_1803_u1} supports the later suggestion. 
Meanwhile the transparent jet posses a polarization signature close to the model of anomalous and inverse depolarization \citep{sokoloff_etal98,homan_12}.
The source exhibits a significant RM gradient across its jet over 4.6~GHz to 15.4~GHz. 
Together with a peaked $m-\lambda^2$ behaviour it may point to a helical magnetic field in the source.
At a distance of $\sim$3~mas from the core, the fractional polarization (see Fig.~\ref{fig_1803_u1}) brightens towards the left edge and fades towards the right jet edge, accompanied by rotation of the E-vectors. 
Such behaviour is explained by relativistic abberation in helical fields \citep{clausen_etal11}, suggesting the sheath magnetic field is more aligned with the line of sight on the left side of the jet.
Polarization vectors are aligned with the jet direction in the core, and are aligned perpendicular in the jet, which implies dominance of the poloidal magnetic field component in the 0148$+$274 parsec-scale jet.

\subsubsection{0342$+$147}
The core of this quasar shows a slightly inverted spectrum and a non-linear EVPA vs. $\lambda^2$, resulting in an RM of $-431\pm51$~\rmu at 8.1~GHz to 15.4~GHz (Fig~\ref{fig_0148}.2). 
The jet EVPA is consistent with a linear $\lambda^2$ law and posses low values of RMs, consistent with 0~\rmu. 
All of the transverse RM slices have width less then 1.5 beam, and do not show any signs of a gradient.
The power-law model fits well with the observed degree of polarization both in the core and jet components, implying the Faraday rotation occurs in an external screen. 
High $m$ at $\sim$7~mas from the core ($\sim$ 15 per~cent) might be an indication of a relativistic shock \citep{2009MNRAS.395..524N} at this position (Fig.~\ref{fig_1803_u1}). 
Nevertheless, the overall EVPA orientation is consistent with the magnetic field being aligned with the local jet direction (Fig~\ref{mfcore}.2).

\subsubsection{0425$+$048 (OF~42)}
The core of this AGN shows a inverted spectrum and breaks the linear EVPA vs. $\lambda^2$ relation over all frequencies, as seen in Fig~\ref{fig_0148}.3. 
Estimates of RM in this region are a few hundreds of \rmu\, while the RM fit in the optically thin region gives a value of a few \rmu.
The core polarization degree behaviour with wavelength is consistent with external Faraday rotation. 
At the same time the jet $m-\lambda^2$ shows a slow fall-off with decreasing frequency, which may be explained by a randomly oriented magnetic field.

The Faraday-corrected EVPA is oriented perpendicularly in the core and changes to parallel direction downstream in the jet (Fig.~\ref{fig_1803_u1}).
The active state of the source or decrease in the opacity may account for such behaviour.
Downstream, the jet EVPA is oriented $\sim40^{\circ}$ to the jet and rotates by almost 90\degr\ at the position of model-fitted jet component (Fig.~\ref{mfcore}.3) where the jet bends.
No RM enhancement is observed at the position where the jet makes the turn, which might indicate an interaction with the surrounding medium or presence of instability driven structures. 
We were unable to conclude about the overall direction of the magnetic field in the quasar. 
It seems, that the contribution of a randomly oriented magnetic field is significant there.

\subsubsection{0507$+$179}
This quasar shows inverted spectrum in the core, as seen in Fig.~\ref{fig_0148}.4. 
The fractional polarization there does not exceed 2~per~cent and shows a tendency to increase its value towards longer wavelengths. 
\citet{trippe_etal12} observed the maximum degree of polarization of $\sim$ 11~per~cent at 80~GHz to 253~GHz, and find significant temporal variations in this value.
The core EVPA changes non-linearly with $\lambda^2$ with the indication of increasing RM value towards shorter wavelengths. 
We estimated an RM of 1055$\pm$53~\rmu\ in 8.1~GHz to 15.4~GHz, which is consistent with their 3$\sigma$ upper limit of $14000$~\rmu\ over 80~GHz to 253~GHz provided by \citeauthor{trippe_etal12}.
The polarization degree in the optically thin jet component is consistent with the model of an external, filamentary Faraday screen, supported by the linear EVPA behaviour with $\lambda^2$.
Even though the width of the transverse RM slices is $\geq$1.5~HPBW, we see no evidence of RM gradients. 
The Faraday--corrected polarization vectors (Fig.~\ref{mfcore}.4 and ~\ref{fig_1803_u1}) change direction along the jet from $-$72\degr\ in the core to $\sim-$90\degr at a 2.5~mas, and 62\degr at 4~mas, implying poloidally dominated magnetic field in the quasar.

\subsubsection{0610$+$260 (3C~154)}
This quasar does not show strong linearly polarized flux at most of the frequencies.
Observable orientation of the polarization angles is presented in Fig.~\ref{mfcore}.5.

\subsubsection{0839$+$187}
This quasar has a steep--spectrum core, as shown in Fig.~\ref{fig_0148}.6. 
Fractional polarization in the core changes with frequency following the model of an external Faraday screen. 
EVPA is not linear with $\lambda^2$ and exhibits large rotation towards short wavelengths. The core RM can be determined in a few pixels only and is of $-607\pm18$~\rmu over 4.6~GHz to 8.4~GHz. Observations of the source at higher frequencies may result in higher values of the rotation measure in its core.
The transparent jet region shows frequency--independent fractional polarization, which is expected if either randomly oriented or regular component of the magnetic field dominates in the Faraday rotating medium \citep[e.g.][]{burn66}.
RM gradients are not observed, although the jet is resolved.
The orientation of the polarization angle (Fig.~\ref{mfcore}.6 and ~\ref{fig_1803_u1}) maintains to transverse direction along the extended jet, which is consistent with a poloidally dominated magnetic field.

\subsubsection{0952$+$179}
The flat--spectrum core of this source is weakly polarized: polarization degree does not exceed 2~per~cent. 
The fractional polarization vs. $\lambda^2$ shows a complex behaviour, which is close to the model of a spectral repolarizer \citep{conway_etal_74}, when a two jet components with different polarization properties are blended.
Assuming a linear fit over 8.1~GHz to 15.4~GHz, we estimated an RM of $-2767\pm53$~\rmu\ in the quasar's core.
This rotation measure translates to $-16900\pm300$~\rmu\ in the rest frame of the source, and thus represents the highest observed rotation measure in our sample.

The optically thin jet shows a complex polarized structure. 
While fractional polarization decreases with $\lambda^2$, the observed EVPA experiences a 40\degr\ jump between 2.4~GHz and 4.6~GHz. 
Differential Faraday rotation \citep{gardner_whiteoak_66,sokoloff_etal98} may account for such polarization signatures, implying co-spatiality of emitting and rotating regions.
Faraday--corrected polarization angles oriented identically over all observed frequencies anywhere in the transparent jet.
This supports the suggestion about an internal Faraday rotation in this jet region, and explains frequency--dependency of the RM.
Significant rotation measure gradients are observed over 1.4~GHz to 2.4~GHz, 4.6~GHz to 8.4~GHz and 8.1~GHz to 15.4~GHz RM maps (Fig.~\ref{fig_0148}.7).
RM--corrected electric vectors are consistent across frequencies and aligned with the jet in this region, meanwhile the EVPA is perpendicular to the jet downtream and upstream of this region.
The apparent bend of the jet at the position of the jet component and change in direction of electric vectors suggests that the jet experiences interaction with the ambient medium, resulting in a local enhancement of the toroidal field component and RM value ~\citep[see, e.g.][]{2008ApJ...681L..69G}.
In this context, transverse RM gradients may arise from the variations in magneto-ionic density and path along the Faraday rotating medium, rather than in changes in the direction of the helical magnetic field. 
The magnetic field in the quasar 0952$+$179 jet may be dominated by poloidal component.

\subsubsection{1004$+$141}
The core EVPA breaks the $\lambda^2$ law and increases its slope towards shorter wavelengths (Fig.~\ref{fig_0148}.8).
A fit to a $\lambda^2$ law over 8.1~GHz to 15.4~GHz is not reliable, while a fit across 4.6~GHz to 8.4~GHz agrees with a linear law and gives RM of $-12\pm17$~\rmu.
RM decreases to $-32\pm17$~\rmu\ in the transparent jet. 
The degree of polarization both in the core and jet is consistent with depolarization in an external medium, which may partially cover the emitting jet region \citep{2008AA...487..865R,2009AA...502...61M}.
Transverse RM slices exhibit a gradient with 2.5$\sigma$ significance and width of 1.8~HPBW.
The EVPAs keep their relative orientation with respect to the jet direction for the whole detected parsec-scale jet showing a very good alignment between the magnetic field and the local jet direction (see Fig.~\ref{fig_1803_u1} and Fig.~\ref{mfcore_3d}).
This implies dominance of the toroidal field component.
Together with the RM gradient this may point to a helical magnetic field in the 1004$+$141 jet.

\subsubsection{1011$+$250}
This quasar exhibits significant linear polarization only in the region located $\sim$ 7~mas from the core as shown in Fig.~\ref{mfcore}.9. 
A reliable RM fit can be made over 2.2~GHz to 8.4~GHz and results in the value of $-157\pm29$~\rmu (see Fig.~\ref{fig_0148}.9). 
The RM-corrected EVPA of the jet is consistent with a poloidal magnetic field.

\subsubsection{1049$+$215}
The core RM increases from $-4\pm4$~\rmu over 1.4~GHz to 2.4~GHz, to $-107\pm17$~\rmu\ and $-547\pm52$~\rmu\ across 4.6~GHz to 8.4~GHz and 8.1~GHz to 15.4~GHz respectively (Fig.~\ref{fig_0148}.10). 
Fractional polarization is complex with $\lambda^2$ in this region, while the spectrum shows partial opaqueness.
The jet RM fit across 1.4~GHz to 8.4~GHz is consistent with 0, however the 15.4~GHz EVPA shows significant rotation in its value, giving rise to an RM of $-656\pm239$~\rmu\ at 8.1~GHz to 15.4~GHz. 
Most probably this value is not real, resulting from the convolution of the 15.4~GHz map with the beam size of the 8.1~GHz map.
The jet fractional polarization is frequency--independent, which is an indication of tangled magnetic fields.
Corrected for Faraday rotation, the polarization vectors (Fig.~\ref{mfcore}.10) are transverse to the local jet direction in the transparent component located 8~mas from the core. 
At the same time, the EVPA varies significantly near the core region, as does the degree of polarization (given in Fig.~\ref{fig_1803_u1}). 
Kinematic study by \citet{2016AJ....152...12L} indicates a stationary feature at a position of 2~mas from the core of the source.
Such behaviour of the polarization properties resembles a relativistic shock or bend of a jet.

\subsubsection{1219$+$285 (W~Comae)}
\begin{figure}
 \centering
 \includegraphics[width=0.3\textwidth,angle=270]{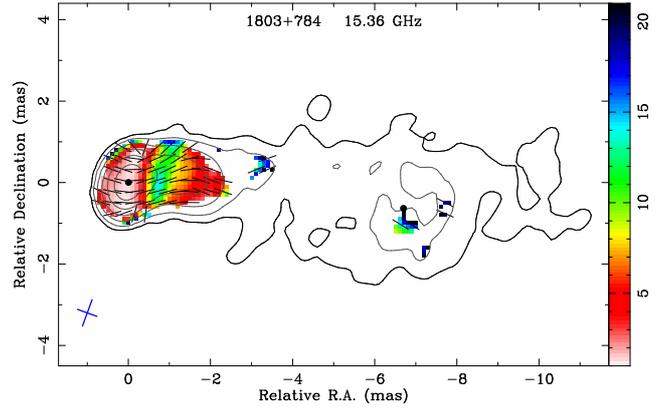}
 \caption{Faraday-corrected map of E-vectors supplemented by linear fractional polarization colour map for 1803$+$784 at 15.4~GHz.
 Black dots mark the positions of the modelled transparent jet and opaque core components.
 Stokes $I$ contours begin at the $4\sigma$ level and are plotted in $\times4$ steps. The color scheme is in mJy.
 Results for other targets are presented in the electronic version of the article.
  \label{fig_1803_u1}
 }
\end{figure}

The blazar 1219$+$285 shows a typical flat-spectrum core and steep-spectrum jet (Fig.~\ref{fig_0148}.11). 
Significant repolarization is seen in the core towards long wavelengths.
The core offset in the EVPA between 1.4~GHz to 2.4~GHz and 4.6~GHz to 15.4~GHz is 50\degr, and may arise from the opacity change.
The estimated RM value does not change significantly through jet and is $-27\pm15$~\rmu, while the transverse RM gradient ($\sim$8~mas downstream) is significant at the $2.7\sigma$ level and has a width of 2.3~HPBW.
\citet{2015MNRAS.450.2441G} detected a $3\sigma$ RM gradient in the blazar across 5~GHz to 15~GHz, with an RM slice about 2.5~mas from the core. 
This differs from our measures in sign of the gradient though our absolute values are similar. 
Our RM-corrected orientation of linear polarization (given in Fig.~\ref{fig_1803_u1}) is coaligned and misaligned with the jet in the core and jet ($\sim$~6~mas) components respectively. 
The VLBI observations of \citet{1994ApJ...435..140G} at 5~GHz show parallel EVPA orientation at the $\sim$~5~mas component, while their core EVPA rotates from perpendicular to parallel orientation within a 2~yr interval.
Kinematic study \citep{2016AJ....152...12L} reveals two stationary features in the blazar at $\sim$~0.2 and 1~mas downstream from the core.
The flaring activity and consequent interaction of ejected material with these feature may account for time variations of the blazar polarized properties.
At 8~mas from the core the jet turns (see Fig.~\ref{mfcore}.11), accompanied by EVPA change from parallel to perpendicular orientation relative to the jet direction.
This complex polarized structure of the blazar does not allow us to make a firm conclusion about its overall magnetic field structure.

\subsubsection{1406$-$076}
Small depolarization in the quasar's core and jet at longer wavelengths (Fig.~\ref{fig_0148}.12) points to external Faraday screen with randomly oriented magnetic field. 
The core EVPA steady rotates towards short wavelengths, resulting in an increase of RM from $30\pm4$~\rmu, to $-47\pm5$~\rmu and $-235\pm17$~\rmu at 1.4~GHz to 2.4~GHz, 2.2~GHz to 5.0~GHz and 5.0~GHz to 15.4~GHz respectively.
\citet{hovatta_etal12} measure a core RM of $-505\pm108$ across 8~GHz to 15~GHz (epoch 2006--04--05), which is consistent with our value.
\citet{2011ARep...55..400V} report RM estimates of 64~\rmu at 5.0~GHz to 15~GHz (epoch 1997--04--05) and a degree of polarization twice as high (6.1 per~cent at 5~GHz). 
As shown in Fig.~\ref{mfcore}.12 and ~\ref{fig_1803_u1}, the corrected electric vector in the core is 67\degr and rotates by 90\degr\ at 0.5~mas downstream due to opacity changes.
Such complex behaviour in the core might be explained by the emergence of a new component, which smears out the parallel orientation of the core EVPA. 
Nevertheless, this complex polarized structure of the quasar is consistent with an overall poloidal magnetic field.

\subsubsection{1458$+$718 (3C~309.1)}
The quasar shows steep spectra both in the core and jet, as seen in Fig.~\ref{fig_0148}.13. 
The fractional polarization has a peaked shape both in the core and jet. 
Meanwhile, this behaviour in the core may be explained by smearing of two jet components within the beam, this is unlikely to be the cause of anomalous depolarization in the optically thin region. 
A helical field may account for this behaviour, which can be applied to the core as well.
\citet{2009AA...502...61M} observed 3C~309.1 with 100--m Effelsberg telescope in 2.64~GHz -- 10.45~GHz range in the period 2006--01--26 -- 2006--06--26. \citeauthor{2009AA...502...61M} obtained the same spectral distribution of the fractional polarization, though they obsered with the single--dish telescope.
Rotation measure steadly increases towards high frequencies, up to $1153\pm56$~\rmu\ across 8.1~GHz to 15.4~GHz.
The observation of \citet{hovatta_etal12} on 2006--09--06 over 8~GHz to 15~GHz range gives a core RM of $628\pm125$~\rmu, and is observed over few pixels only.
Measures of \citet{1997VA.....41..225A}, made on 1995--03--24 and 1995--12--11 with the VLBA at 5~GHz and 8.4~GHz, account the core RM of $\approx$450 \rmu.
The jet RM is $-31\pm53$~\rmu\ compared to $231\pm103$~\rmu\ of \citeauthor{hovatta_etal12}
The jet does not exhibit RM gradient at high frequencies, though the 1.4~GHz to 2.4~GHz RM map shows complex Faraday structure: three transverse RM profiles taken at different jet locations (see Fig.~\ref{fig_trs}) present significant RM gradients. 
These gradients are sharp and may reflect changes in the density and thickness of magneto-ionic media rather than in magnetic field.
RM--corrected polarization angles are presented in Fig.~\ref{mfcore}.13 and Fig.~\ref{fig_1803_u1}.
The core and jet EVPAs for regions up to 30~mas are well--ordered and consistent with an apparent poloidal magnetic field in the quasar jet.
Meanwhile, the polarization signature 30~mas downstream resembles a `spine-sheath' structure, indicating dominance of the toroidal field component in the jet spine and poloidal in its sheath.
Such polarization structure has previously been observed, e.g. in the jet of 1652$+$398 \citep[Mrk~501,][]{2005MNRAS.356..859P}. 
Moreover, both 1458$+$718 and 1652$+$398 have similar kinematic picture in their jets \citep{2016AJ....152...12L}.
If we assume that the jet of 1458$+$718 indeed has a spine-sheath structure, then the transition from apparent $\mathrm{B}_{\parallel}$ to $\mathrm{B}_{\perp}$ at 30~mas distance from the core may be explained by a change in the viewing angle.
An increase in the fractional polarization at the model-fitted jet component support this assumption, since the sheath is expected to have a more uniform magnetic field \citep{lyutikov_etal05}.

\subsubsection{1642$+$690}
Behaviour of the EVPA and $m$ in the core of this quasar (Fig.~\ref{fig_0148}.14) is similar to that descibed by a model of multiple Faraday components. 
Therefore our linear RM fits may differ from the real values. 
The fractional polarization at 86~GHz is $1.1\pm0.3$ per~cent (2010--08--15, \citealt{agudo_etal14}) and is comparable to our measures.
The core RM changes sign between 1.4~GHz to 8.4~GHz and 8.1~GHz to 15.4~GHz, while downstream the jet RMs have comparable values.
\citet{hovatta_etal12} observe a core RM of $148\pm75$~\rmu\ over 8~GHz to 15~GHz (epoch 2006--03--09), which is smaller than our measure ($272\pm50$~\rmu).
The jet fractional polarization and electric vector behaviour with $\lambda^2$ are consistent with the inverse depolarization law taking place in a helical magnetic field \citep{homan_12}.
Transverse RM slices are resolved and show gradients across the 4.6~GHz to 8.4~GHz and 8.1~GHz to 15.4~GHz bands.
RM--corrected polarization structure of the jet (Fig,~\ref{mfcore}.14 and Fig.~\ref{fig_1803_u1}) is complex: electric vectors are perpendicular to the jet from the jet base up to $\sim7$~mas downstream, while the jet component located 10~mas downstream from the core has signatures of the `spine-sheath' structure. 
Moreover, the high degree of polarization there ($\sim$~30 per~cent) indicates well-ordered magnetic field.
Overall, this result is consistent with a large--scale helical magnetic field in this quasar.
 
\subsubsection{1655$+$077}
The core degree of polarization slightly increases towards long wavelengths, as shown in Fig.~\ref{fig_0148}.15.
The observations of the source at sub~mm wavelengths \citep{agudo_etal14} provide measures of $m$ of $9.1\pm0.3$ per~cent (86~GHz) and $7.2\pm2.1$ per~cent (229~GHz).
Taken together with our measures, these results are consistent with the external Faraday screen in the core of 1655$+$077.
At the same time, the jet fractional polarization also follows this law.
EVPA in the core rotates at a high rate, which resulted in an increase of RM from $24\pm4$~\rmu\ over 1.4~GHz to 2.4~GHz to $-491\pm13$~\rmu\ over 4.6~GHz to 15.4~GHz. 
Moreover, \citeauthor{agudo_etal14} estimate an upper limit on the quasar's RM of 41900~\rmu across 86~GHz to 229~GHz, which agrees with the power-law increase in RM value with frequency.
\citet{hovatta_etal12} measured the core RM of $\-1286\pm91$~\rmu\ on 2006--11--10.
This value significantly differs from our RM, though the time difference between our observations is six months. Such rapid time variations of the rotation measure suggests a magnetized plasma associated with the jet acts like a Faraday screen and is responsible for observed RM \citep[see, e.g.][]{asada_etal08}.
Electric vectors in the core and jet are coaligned with jet direction (see Fig.~\ref{mfcore}.15 and Fig.~\ref{fig_1803_u1}). 
If the toroidal component of the magnetic field prevails, then the core EVPA may be altered by ongoing activity in this region.
Indeed, \citet{2016AJ....152...12L} analysis shows that jet component is located within 1~mas of the core at this epoch.
Otherwise, the toroidal component of the field may lie either transverse or parallel to the jet in the plane of the sky \citep{lyutikov_etal05}.
Changes in the EVPA orientation in the core and jet components then can be attributed to the variations of the jet viewing angle.

\subsubsection{1803$+$784}
This BL~Lac object displays (Fig.~\ref{fig_0148}.16) a flat RM through the jet (from $28\pm17$~\rmu\ in the core to $11\pm24$~\rmu\ in the jet). 
While synchrotron emission is significantly depolarized and wavelength--independent in the jet, the core exhibits strong repolarization. 
Magnetic field in the core should be regular with a random line--of--sight component and transverse component, changing its direction.
In the same time a model of random magnetic field is consistent with the jet region with no transverse RM gradients.
Considering observations of the source made by different authors \citep{2003MNRAS.339..669G,jorstad_etal07,zavala_taylor_03,mahmud_etal09,hovatta_etal12}, the core region shows temporal RM variations in its value and sign (Table~\ref{t:1803}).
\citet{2013ApJ...772...14C} associate the 1803$+$784 core with a conical shock, assuming combination of random and poloidal magnetic fields, and reproduce the observed polarization structure numerically. 
Then within this model, RM changes my arise from interaction of ejected material with the shock front.
The kinematic structure of the source in this region reveals \citep{2016AJ....152...12L} four persistent features within 1~mas of the core.
Enhancement of the fractional polarization (see Fig.~\ref{fig_1803_u1}) 1~mas downstream of the core and alignment of E-vectors with the jet  also support this suggestion.
Large-scale polarization structure of the source (Fig.~\ref{mfcore}.16) shows that the parallel orientation of the EVPA is maintained along 50~mas, which is approximately 300~pc in the projection on the sky plane.
We suggest that a global toroidal magnetic field threads the jet of this source.
Other authors also favour this idea \citep[e.g.][]{1999NewAR..43..691G,2003MNRAS.339..669G}.

\begin{table}
  \caption{RM measures of 1803$+$784.\label{t:1803}}
  \begin{center}
  \begin{tabular}{lccc}
  \hline
   Epoch & $\nu$ & RM & Reference\\
   & (GHz) & \rmu & \\
\hline   
1997--04--06 & 5--22 & $-237\pm64$ & 1\\
1998--2001 & 43--222 & $4360\pm3740$ & 2\\
2000--06--27 & 8.1--15.2 & $-201\pm55$ & 3\\
2002--08--24 & 15.3--43.1 & $\sim600$ & 4\\
2003--08--22 & 4.6--15.4 & $\sim-300$ & 4\\
2006--09--06 & 8--15 & $112\pm74$ & 5\\
2007--05--03 & 5.0--15.4& $28\pm17$ & 6\\
\hline
\end{tabular}
\end{center}
\medskip

References: 1 -- \citet{2003MNRAS.339..669G}; 2 -- \citet{jorstad_etal07}; 3 -- \citet{zavala_taylor_03}; 4 -- \citet{mahmud_etal09}; 5 -- \citet{hovatta_etal12}; 6 -- this work.
\end{table}

\subsubsection{1830$+$285}
This quasar is weakly polarized and reliable RM fits to $\lambda^2$ were obtained across 2.2~GHz to 5.0~GHz only (Fig.~\ref{fig_0148}.17). 
The core RM is $-50\pm8$~\rmu\ and decreases downstream along the jet to 0~\rmu.
Faraday-corrected electric vectors are perpendicular to the jet (Fig.~\ref{mfcore}.17).

\subsubsection{1845$+$797 (3C~390.3)}
This object does not show strong linearly polarized flux density at the majority of frequencies (see Fig.~\ref{mfcore}.18).

\subsubsection{2201$+$315}
The quasar's core and jet are weakly polarized and both show fall-off in fractional polarization with $\lambda^2$ (Fig.~\ref{fig_0148}.19), being consistent with the model of external rotation.
An estimate of the core fractional polarization at 86~GHz ($\leq1$ per~cent) and 228~GHz ($\leq4.7$ per~cent, \citealt{agudo_etal14}) agree with our measures and with external Faraday rotation.
The core RM value increases towards short wavelengths, reaching $-242\pm17$~\rmu\ across 5.0~GHz to 15.4~GHz.
Two mas downstream from the core the rotation measure is $602\pm32$~\rmu\ and decreases to $15\pm27$~\rmu further from the core.
\citet{zavala_taylor_04} and \citet{hovatta_etal12} observed the 2201$+$315 jet across similar bandwidths on 2001--06--20 and 2006--10--06 accordingly. \citeauthor{zavala_taylor_04} measured the RM of 612~\rmu and 5~\rmu 2~mas and 3~mas of the core. \citeauthor{hovatta_etal12} report the RM values of $528\pm91$~\rmu and $-21\pm168$~\rmu at the same positions in the jet. Hereby, the whole 2201$+$315 jet shows no signs of temporal variations in RM value and maintains complex Faraday structure over time.
Passage of jet component \citep[as seen in][]{2016AJ....152...12L} through core region may be responsible for observed complex structure in all observations.
However absence of the temporal RM variations makes this suggestion to be unlikely.
A relativistic shock may account for this RM enhancement, as, e.g. has been observed by \citet{2016ApJ...817...96G} in BL~Lac.
Kinematic analysis shows a stationary feature in the 2201$+$315 jet \citep{2016AJ....152...12L} 0.5~mas downstream from the core, far from the region where rotation measure increases. 
The corrected core EVPAs (see Fig.~\ref{mfcore}.19, Fig.~\ref{fig_1803_u1} and abovementioned works) change longitudinal direction (epoch 2001--06--20) to transversal (2006--10--06 and 2007--04--30), meaning that this region is affected by source activity.
The magnetic field shows an apparent poloidal structure downstream from the core.
In turn, RM gradients are observed in the 2201$+$315 jet (Table~\ref{t:grads} and Figure~\ref{fig_trs}), which is an indicator of a helical magnetic field.

\subsubsection{2320$+$506}
Weak linear polarization in the core and jet of this quasar, given in Fig.~\ref{fig_0148}.20, follows the model of external Faraday rotation. 
The core RM increases from $44\pm4$~\rmu\ over 1.4~GHz to 2.4~GHz to $-1627\pm18$~\rmu\ over 5.0~GHz to 15.4~GHz, while jet RM is $19\pm4$~\rmu.
The linear polarization (Fig.\ref{mfcore}.20) is aligned and misaligned in the core and jet accordingly, which is a signature of a poloidal magnetic field. 
At the the same time E-vectors show (Fig.~\ref{fig_1803_u1}) complex structure 1--3~mas downstream from the core: vectors are parallel at one side and turn to perpendicular direction at another side of the jet, accompanied by asymmetry in fractional polarization across this region.
Such geometry arises in a model of helical magnetic fields \citep{lyutikov_etal05,2011ApJ...737...42P}.
In this context the observed poloidal magnetic field may originate in a jet sheath, while the spine holds toroidal field.

\section{Summary}
\label{sum}

We have performed Faraday rotation measure and polarization analysis of 20 AGN jets through VLBI imaging at nine frequency bands across the 1.4~GHz to 15.4~GHz range. 
Fractional linear polarization of opaque cores is typically found on a level of a few per~cent and can rise above 20~per~cent for jet regions, similar to results of other studies.
Wavelength-dependent variation of the fractional polarization in both the core and jet regions is consistent with the models of external Faraday rotation. 
Fractional polariation in the majority optically thin jet components decreases with increasing wavelength.
Though a few sources show low level of depolarization in these regions, which is an indication of the presence of either randomly oriented or regular magnetic fields.
An anomalous depolarization is observed in a few transparent jet components, indicating presence of the internal Faraday rotation, taking place when emitting region is co-spatial with the rotating environment.
Besides, a number of opaque cores exhibit polarized structure, produced by smearing of a multiple jet or RM components within the telescope beam, or show signatures of internal Faraday rotation.

The EVPAs in jet regions with optically thin synchrotron emission follow a linear $\lambda^2$ law, resulting in an absolute observed rotation measure in the range from 0 to 167~\rmu\ with a median value of 22~\rmu. 
RMs in the opaque cores systematically increase with frequency, breaking the $\lambda^2$ law, with the maximum value reaching $-16900$~\rmu\ in the rest frame of the source 0952$+$179.
Such behaviour is expected within an assumption that the magnetic field strength and particle density increase closer to the central super massive black hole where more Faraday active material lies between the observer and the source.
Since our sources exhibit large apparent core shifts, this effect is considerable.
The measured dependence in the core region, $|\mathrm{RM}_\mathrm{core}|\sim\nu^a$, supports the hypothesis of a circum-jet location for the Faraday screen and a dominant contribution of the toroidal magnetic field component.

For seven sources from our sample, RM measures are available in the literature. 
Rotation measure values for six of the sources agree with our estimates (at the $3\sigma$ level). While one source (1655$+$077) exhibits significant temporal variations in the RM over time ($>3\sigma$).
Simultaneous RM variability with the active state of the source points to a tight connection of the innermost jet regions and Faraday screen.

We detect significant transverse RM gradients in seven AGNs.
The behaviour of the transverse RM profiles agrees with models of relativistic jets with helically-shaped magnetic fields or helical motion.
Meanwhile, a few RM gradients across jets may arise due to change in the density and thickness of magneto-ionic Faraday medium, rather than in the strength and orientation of magnetic field.
While 70 per~cent of the considered RM slices resolve the jet (wider than 1.5~HPBW), significant RM gradients are observed in only 20 per~cent of all slices.
A lack of detected RM gradients, anomalous and low level of depolarization indicate the presence of randomly oriented magnetic fields in the Faraday medium. 

Faraday--corrected rotation polarization vectors in the model-fitted core and jet components are oriented at any angle between 0\degr\ and 90\degr\ relative to the jet direction.
Transparent components show a weak excess of sources having transverse intrinsic EVPAs.
The detailed analysis of individual sources shows that the direction of the electric field in the core region may be affected by ongoing flaring activity or by different types of jet instabilities.
In this consideration, our sources prefer to align their EVPAs with the local jet flow in the jet bases.
Overall, this implies dominance of the poloidal magnetic field component in the studied AGN jets.

We proposed and applied a combined analysis of the Faraday RMs and apparent shift of the opaque cores with frequency. This enables study of the spatial, pc--scale magnetic field geometry, and connects field orientation with its magnitude.

Considering the large-scale EVPA maps, we see a complex structure of the electric field orientation in half of the sources.
These are, for example, spine-sheath structure or rotation of the EVPA at near side of the jet, accompanied by an increase in fractional polarization.
A few sources (e.g. 1458$+$718) show a transition from apparent poloidal to helical magnetic field structures, likely caused by a jet bend.
We assume that the spine--sheath structure is universal for the all observed jets. Then the observed RMs originate in the jet sheath located foreground to the jet spine, while in some cases they can be cospatial. There is no preferred orientation of the magnetic field in the sheath: it can be randomly oriented, be regular and have helical shape. Meanwhile the spine contains well ordered magnetic field.
In this context, we can observe only the spine in some sources, while only the jet sheath may be visible in other sources.
Partially this may explain the overall preference of our sources to orient their magnetic fields either transverse or along the jet direction.
To reconstruct real geometry of magnetic fields in AGN jets, a combined polarimetric and kinematic analysis should be done, as well as measurements with both high sensitivity and resolution, allowing to study polarized structure of AGN jets in more detail.

\section*{Acknowledgments}
EVK thanks Talvikki Hovatta for profound introduction in the RM analysis.
We thank the anonymous referee, Eduardo Ros, and Allan Roy for comprehensive comments and suggestions which helped to improve the manuscript.
This research was supported by Russian Foundation for Basic Research (projects 14-02-31789 and 17-02-00197). 
The VLBA is an instrument of the National Radio Astronomy Observatory, a facility of the National
Science Foundation operated under cooperative agreement by Associated Universities, Inc. 
This research has made use of data from the MOJAVE database that is maintained by the MOJAVE 
team \citep{lister_etal09} and the University of Michigan Radio Astronomy Observatory which has been supported by the University of Michigan and by a series of grants from the National Science 
Foundation, most recently AST-0607523. This research has made use of NASA's Astrophysics Data System Bibliographic Services.

\bibliographystyle{mnras}
\bibliography{evk_bk134}

\begin{thebibliography}{}
\makeatletter
\relax
\def\mn@urlcharsother{\let\do\@makeother \do\$\do\&\do\#\do\^\do\_\do\%\do\~}
\def\mn@doi{\begingroup\mn@urlcharsother \@ifnextchar [ {\mn@doi@}
  {\mn@doi@[]}}
\def\mn@doi@[#1]#2{\def\@tempa{#1}\ifx\@tempa\@empty \href
  {http://dx.doi.org/#2} {doi:#2}\else \href {http://dx.doi.org/#2} {#1}\fi
  \endgroup}
\def\mn@eprint#1#2{\mn@eprint@#1:#2::\@nil}
\def\mn@eprint@arXiv#1{\href {http://arxiv.org/abs/#1} {{\tt arXiv:#1}}}
\def\mn@eprint@dblp#1{\href {http://dblp.uni-trier.de/rec/bibtex/#1.xml}
  {dblp:#1}}
\def\mn@eprint@#1:#2:#3:#4\@nil{\def\@tempa {#1}\def\@tempb {#2}\def\@tempc
  {#3}\ifx \@tempc \@empty \let \@tempc \@tempb \let \@tempb \@tempa \fi \ifx
  \@tempb \@empty \def\@tempb {arXiv}\fi \@ifundefined
  {mn@eprint@\@tempb}{\@tempb:\@tempc}{\expandafter \expandafter \csname
  mn@eprint@\@tempb\endcsname \expandafter{\@tempc}}}

\bibitem[\protect\citeauthoryear{{Aaron}, {Wardle}  \& {Roberts}}{{Aaron}
  et~al.}{1997}]{1997VA.....41..225A}
{Aaron} S.~E.,  {Wardle} J.~F.~C.,   {Roberts} D.~H.,  1997, \mn@doi [Vistas in
  Astronomy] {10.1016/S0083-6656(97)00010-X}, \href
  {http://adsabs.harvard.edu/abs/1997VA.....41..225A} {41, 225}

\bibitem[\protect\citeauthoryear{{Afanas'ev}, {Dodonov}, {Moiseev}, {Gorshkov},
  {Konnikova}  \& {Mingaliev}}{{Afanas'ev} et~al.}{2003}]{afanasiev_etal03}
{Afanas'ev} V.~L.,  {Dodonov} S.~N.,  {Moiseev} A.~V.,  {Gorshkov} A.~G.,
  {Konnikova} V.~K.,   {Mingaliev} M.~G.,  2003, \mn@doi [\arep]
  {10.1134/1.1583772}, \href
  {http://adsabs.harvard.edu/abs/2003ARep...47..458A} {47, 458}

\bibitem[\protect\citeauthoryear{{Agudo}, {Thum}, {G{\'o}mez}  \&
  {Wiesemeyer}}{{Agudo} et~al.}{2014}]{agudo_etal14}
{Agudo} I.,  {Thum} C.,  {G{\'o}mez} J.~L.,   {Wiesemeyer} H.,  2014, \mn@doi
  [\aap] {10.1051/0004-6361/201423366}, \href
  {http://adsabs.harvard.edu/abs/2014A%26A...566A..59A} {566, A59}

\bibitem[\protect\citeauthoryear{{Akahori} \& {Ryu}}{{Akahori} \&
  {Ryu}}{2010}]{akahori_ryu_10}
{Akahori} T.,  {Ryu} D.,  2010, \mn@doi [\apj] {10.1088/0004-637X/723/1/476},
  \href {http://adsabs.harvard.edu/abs/2010ApJ...723..476A} {723, 476}

\bibitem[\protect\citeauthoryear{{Akahori} \& {Ryu}}{{Akahori} \&
  {Ryu}}{2011}]{akahori_ryu_11}
{Akahori} T.,  {Ryu} D.,  2011, \mn@doi [\apj] {10.1088/0004-637X/738/2/134},
  \href {http://adsabs.harvard.edu/abs/2011ApJ...738..134A} {738, 134}

\bibitem[\protect\citeauthoryear{{Algaba}}{{Algaba}}{2013}]{algaba13}
{Algaba} J.~C.,  2013, \mn@doi [\mnras] {10.1093/mnras/sts624}, \href
  {http://adsabs.harvard.edu/abs/2013MNRAS.429.3551A} {429, 3551}

\bibitem[\protect\citeauthoryear{{Anderson}, {Gaensler}, {Feain}  \&
  {Franzen}}{{Anderson} et~al.}{2015}]{2015ApJ...815...49A}
{Anderson} C.~S.,  {Gaensler} B.~M.,  {Feain} I.~J.,   {Franzen} T.~M.~O.,
  2015, \mn@doi [\apj] {10.1088/0004-637X/815/1/49}, \href
  {http://adsabs.harvard.edu/abs/2015ApJ...815...49A} {815, 49}

\bibitem[\protect\citeauthoryear{{Anderson}, {Gaensler}  \& {Feain}}{{Anderson}
  et~al.}{2016}]{2016ApJ...825...59A}
{Anderson} C.~S.,  {Gaensler} B.~M.,   {Feain} I.~J.,  2016, \mn@doi [\apj]
  {10.3847/0004-637X/825/1/59}, \href
  {http://adsabs.harvard.edu/abs/2016ApJ...825...59A} {825, 59}

\bibitem[\protect\citeauthoryear{{Asada}, {Inoue}, {Uchida}, {Kameno},
  {Fujisawa}, {Iguchi}  \& {Mutoh}}{{Asada} et~al.}{2002}]{asada_etal02}
{Asada} K.,  {Inoue} M.,  {Uchida} Y.,  {Kameno} S.,  {Fujisawa} K.,  {Iguchi}
  S.,   {Mutoh} M.,  2002, \mn@doi [\pasj] {10.1093/pasj/54.3.L39}, \href
  {http://adsabs.harvard.edu/abs/2002PASJ...54L..39A} {54, L39}

\bibitem[\protect\citeauthoryear{{Asada}, {Inoue}, {Kameno}  \&
  {Nagai}}{{Asada} et~al.}{2008a}]{asada_etal08}
{Asada} K.,  {Inoue} M.,  {Kameno} S.,   {Nagai} H.,  2008a, \mn@doi [\apj]
  {10.1086/524000}, \href {http://adsabs.harvard.edu/abs/2008ApJ...675...79A}
  {675, 79}

\bibitem[\protect\citeauthoryear{{Asada}, {Inoue}, {Nakamura}, {Kameno}  \&
  {Nagai}}{{Asada} et~al.}{2008b}]{2008ApJ...682..798A}
{Asada} K.,  {Inoue} M.,  {Nakamura} M.,  {Kameno} S.,   {Nagai} H.,  2008b,
  \mn@doi [\apj] {10.1086/588573}, \href
  {http://adsabs.harvard.edu/abs/2008ApJ...682..798A} {682, 798}

\bibitem[\protect\citeauthoryear{{Asada}, {Nakamura}, {Inoue}, {Kameno}  \&
  {Nagai}}{{Asada} et~al.}{2010}]{asada_etal10}
{Asada} K.,  {Nakamura} M.,  {Inoue} M.,  {Kameno} S.,   {Nagai} H.,  2010,
  \mn@doi [\apj] {10.1088/0004-637X/720/1/41}, \href
  {http://adsabs.harvard.edu/abs/2010ApJ...720...41A} {720, 41}

\bibitem[\protect\citeauthoryear{{Attridge}, {Roberts}  \& {Wardle}}{{Attridge}
  et~al.}{1999}]{1999ApJ...518L..87A}
{Attridge} J.~M.,  {Roberts} D.~H.,   {Wardle} J.~F.~C.,  1999, \mn@doi [\apjl]
  {10.1086/312078}, \href {http://adsabs.harvard.edu/abs/1999ApJ...518L..87A}
  {518, L87}

\bibitem[\protect\citeauthoryear{{Bernet}, {Miniati}  \& {Lilly}}{{Bernet}
  et~al.}{2012}]{bernet_etal12}
{Bernet} M.~L.,  {Miniati} F.,   {Lilly} S.~J.,  2012, \mn@doi [\apj]
  {10.1088/0004-637X/761/2/144}, \href
  {http://adsabs.harvard.edu/abs/2012ApJ...761..144B} {761, 144}

\bibitem[\protect\citeauthoryear{{Blandford}}{{Blandford}}{1993}]{blandford93}
{Blandford} R.,  1993, in Astrophysics and Space Science Library. pp 15--33

\bibitem[\protect\citeauthoryear{{Blandford} \& {K{\"o}nigl}}{{Blandford} \&
  {K{\"o}nigl}}{1979}]{blandford_konigl_79}
{Blandford} R.~D.,  {K{\"o}nigl} A.,  1979, \mn@doi [\apj] {10.1086/157262},
  \href {http://adsabs.harvard.edu/abs/1979ApJ...232...34B} {232, 34}

\bibitem[\protect\citeauthoryear{{Blandford} \& {Znajek}}{{Blandford} \&
  {Znajek}}{1977}]{blandford_znajek_77}
{Blandford} R.~D.,  {Znajek} R.~L.,  1977, \mnras, \href
  {http://adsabs.harvard.edu/abs/1977MNRAS.179..433B} {179, 433}

\bibitem[\protect\citeauthoryear{{Bridle} \& {Greisen}}{{Bridle} \&
  {Greisen}}{1994}]{bridle_greisen_94}
{Bridle} A.~H.,  {Greisen} E.~W.,  1994, AIPS Memo, 87

\bibitem[\protect\citeauthoryear{{Broderick} \& {Loeb}}{{Broderick} \&
  {Loeb}}{2009}]{broderick_loeb_09}
{Broderick} A.~E.,  {Loeb} A.,  2009, \mn@doi [\apjl]
  {10.1088/0004-637X/703/2/L104}, \href
  {http://adsabs.harvard.edu/abs/2009ApJ...703L.104B} {703, L104}

\bibitem[\protect\citeauthoryear{{Broderick} \& {McKinney}}{{Broderick} \&
  {McKinney}}{2010}]{broderick_mckinney_10}
{Broderick} A.~E.,  {McKinney} J.~C.,  2010, \mn@doi [\apj]
  {10.1088/0004-637X/725/1/750}, \href
  {http://adsabs.harvard.edu/abs/2010ApJ...725..750B} {725, 750}

\bibitem[\protect\citeauthoryear{{Brown}, {Roberts}  \& {Wardle}}{{Brown}
  et~al.}{1994}]{1994ApJ...437..108B}
{Brown} L.~F.,  {Roberts} D.~H.,   {Wardle} J.~F.~C.,  1994, \mn@doi [\apj]
  {10.1086/174979}, \href {http://adsabs.harvard.edu/abs/1994ApJ...437..108B}
  {437, 108}

\bibitem[\protect\citeauthoryear{{Burn}}{{Burn}}{1966}]{burn66}
{Burn} B.~J.,  1966, \mnras, \href
  {http://adsabs.harvard.edu/abs/1966MNRAS.133...67B} {133, 67}

\bibitem[\protect\citeauthoryear{{Cawthorne}}{{Cawthorne}}{2006}]{2006MNRAS.367..851C}
{Cawthorne} T.~V.,  2006, \mn@doi [\mnras] {10.1111/j.1365-2966.2006.10019.x},
  \href {http://adsabs.harvard.edu/abs/2006MNRAS.367..851C} {367, 851}

\bibitem[\protect\citeauthoryear{{Cawthorne}, {Jorstad}  \&
  {Marscher}}{{Cawthorne} et~al.}{2013}]{2013ApJ...772...14C}
{Cawthorne} T.~V.,  {Jorstad} S.~G.,   {Marscher} A.~P.,  2013, \mn@doi [\apj]
  {10.1088/0004-637X/772/1/14}, \href
  {http://adsabs.harvard.edu/abs/2013ApJ...772...14C} {772, 14}

\bibitem[\protect\citeauthoryear{{Clausen-Brown}, {Lyutikov}  \&
  {Kharb}}{{Clausen-Brown} et~al.}{2011}]{clausen_etal11}
{Clausen-Brown} E.,  {Lyutikov} M.,   {Kharb} P.,  2011, \mn@doi [\mnras]
  {10.1111/j.1365-2966.2011.18757.x}, \href
  {http://adsabs.harvard.edu/abs/2011MNRAS.415.2081C} {415, 2081}

\bibitem[\protect\citeauthoryear{{Conway}, {Haves}, {Kronberg}, {Stannard},
  {Vallee}  \& {Wardle}}{{Conway} et~al.}{1974}]{conway_etal_74}
{Conway} R.~G.,  {Haves} P.,  {Kronberg} P.~P.,  {Stannard} D.,  {Vallee}
  J.~P.,   {Wardle} J.~F.~C.,  1974, \mnras, \href
  {http://adsabs.harvard.edu/abs/1974MNRAS.168..137C} {168, 137}

\bibitem[\protect\citeauthoryear{{Farnes}, {Gaensler}  \& {Carretti}}{{Farnes}
  et~al.}{2014}]{farnes_etal14}
{Farnes} J.~S.,  {Gaensler} B.~M.,   {Carretti} E.,  2014, \mn@doi [\apjs]
  {10.1088/0067-0049/212/1/15}, \href
  {http://adsabs.harvard.edu/abs/2014ApJS..212...15F} {212, 15}

\bibitem[\protect\citeauthoryear{{Fey} \& {Charlot}}{{Fey} \&
  {Charlot}}{1997}]{fey_charlot_97}
{Fey} A.~L.,  {Charlot} P.,  1997, \mn@doi [\apjs] {10.1086/313017}, \href
  {http://adsabs.harvard.edu/abs/1997ApJS..111...95F} {111, 95}

\bibitem[\protect\citeauthoryear{{Finke}, {Shields}, {B{\"o}ttcher}  \&
  {Basu}}{{Finke} et~al.}{2008}]{finke_etal08}
{Finke} J.~D.,  {Shields} J.~C.,  {B{\"o}ttcher} M.,   {Basu} S.,  2008,
  \mn@doi [\aap] {10.1051/0004-6361:20078492}, \href
  {http://adsabs.harvard.edu/abs/2008A%26A...477..513F} {477, 513}

\bibitem[\protect\citeauthoryear{{Fuhrmann} et~al.,}{{Fuhrmann}
  et~al.}{2014}]{2014MNRAS.441.1899F}
{Fuhrmann} L.,  et~al., 2014, \mn@doi [\mnras] {10.1093/mnras/stu540}, \href
  {http://adsabs.harvard.edu/abs/2014MNRAS.441.1899F} {441, 1899}

\bibitem[\protect\citeauthoryear{{Fukui}}{{Fukui}}{1973}]{fukui_73}
{Fukui} M.,  1973, \pasj, \href
  {http://adsabs.harvard.edu/abs/1973PASJ...25..181F} {25, 181}

\bibitem[\protect\citeauthoryear{{Gabuzda}}{{Gabuzda}}{1999}]{1999NewAR..43..691G}
{Gabuzda} D.~C.,  1999, \mn@doi [\nar] {10.1016/S1387-6473(99)00079-2}, \href
  {http://adsabs.harvard.edu/abs/1999NewAR..43..691G} {43, 691}

\bibitem[\protect\citeauthoryear{{Gabuzda} \& {Chernetskii}}{{Gabuzda} \&
  {Chernetskii}}{2003}]{2003MNRAS.339..669G}
{Gabuzda} D.~C.,  {Chernetskii} V.~A.,  2003, \mn@doi [\mnras]
  {10.1046/j.1365-8711.2003.06204.x}, \href
  {http://adsabs.harvard.edu/abs/2003MNRAS.339..669G} {339, 669}

\bibitem[\protect\citeauthoryear{{Gabuzda}, {Mullan}, {Cawthorne}, {Wardle}  \&
  {Roberts}}{{Gabuzda} et~al.}{1994}]{1994ApJ...435..140G}
{Gabuzda} D.~C.,  {Mullan} C.~M.,  {Cawthorne} T.~V.,  {Wardle} J.~F.~C.,
  {Roberts} D.~H.,  1994, \mn@doi [\apj] {10.1086/174801}, \href
  {http://adsabs.harvard.edu/abs/1994ApJ...435..140G} {435, 140}

\bibitem[\protect\citeauthoryear{{Gabuzda}, {Knuettel}  \& {Reardon}}{{Gabuzda}
  et~al.}{2015}]{2015MNRAS.450.2441G}
{Gabuzda} D.~C.,  {Knuettel} S.,   {Reardon} B.,  2015, \mn@doi [\mnras]
  {10.1093/mnras/stv555}, \href
  {http://adsabs.harvard.edu/abs/2015MNRAS.450.2441G} {450, 2441}

\bibitem[\protect\citeauthoryear{{Gardner} \& {Whiteoak}}{{Gardner} \&
  {Whiteoak}}{1966}]{gardner_whiteoak_66}
{Gardner} F.~F.,  {Whiteoak} J.~B.,  1966, \mn@doi [\araa]
  {10.1146/annurev.aa.04.090166.001333}, \href
  {http://adsabs.harvard.edu/abs/1966ARA%26A...4..245G} {4, 245}

\bibitem[\protect\citeauthoryear{{Giroletti} et~al.,}{{Giroletti}
  et~al.}{2015}]{giroletti_etal15}
{Giroletti} M.,  et~al., 2015, 2014 Fermi Symposium proceedings, \href
  {http://adsabs.harvard.edu/abs/2015arXiv150304597G} {eConf C141020.1}

\bibitem[\protect\citeauthoryear{{Goldstein} \& {Reed}}{{Goldstein} \&
  {Reed}}{1984}]{gl_84}
{Goldstein} Jr. S.~J.,  {Reed} J.~A.,  1984, \mn@doi [\apj] {10.1086/162337},
  \href {http://adsabs.harvard.edu/abs/1984ApJ...283..540G} {283, 540}

\bibitem[\protect\citeauthoryear{{G{\'o}mez}, {Marscher}, {Alberdi}, {Jorstad}
  \& {Agudo}}{{G{\'o}mez} et~al.}{2002}]{gomez_etal02}
{G{\'o}mez} J.~L.,  {Marscher} A.~P.,  {Alberdi} A.,  {Jorstad} S.~G.,
  {Agudo} I.,  2002, VLBA Scientific Memo, 30

\bibitem[\protect\citeauthoryear{{G{\'o}mez}, {Marscher}, {Jorstad}, {Agudo}
  \& {Roca-Sogorb}}{{G{\'o}mez} et~al.}{2008}]{2008ApJ...681L..69G}
{G{\'o}mez} J.~L.,  {Marscher} A.~P.,  {Jorstad} S.~G.,  {Agudo} I.,
  {Roca-Sogorb} M.,  2008, \mn@doi [\apjl] {10.1086/590388}, \href
  {http://adsabs.harvard.edu/abs/2008ApJ...681L..69G} {681, L69}

\bibitem[\protect\citeauthoryear{{G{\'o}mez} et~al.,}{{G{\'o}mez}
  et~al.}{2016}]{2016ApJ...817...96G}
{G{\'o}mez} J.~L.,  et~al., 2016, \mn@doi [\apj] {10.3847/0004-637X/817/2/96},
  \href {http://adsabs.harvard.edu/abs/2016ApJ...817...96G} {817, 96}

\bibitem[\protect\citeauthoryear{{Hada}, {Doi}, {Kino}, {Nagai}, {Hagiwara}  \&
  {Kawaguchi}}{{Hada} et~al.}{2011}]{2011Natur.477..185H}
{Hada} K.,  {Doi} A.,  {Kino} M.,  {Nagai} H.,  {Hagiwara} Y.,   {Kawaguchi}
  N.,  2011, \mn@doi [\nat] {10.1038/nature10387}, \href
  {http://adsabs.harvard.edu/abs/2011Natur.477..185H} {477, 185}

\bibitem[\protect\citeauthoryear{{Han}}{{Han}}{2013}]{han13}
{Han} J.,  2013, in {van Leeuwen} J.,  ed.,  IAU Symposium Vol. 291, IAU
  Symposium. pp 223--228 (\mn@eprint {arXiv} {1210.7029}),
  \mn@doi{10.1017/S174392131202371X}

\bibitem[\protect\citeauthoryear{{Hardee}}{{Hardee}}{2004}]{2004ApSS.293..117H}
{Hardee} P.~E.,  2004, \mn@doi [\apss] {10.1023/B:ASTR.0000044659.76968.60},
  \href {http://adsabs.harvard.edu/abs/2004Ap%26SS.293..117H} {293, 117}

\bibitem[\protect\citeauthoryear{{Homan}}{{Homan}}{2012}]{homan_12}
{Homan} D.~C.,  2012, \mn@doi [\apjl] {10.1088/2041-8205/747/2/L24}, \href
  {http://adsabs.harvard.edu/abs/2012ApJ...747L..24H} {747, L24}

\bibitem[\protect\citeauthoryear{{Hovatta}, {Lister}, {Aller}, {Aller},
  {Homan}, {Kovalev}, {Pushkarev}  \& {Savolainen}}{{Hovatta}
  et~al.}{2012}]{hovatta_etal12}
{Hovatta} T.,  {Lister} M.~L.,  {Aller} M.~F.,  {Aller} H.~D.,  {Homan} D.~C.,
  {Kovalev} Y.~Y.,  {Pushkarev} A.~B.,   {Savolainen} T.,  2012, \mn@doi [\aj]
  {10.1088/0004-6256/144/4/105}, \href
  {http://adsabs.harvard.edu/abs/2012AJ....144..105H} {144, 105}

\bibitem[\protect\citeauthoryear{{Istomin} \& {Pariev}}{{Istomin} \&
  {Pariev}}{1994}]{1994MNRAS.267..629I}
{Istomin} Y.~N.,  {Pariev} V.~I.,  1994, \mn@doi [\mnras]
  {10.1093/mnras/267.3.629}, \href
  {http://adsabs.harvard.edu/abs/1994MNRAS.267..629I} {267, 629}

\bibitem[\protect\citeauthoryear{{Jansson} \& {Farrar}}{{Jansson} \&
  {Farrar}}{2012}]{jansson_farrar_12}
{Jansson} R.,  {Farrar} G.~R.,  2012, \mn@doi [\apj]
  {10.1088/0004-637X/757/1/14}, \href
  {http://adsabs.harvard.edu/abs/2012ApJ...757...14J} {757, 14}

\bibitem[\protect\citeauthoryear{{Jones} \& {Odell}}{{Jones} \&
  {Odell}}{1977}]{jones_odell_77b}
{Jones} T.~W.,  {Odell} S.~L.,  1977, \mn@doi [\apj] {10.1086/155353}, \href
  {http://adsabs.harvard.edu/abs/1977ApJ...215..236J} {215, 236}

\bibitem[\protect\citeauthoryear{{Jorstad} et~al.,}{{Jorstad}
  et~al.}{2007}]{jorstad_etal07}
{Jorstad} S.~G.,  et~al., 2007, \mn@doi [\aj] {10.1086/519996}, \href
  {http://adsabs.harvard.edu/abs/2007AJ....134..799J} {134, 799}

\bibitem[\protect\citeauthoryear{{Junor}, {Salter}, {Saikia}, {Mantovani}  \&
  {Peck}}{{Junor} et~al.}{1999}]{1999MNRAS.308..955J}
{Junor} W.,  {Salter} C.~J.,  {Saikia} D.~J.,  {Mantovani} F.,   {Peck} A.~B.,
  1999, \mn@doi [\mnras] {10.1046/j.1365-8711.1999.02774.x}, \href
  {http://adsabs.harvard.edu/abs/1999MNRAS.308..955J} {308, 955}

\bibitem[\protect\citeauthoryear{{Konigl}}{{Konigl}}{1981}]{konigl81}
{Konigl} A.,  1981, \mn@doi [\apj] {10.1086/158638}, \href
  {http://adsabs.harvard.edu/abs/1981ApJ...243..700K} {243, 700}

\bibitem[\protect\citeauthoryear{{Kovalev}, {Lobanov}, {Pushkarev}  \&
  {Zensus}}{{Kovalev} et~al.}{2008}]{kovalev_etal08}
{Kovalev} Y.~Y.,  {Lobanov} A.~P.,  {Pushkarev} A.~B.,   {Zensus} J.~A.,  2008,
  \mn@doi [\aap] {10.1051/0004-6361:20078679}, \href
  {http://adsabs.harvard.edu/abs/2008A%26A...483..759K} {483, 759}

\bibitem[\protect\citeauthoryear{{Kravchenko}, {Cotton}  \&
  {Kovalev}}{{Kravchenko} et~al.}{2015}]{kravchenko_etal15}
{Kravchenko} E.~V.,  {Cotton} W.~D.,   {Kovalev} Y.~Y.,  2015, in {Massaro} F.,
   {Cheung} C.~C.,  {Lopez} E.,   {Siemiginowska} A.,  eds,  IAU Symposium Vol.
  313, IAU Symposium. pp 128--132 (\mn@eprint {arXiv} {1412.1728}),
  \mn@doi{10.1017/S1743921315002057}

\bibitem[\protect\citeauthoryear{{Kravchenko}, {Kovalev}, {Hovatta}  \&
  {Ramakrishnan}}{{Kravchenko} et~al.}{2016}]{2016MNRAS.462.2747K}
{Kravchenko} E.~V.,  {Kovalev} Y.~Y.,  {Hovatta} T.,   {Ramakrishnan} V.,
  2016, \mn@doi [\mnras] {10.1093/mnras/stw1776}, \href
  {http://adsabs.harvard.edu/abs/2016MNRAS.462.2747K} {462, 2747}

\bibitem[\protect\citeauthoryear{{Kutkin} et~al.,}{{Kutkin}
  et~al.}{2014}]{kutkin_etal14}
{Kutkin} A.~M.,  et~al., 2014, \mn@doi [\mnras] {10.1093/mnras/stt2133}, \href
  {http://adsabs.harvard.edu/abs/2014MNRAS.437.3396K} {437, 3396}

\bibitem[\protect\citeauthoryear{{Laing}}{{Laing}}{1980}]{1980MNRAS.193..439L}
{Laing} R.~A.,  1980, \mn@doi [\mnras] {10.1093/mnras/193.3.439}, \href
  {http://adsabs.harvard.edu/abs/1980MNRAS.193..439L} {193, 439}

\bibitem[\protect\citeauthoryear{{Laing}}{{Laing}}{1981}]{laing81}
{Laing} R.~A.,  1981, \mn@doi [\apj] {10.1086/159132}, \href
  {http://adsabs.harvard.edu/abs/1981ApJ...248...87L} {248, 87}

\bibitem[\protect\citeauthoryear{{Laing}}{{Laing}}{1996}]{1996ASPC..100..241L}
{Laing} R.~A.,  1996, in {Hardee} P.~E.,  {Bridle} A.~H.,   {Zensus} J.~A.,
  eds,  Astronomical Society of the Pacific Conference Series Vol. 100, Energy
  Transport in Radio Galaxies and Quasars. p.~241

\bibitem[\protect\citeauthoryear{{Lamee}, {Rudnick}, {Farnes}, {Carretti},
  {Gaensler}, {Haverkorn}  \& {Poppi}}{{Lamee}
  et~al.}{2016}]{2016ApJ...829....5L}
{Lamee} M.,  {Rudnick} L.,  {Farnes} J.~S.,  {Carretti} E.,  {Gaensler} B.~M.,
  {Haverkorn} M.,   {Poppi} S.,  2016, \mn@doi [\apj]
  {10.3847/0004-637X/829/1/5}, \href
  {http://adsabs.harvard.edu/abs/2016ApJ...829....5L} {829, 5}

\bibitem[\protect\citeauthoryear{{Leppanen}, {Zensus}  \& {Diamond}}{{Leppanen}
  et~al.}{1995}]{1995AJ....110.2479L}
{Leppanen} K.~J.,  {Zensus} J.~A.,   {Diamond} P.~J.,  1995, \mn@doi [\aj]
  {10.1086/117706}, \href {http://adsabs.harvard.edu/abs/1995AJ....110.2479L}
  {110, 2479}

\bibitem[\protect\citeauthoryear{{Lico} et~al.,}{{Lico}
  et~al.}{2014}]{lico_etal14}
{Lico} R.,  et~al., 2014, \mn@doi [\aap] {10.1051/0004-6361/201424341}, \href
  {http://adsabs.harvard.edu/abs/2014A%26A...571A..54L} {571, A54}

\bibitem[\protect\citeauthoryear{{Lister} \& {Homan}}{{Lister} \&
  {Homan}}{2005}]{lister_homan_05}
{Lister} M.~L.,  {Homan} D.~C.,  2005, \mn@doi [\aj] {10.1086/432969}, \href
  {http://adsabs.harvard.edu/abs/2005AJ....130.1389L} {130, 1389}

\bibitem[\protect\citeauthoryear{{Lister} et~al.,}{{Lister}
  et~al.}{2009}]{lister_etal09}
{Lister} M.~L.,  et~al., 2009, \mn@doi [\aj] {10.1088/0004-6256/137/3/3718},
  \href {http://adsabs.harvard.edu/abs/2009AJ....137.3718L} {137, 3718}

\bibitem[\protect\citeauthoryear{{Lister} et~al.,}{{Lister}
  et~al.}{2013}]{lister_etal13}
{Lister} M.~L.,  et~al., 2013, \mn@doi [\aj] {10.1088/0004-6256/146/5/120},
  \href {http://adsabs.harvard.edu/abs/2013AJ....146..120L} {146, 120}

\bibitem[\protect\citeauthoryear{{Lister} et~al.,}{{Lister}
  et~al.}{2016}]{2016AJ....152...12L}
{Lister} M.~L.,  et~al., 2016, \mn@doi [\aj] {10.3847/0004-6256/152/1/12},
  \href {http://adsabs.harvard.edu/abs/2016AJ....152...12L} {152, 12}

\bibitem[\protect\citeauthoryear{{Lobanov}}{{Lobanov}}{1998}]{lobanov98}
{Lobanov} A.~P.,  1998, \aap, \href
  {http://adsabs.harvard.edu/abs/1998A%26A...330...79L} {330, 79}

\bibitem[\protect\citeauthoryear{{Lobanov} \& {Zensus}}{{Lobanov} \&
  {Zensus}}{2001}]{lobanov_zensus_01}
{Lobanov} A.~P.,  {Zensus} J.~A.,  2001, \mn@doi [Science]
  {10.1126/science.1063239}, \href
  {http://adsabs.harvard.edu/abs/2001Sci...294..128L} {294, 128}

\bibitem[\protect\citeauthoryear{{Lobanov} et~al.,}{{Lobanov}
  et~al.}{2015}]{lobanov_etal15}
{Lobanov} A.~P.,  et~al., 2015, preprint, \href
  {http://adsabs.harvard.edu/abs/2015arXiv150404273L} {} (\mn@eprint {arXiv}
  {1504.04273})

\bibitem[\protect\citeauthoryear{{Lyutikov}, {Pariev}  \& {Gabuzda}}{{Lyutikov}
  et~al.}{2005}]{lyutikov_etal05}
{Lyutikov} M.,  {Pariev} V.~I.,   {Gabuzda} D.~C.,  2005, \mn@doi [\mnras]
  {10.1111/j.1365-2966.2005.08954.x}, \href
  {http://adsabs.harvard.edu/abs/2005MNRAS.360..869L} {360, 869}

\bibitem[\protect\citeauthoryear{{Mahmud}, {Gabuzda}  \& {Bezrukovs}}{{Mahmud}
  et~al.}{2009}]{mahmud_etal09}
{Mahmud} M.,  {Gabuzda} D.~C.,   {Bezrukovs} V.,  2009, \mn@doi [\mnras]
  {10.1111/j.1365-2966.2009.15013.x}, \href
  {http://adsabs.harvard.edu/abs/2009MNRAS.400....2M} {400, 2}

\bibitem[\protect\citeauthoryear{{Mahmud}, {Coughlan}, {Murphy}, {Gabuzda}  \&
  {Hallahan}}{{Mahmud} et~al.}{2013}]{mahmud_etal13}
{Mahmud} M.,  {Coughlan} C.~P.,  {Murphy} E.,  {Gabuzda} D.~C.,   {Hallahan}
  D.~R.,  2013, \mn@doi [\mnras] {10.1093/mnras/stt201}, \href
  {http://adsabs.harvard.edu/abs/2013MNRAS.431..695M} {431, 695}

\bibitem[\protect\citeauthoryear{{Mantovani}, {Mack}, {Montenegro-Montes},
  {Rossetti}  \& {Kraus}}{{Mantovani} et~al.}{2009}]{2009AA...502...61M}
{Mantovani} F.,  {Mack} K.-H.,  {Montenegro-Montes} F.~M.,  {Rossetti} A.,
  {Kraus} A.,  2009, \mn@doi [\aap] {10.1051/0004-6361/200911815}, \href
  {http://adsabs.harvard.edu/abs/2009A%26A...502...61M} {502, 61}

\bibitem[\protect\citeauthoryear{{Mantovani}, {Rossetti}, {Junor}, {Saikia}  \&
  {Salter}}{{Mantovani} et~al.}{2010}]{2010AA...518A..33M}
{Mantovani} F.,  {Rossetti} A.,  {Junor} W.,  {Saikia} D.~J.,   {Salter} C.~J.,
   2010, \mn@doi [\aap] {10.1051/0004-6361/201014400}, \href
  {http://adsabs.harvard.edu/abs/2010A%26A...518A..33M} {518, A33}

\bibitem[\protect\citeauthoryear{{Mao}, {Gaensler}, {Haverkorn}, {Zweibel},
  {Madsen}, {McClure-Griffiths}, {Shukurov}  \& {Kronberg}}{{Mao}
  et~al.}{2010}]{mao_etal10}
{Mao} S.~A.,  {Gaensler} B.~M.,  {Haverkorn} M.,  {Zweibel} E.~G.,  {Madsen}
  G.~J.,  {McClure-Griffiths} N.~M.,  {Shukurov} A.,   {Kronberg} P.~P.,  2010,
  \mn@doi [\apj] {10.1088/0004-637X/714/2/1170}, \href
  {http://adsabs.harvard.edu/abs/2010ApJ...714.1170M} {714, 1170}

\bibitem[\protect\citeauthoryear{{Marcaide} \& {Shapiro}}{{Marcaide} \&
  {Shapiro}}{1984}]{marcaide_shapiro_84}
{Marcaide} J.~M.,  {Shapiro} I.~I.,  1984, \mn@doi [\apj] {10.1086/161592},
  \href {http://adsabs.harvard.edu/abs/1984ApJ...276...56M} {276, 56}

\bibitem[\protect\citeauthoryear{{Marscher}}{{Marscher}}{2008}]{marscher08}
{Marscher} A.~P.,  2008, in {Rector} T.~A.,  {De Young} D.~S.,  eds,
  Astronomical Society of the Pacific Conference Series Vol. 386, Extragalactic
  Jets: Theory and Observation from Radio to Gamma Ray. p.~437

\bibitem[\protect\citeauthoryear{{Meier}, {Koide}  \& {Uchida}}{{Meier}
  et~al.}{2001}]{meier_etal01}
{Meier} D.~L.,  {Koide} S.,   {Uchida} Y.,  2001, \mn@doi [Science]
  {10.1126/science.291.5501.84}, \href
  {http://adsabs.harvard.edu/abs/2001Sci...291...84M} {291, 84}

\bibitem[\protect\citeauthoryear{{Mizuno}, {Hardee}  \& {Nishikawa}}{{Mizuno}
  et~al.}{2007}]{2007ApJ...662..835M}
{Mizuno} Y.,  {Hardee} P.,   {Nishikawa} K.-I.,  2007, \mn@doi [\apj]
  {10.1086/518106}, \href {http://adsabs.harvard.edu/abs/2007ApJ...662..835M}
  {662, 835}

\bibitem[\protect\citeauthoryear{{Nalewajko}}{{Nalewajko}}{2009}]{2009MNRAS.395..524N}
{Nalewajko} K.,  2009, \mn@doi [\mnras] {10.1111/j.1365-2966.2009.14559.x},
  \href {http://adsabs.harvard.edu/abs/2009MNRAS.395..524N} {395, 524}

\bibitem[\protect\citeauthoryear{{Nishikawa}, {Richardson}, {Koide}, {Shibata},
  {Kudoh}, {Hardee}  \& {Fishman}}{{Nishikawa}
  et~al.}{2005}]{2005ApJ...625...60N}
{Nishikawa} K.-I.,  {Richardson} G.,  {Koide} S.,  {Shibata} K.,  {Kudoh} T.,
  {Hardee} P.,   {Fishman} G.~J.,  2005, \mn@doi [\apj] {10.1086/429360}, \href
  {http://adsabs.harvard.edu/abs/2005ApJ...625...60N} {625, 60}

\bibitem[\protect\citeauthoryear{{O'Sullivan} \& {Gabuzda}}{{O'Sullivan} \&
  {Gabuzda}}{2009a}]{osullivan_gabuzda_09}
{O'Sullivan} S.~P.,  {Gabuzda} D.~C.,  2009a, \mn@doi [\mnras]
  {10.1111/j.1365-2966.2008.14213.x}, \href
  {http://adsabs.harvard.edu/abs/2009MNRAS.393..429O} {393, 429}

\bibitem[\protect\citeauthoryear{{O'Sullivan} \& {Gabuzda}}{{O'Sullivan} \&
  {Gabuzda}}{2009b}]{2009MNRAS.400...26O}
{O'Sullivan} S.~P.,  {Gabuzda} D.~C.,  2009b, \mn@doi [\mnras]
  {10.1111/j.1365-2966.2009.15428.x}, \href
  {http://adsabs.harvard.edu/abs/2009MNRAS.400...26O} {400, 26}

\bibitem[\protect\citeauthoryear{{O'Sullivan} et~al.,}{{O'Sullivan}
  et~al.}{2012}]{osullivan_etal12}
{O'Sullivan} S.~P.,  et~al., 2012, \mn@doi [\mnras]
  {10.1111/j.1365-2966.2012.20554.x}, \href
  {http://adsabs.harvard.edu/abs/2012MNRAS.421.3300O} {421, 3300}

\bibitem[\protect\citeauthoryear{{Oppermann} et~al.,}{{Oppermann}
  et~al.}{2012}]{opperman_etal12}
{Oppermann} N.,  et~al., 2012, \mn@doi [\aap] {10.1051/0004-6361/201118526},
  \href {http://adsabs.harvard.edu/abs/2012A%26A...542A..93O} {542, A93}

\bibitem[\protect\citeauthoryear{{Pacholczyk} \& {Swihart}}{{Pacholczyk} \&
  {Swihart}}{1967}]{pacholczyk_swihart_67}
{Pacholczyk} A.~G.,  {Swihart} T.~L.,  1967, \mn@doi [\apj] {10.1086/149364},
  \href {http://adsabs.harvard.edu/abs/1967ApJ...150..647P} {150, 647}

\bibitem[\protect\citeauthoryear{{Pasetto}, {Kraus}, {Mack}, {Bruni}  \&
  {Carrasco-Gonz{\'a}lez}}{{Pasetto} et~al.}{2016}]{2016A...586A.117P}
{Pasetto} A.,  {Kraus} A.,  {Mack} K.-H.,  {Bruni} G.,
  {Carrasco-Gonz{\'a}lez} C.,  2016, \mn@doi [\aap]
  {10.1051/0004-6361/201526963}, \href
  {http://adsabs.harvard.edu/abs/2016A%26A...586A.117P} {586, A117}

\bibitem[\protect\citeauthoryear{{Petrov}, {Gordon}, {Gipson}, {MacMillan},
  {Ma}, {Fomalont}, {Walker}  \& {Carabajal}}{{Petrov}
  et~al.}{2009}]{petrov_etal09}
{Petrov} L.,  {Gordon} D.,  {Gipson} J.,  {MacMillan} D.,  {Ma} C.,  {Fomalont}
  E.,  {Walker} R.~C.,   {Carabajal} C.,  2009, \mn@doi [Journal of Geodesy]
  {10.1007/s00190-009-0304-7}, \href
  {http://adsabs.harvard.edu/abs/2009JGeod..83..859P} {83, 859}

\bibitem[\protect\citeauthoryear{{Planck Collaboration} et~al.,}{{Planck
  Collaboration} et~al.}{2015}]{planck_15}
{Planck Collaboration} et~al., 2015, preprint, \href
  {http://adsabs.harvard.edu/abs/2015arXiv150201589P} {} (\mn@eprint {arXiv}
  {1502.01589})

\bibitem[\protect\citeauthoryear{{Pollack}, {Taylor}  \& {Zavala}}{{Pollack}
  et~al.}{2003}]{pollack_etal03}
{Pollack} L.~K.,  {Taylor} G.~B.,   {Zavala} R.~T.,  2003, \mn@doi [\apj]
  {10.1086/374712}, \href {http://adsabs.harvard.edu/abs/2003ApJ...589..733P}
  {589, 733}

\bibitem[\protect\citeauthoryear{{Porth}, {Fendt}, {Meliani}  \&
  {Vaidya}}{{Porth} et~al.}{2011}]{2011ApJ...737...42P}
{Porth} O.,  {Fendt} C.,  {Meliani} Z.,   {Vaidya} B.,  2011, \mn@doi [\apj]
  {10.1088/0004-637X/737/1/42}, \href
  {http://adsabs.harvard.edu/abs/2011ApJ...737...42P} {737, 42}

\bibitem[\protect\citeauthoryear{{Pushkarev}, {Gabuzda}, {Vetukhnovskaya}  \&
  {Yakimov}}{{Pushkarev} et~al.}{2005}]{2005MNRAS.356..859P}
{Pushkarev} A.~B.,  {Gabuzda} D.~C.,  {Vetukhnovskaya} Y.~N.,   {Yakimov}
  V.~E.,  2005, \mn@doi [\mnras] {10.1111/j.1365-2966.2004.08535.x}, \href
  {http://adsabs.harvard.edu/abs/2005MNRAS.356..859P} {356, 859}

\bibitem[\protect\citeauthoryear{{Pushkarev}, {Hovatta}, {Kovalev}, {Lister},
  {Lobanov}, {Savolainen}  \& {Zensus}}{{Pushkarev}
  et~al.}{2012}]{pushkarev_etal12}
{Pushkarev} A.~B.,  {Hovatta} T.,  {Kovalev} Y.~Y.,  {Lister} M.~L.,  {Lobanov}
  A.~P.,  {Savolainen} T.,   {Zensus} J.~A.,  2012, \mn@doi [\aap]
  {10.1051/0004-6361/201219173}, \href
  {http://adsabs.harvard.edu/abs/2012A%26A...545A.113P} {545, A113}

\bibitem[\protect\citeauthoryear{{Pushkarev}, {Lister}, {Kovalev}  \&
  {Savolainen}}{{Pushkarev} et~al.}{2014}]{pushkarev_etal15}
{Pushkarev} A.,  {Lister} M.,  {Kovalev} Y.,   {Savolainen} T.,  2014, in
  Proceedings of the 12th European VLBI Network Symposium and Users Meeting
  (EVN 2014). 7-10 October 2014. Cagliari, Italy. p.~104 (\mn@eprint {arXiv}
  {1502.00283})

\bibitem[\protect\citeauthoryear{{Rani}, {Krichbaum}, {Marscher}, {Jorstad},
  {Hodgson}, {Fuhrmann}  \& {Zensus}}{{Rani} et~al.}{2014}]{2014AA...571L...2R}
{Rani} B.,  {Krichbaum} T.~P.,  {Marscher} A.~P.,  {Jorstad} S.~G.,  {Hodgson}
  J.~A.,  {Fuhrmann} L.,   {Zensus} J.~A.,  2014, \mn@doi [\aap]
  {10.1051/0004-6361/201424796}, \href
  {http://adsabs.harvard.edu/abs/2014A%26A...571L...2R} {571, L2}

\bibitem[\protect\citeauthoryear{{Roberts}, {Wardle}  \& {Brown}}{{Roberts}
  et~al.}{1994}]{roberts_etal94}
{Roberts} D.~H.,  {Wardle} J.~F.~C.,   {Brown} L.~F.,  1994, \mn@doi [\apj]
  {10.1086/174180}, \href {http://adsabs.harvard.edu/abs/1994ApJ...427..718R}
  {427, 718}

\bibitem[\protect\citeauthoryear{{Rossetti}, {Dallacasa}, {Fanti}, {Fanti}  \&
  {Mack}}{{Rossetti} et~al.}{2008}]{2008AA...487..865R}
{Rossetti} A.,  {Dallacasa} D.,  {Fanti} C.,  {Fanti} R.,   {Mack} K.-H.,
  2008, \mn@doi [\aap] {10.1051/0004-6361:20079047}, \href
  {http://adsabs.harvard.edu/abs/2008A%26A...487..865R} {487, 865}

\bibitem[\protect\citeauthoryear{{Savolainen}, {Homan}, {Hovatta}, {Kadler},
  {Kovalev}, {Lister}, {Ros}  \& {Zensus}}{{Savolainen}
  et~al.}{2010}]{savolainen_etal10}
{Savolainen} T.,  {Homan} D.~C.,  {Hovatta} T.,  {Kadler} M.,  {Kovalev} Y.~Y.,
   {Lister} M.~L.,  {Ros} E.,   {Zensus} J.~A.,  2010, \mn@doi [\aap]
  {10.1051/0004-6361/200913740}, \href
  {http://adsabs.harvard.edu/abs/2010A%26A...512A..24S} {512, A24}

\bibitem[\protect\citeauthoryear{Schwarz}{Schwarz}{1978}]{schwarz_78}
Schwarz G.,  1978, The annals of statistics, 6, 461

\bibitem[\protect\citeauthoryear{{Shepherd}}{{Shepherd}}{1997}]{shepherd_97}
{Shepherd} M.~C.,  1997, in {Hunt} G.,  {Payne} H.,  eds,  Astronomical Society
  of the Pacific Conference Series Vol. 125, Astronomical Data Analysis
  Software and Systems VI. p.~77

\bibitem[\protect\citeauthoryear{{Shepherd}, {Pearson}  \& {Taylor}}{{Shepherd}
  et~al.}{1994}]{shepherd_etal94}
{Shepherd} M.~C.,  {Pearson} T.~J.,   {Taylor} G.~B.,  1994, in Bulletin of the
  American Astronomical Society. pp 987--989

\bibitem[\protect\citeauthoryear{{Sokoloff}, {Bykov}, {Shukurov},
  {Berkhuijsen}, {Beck}  \& {Poezd}}{{Sokoloff} et~al.}{1998}]{sokoloff_etal98}
{Sokoloff} D.~D.,  {Bykov} A.~A.,  {Shukurov} A.,  {Berkhuijsen} E.~M.,  {Beck}
  R.,   {Poezd} A.~D.,  1998, \mn@doi [\mnras]
  {10.1046/j.1365-8711.1998.01782.x}, \href
  {http://adsabs.harvard.edu/abs/1998MNRAS.299..189S} {299, 189}

\bibitem[\protect\citeauthoryear{{Sokolovsky}, {Kovalev}, {Pushkarev}  \&
  {Lobanov}}{{Sokolovsky} et~al.}{2011}]{sokolovsky_etal11}
{Sokolovsky} K.~V.,  {Kovalev} Y.~Y.,  {Pushkarev} A.~B.,   {Lobanov} A.~P.,
  2011, \mn@doi [\aap] {10.1051/0004-6361/201016072}, \href
  {http://adsabs.harvard.edu/abs/2011A%26A...532A..38S} {532, A38}

\bibitem[\protect\citeauthoryear{{Taylor}}{{Taylor}}{2000}]{taylor_00}
{Taylor} G.~B.,  2000, \mn@doi [\apj] {10.1086/308666}, \href
  {http://adsabs.harvard.edu/abs/2000ApJ...533...95T} {533, 95}

\bibitem[\protect\citeauthoryear{{Taylor} \& {Myers}}{{Taylor} \&
  {Myers}}{2000}]{taylor_myers_00}
{Taylor} G.~B.,  {Myers} S.~T.,  2000, VLBA Scientific Memo, 26

\bibitem[\protect\citeauthoryear{{Taylor} \& {Zavala}}{{Taylor} \&
  {Zavala}}{2010}]{taylor_zavala_10}
{Taylor} G.~B.,  {Zavala} R.,  2010, \mn@doi [\apjl]
  {10.1088/2041-8205/722/2/L183}, \href
  {http://adsabs.harvard.edu/abs/2010ApJ...722L.183T} {722, L183}

\bibitem[\protect\citeauthoryear{{Taylor}, {Stil}  \& {Sunstrum}}{{Taylor}
  et~al.}{2009}]{taylor_etal09}
{Taylor} A.~R.,  {Stil} J.~M.,   {Sunstrum} C.,  2009, \mn@doi [\apj]
  {10.1088/0004-637X/702/2/1230}, \href
  {http://adsabs.harvard.edu/abs/2009ApJ...702.1230T} {702, 1230}

\bibitem[\protect\citeauthoryear{{Tchekhovskoy}, {Narayan}  \&
  {McKinney}}{{Tchekhovskoy} et~al.}{2011}]{2011MNRAS.418L..79T}
{Tchekhovskoy} A.,  {Narayan} R.,   {McKinney} J.~C.,  2011, \mn@doi [\mnras]
  {10.1111/j.1745-3933.2011.01147.x}, \href
  {http://adsabs.harvard.edu/abs/2011MNRAS.418L..79T} {418, L79}

\bibitem[\protect\citeauthoryear{{Tribble}}{{Tribble}}{1991}]{tribble91}
{Tribble} P.~C.,  1991, \mnras, \href
  {http://adsabs.harvard.edu/abs/1991MNRAS.250..726T} {250, 726}

\bibitem[\protect\citeauthoryear{{Trippe}, {Neri}, {Krips}, {Castro-Carrizo},
  {Bremer}, {Pi{\'e}tu}  \& {Winters}}{{Trippe} et~al.}{2012}]{trippe_etal12}
{Trippe} S.,  {Neri} R.,  {Krips} M.,  {Castro-Carrizo} A.,  {Bremer} M.,
  {Pi{\'e}tu} V.,   {Winters} J.~M.,  2012, \mn@doi [\aap]
  {10.1051/0004-6361/201118563}, \href
  {http://adsabs.harvard.edu/abs/2012A%26A...540A..74T} {540, A74}

\bibitem[\protect\citeauthoryear{{Udomprasert}, {Taylor}, {Pearson}  \&
  {Roberts}}{{Udomprasert} et~al.}{1997}]{1997ApJ...483L...9U}
{Udomprasert} P.~S.,  {Taylor} G.~B.,  {Pearson} T.~J.,   {Roberts} D.~H.,
  1997, \mn@doi [\apjl] {10.1086/310725}, \href
  {http://adsabs.harvard.edu/abs/1997ApJ...483L...9U} {483, L9}

\bibitem[\protect\citeauthoryear{{Van Eck} et~al.,}{{Van Eck}
  et~al.}{2011}]{vaneck_etal11}
{Van Eck} C.~L.,  et~al., 2011, \mn@doi [\apj] {10.1088/0004-637X/728/2/97},
  \href {http://adsabs.harvard.edu/abs/2011ApJ...728...97V} {728, 97}

\bibitem[\protect\citeauthoryear{{V{\'e}ron-Cetty} \&
  {V{\'e}ron}}{{V{\'e}ron-Cetty} \& {V{\'e}ron}}{2010}]{veron_veron_10}
{V{\'e}ron-Cetty} M.-P.,  {V{\'e}ron} P.,  2010, \mn@doi [\aap]
  {10.1051/0004-6361/201014188}, \href
  {http://adsabs.harvard.edu/abs/2010A%26A...518A..10V} {518, A10}

\bibitem[\protect\citeauthoryear{{Vetukhnovskaya}, {Gabuzda}  \&
  {Yakimov}}{{Vetukhnovskaya} et~al.}{2011}]{2011ARep...55..400V}
{Vetukhnovskaya} Y.~N.,  {Gabuzda} D.~C.,   {Yakimov} V.~E.,  2011, \mn@doi
  [Astronomy Reports] {10.1134/S1063772911050088}, \href
  {http://adsabs.harvard.edu/abs/2011ARep...55..400V} {55, 400}

\bibitem[\protect\citeauthoryear{{Vlahakis} \& {K{\"o}nigl}}{{Vlahakis} \&
  {K{\"o}nigl}}{2004}]{vlahakis_konigl_04}
{Vlahakis} N.,  {K{\"o}nigl} A.,  2004, \mn@doi [\apj] {10.1086/382670}, \href
  {http://adsabs.harvard.edu/abs/2004ApJ...605..656V} {605, 656}

\bibitem[\protect\citeauthoryear{{Wardle}, {Cawthorne}, {Roberts}  \&
  {Brown}}{{Wardle} et~al.}{1994}]{1994ApJ...437..122W}
{Wardle} J.~F.~C.,  {Cawthorne} T.~V.,  {Roberts} D.~H.,   {Brown} L.~F.,
  1994, \mn@doi [\apj] {10.1086/174980}, \href
  {http://adsabs.harvard.edu/abs/1994ApJ...437..122W} {437, 122}

\bibitem[\protect\citeauthoryear{{Zamaninasab}, {Savolainen}, {Clausen-Brown},
  {Hovatta}, {Lister}, {Krichbaum}, {Kovalev}  \& {Pushkarev}}{{Zamaninasab}
  et~al.}{2013}]{zamaninasab_etal13}
{Zamaninasab} M.,  {Savolainen} T.,  {Clausen-Brown} E.,  {Hovatta} T.,
  {Lister} M.~L.,  {Krichbaum} T.~P.,  {Kovalev} Y.~Y.,   {Pushkarev} A.~B.,
  2013, \mn@doi [\mnras] {10.1093/mnras/stt1816}, \href
  {http://adsabs.harvard.edu/abs/2013MNRAS.436.3341Z} {436, 3341}

\bibitem[\protect\citeauthoryear{{Zamaninasab}, {Clausen-Brown}, {Savolainen}
  \& {Tchekhovskoy}}{{Zamaninasab} et~al.}{2014}]{zamaninasab_etal14}
{Zamaninasab} M.,  {Clausen-Brown} E.,  {Savolainen} T.,   {Tchekhovskoy} A.,
  2014, \mn@doi [\nat] {10.1038/nature13399}, \href
  {http://adsabs.harvard.edu/abs/2014Natur.510..126Z} {510, 126}

\bibitem[\protect\citeauthoryear{{Zavala} \& {Taylor}}{{Zavala} \&
  {Taylor}}{2003}]{zavala_taylor_03}
{Zavala} R.~T.,  {Taylor} G.~B.,  2003, \mn@doi [\apj] {10.1086/374619}, \href
  {http://adsabs.harvard.edu/abs/2003ApJ...589..126Z} {589, 126}

\bibitem[\protect\citeauthoryear{{Zavala} \& {Taylor}}{{Zavala} \&
  {Taylor}}{2004}]{zavala_taylor_04}
{Zavala} R.~T.,  {Taylor} G.~B.,  2004, \mn@doi [\apj] {10.1086/422741}, \href
  {http://adsabs.harvard.edu/abs/2004ApJ...612..749Z} {612, 749}

\makeatother
\end{thebibliography}


\appendix
\section{Supporting information}

Additional information may be found in the online version of the article:
\begin{enumerate}
\item Table 3. Measured properties of the sources on Stokes $I$ and linear polarization maps
\item Figure 2. Faraday rotation measure maps and linear polarizaion properties for all of the sources
\item Figure 5. Faraday-corrected maps of electric vector position angle for all of the sources across the 1.4~GHz to 15.4~GHz range
\item Figure 8. Faraday-corrected maps of electric vector position angle and linear fractional polarization for the individual sources at 15~GHz
\end{enumerate}


\bsp
\label{lastpage}
\end{document}